\documentclass[12pt]{article}

\usepackage{tikz}
\usetikzlibrary{matrix,arrows,decorations.pathmorphing,decorations.markings,patterns}

\tikzstyle arrowstyle=[scale=1]
\tikzstyle directed=[postaction={decorate,decoration={markings,
    mark=at position .65 with {\arrow[arrowstyle]{stealth}}}}]
\tikzstyle reverse directed=[postaction={decorate,decoration={markings,
    mark=at position .65 with {\arrowreversed[arrowstyle]{stealth};}}}]
    
\usepackage[hmargin=3cm,vmargin=2cm]{geometry}
\usepackage{calrsfs}
\usepackage[vcentermath]{youngtab}
\usepackage{ytableau}
\usepackage{comment}

\usepackage{subfig}

\usepackage{jheppub}

\usepackage{amsmath,amssymb,euscript,array,mathrsfs,amsfonts}
\usepackage{epsfig}

\newcommand{\field}[1]{\mathbb{#1}} 
\newcommand*{\Dsl}[0]{{\rlap{\kern2.25pt /}{D}}}
\newcommand*{\Asl}[0]{{\rlap{\kern2.25pt /}{A}}}
\newcommand*{\dsl}[0]{{\rlap{\kern0.5pt /}{\partial}}}
\newcommand*{\xisl}[0]{{\rlap{\kern0.5pt /}{\xi}}}
\newcommand*{\asl}[0]{{\rlap{\kern0.5pt /}{a}}}
\newcommand*{\bsl}[0]{{\rlap{\kern0.5pt /}{b}}}

\newcommand*{\tr}[0]{{\rm tr}}

\newcommand{\gapprox}{\raisebox{-0.5ex}{$\
\stackrel{\textstyle>}{\textstyle\sim}\ $}}

\def\Dslash{\,\,{\raise.15ex\hbox{/}\mkern-12mu D}}

\newcommand{\SP}[1]{\begin{equation}\begin{split} #1
\end{split}\end{equation}}

\def\B0{{\boldsymbol 0}}

\def\tr{{\rm tr}}

\def\det{{\rm det}}

\def\Dbarslash{\,\,{\raise.15ex\hbox{/}\mkern-12mu {\bar D}}}
\def\Dslash{\,\,{\raise.15ex\hbox{/}\mkern-12mu D}}
\def\delslash{\,\,{\raise.15ex\hbox{/}\mkern-9mu \partial}}
\def\delbarslash{\,\,{\raise.15ex\hbox{/}\mkern-9mu {\bar\partial}}}

\def\VEV#1{\left\langle #1\right\rangle}

\newcommand{\proj}[4]{\mathbb{P}_{\foreach \index in {1,...,#2}{#3_\index}}^{#1 \ \ \foreach \index in {1,...,#2}{#4_\index}}}
\newcommand{\projnc}[5]{\mathbb{P}_{\foreach \index in {1,...,#2}{#3_\index}}^{#1 \ \ \foreach \index in #5 {#4_\index}}}

\newcommand{\tens}[4]{#1_{\foreach \index in {1,...,#2}{#3_\index}}^{\ \ \foreach \index in {1,...,#2}{#4_\index}}}

\newcommand{\Useries}[4]{\foreach \index in {1,...,#2}{#1_{#3_\index}^{\ #4_\index}}}

\newcommand{\EQ}[1]{\begin{equation}\begin{split} #1
\end{split}\end{equation}}

\title{Calculating the chiral condensate of QCD at infinite coupling using a generalised lattice diagrammatic approach}
\author[a]{Alexander S. Christensen,}
\author[a]{Joyce C. Myers,}
\author[a]{Peter D. Pedersen,}
\author[b]{Jan Rosseel}
\affiliation[a]{Niels Bohr Institute, Blegdamsvej 17, 2100 Copenhagen, Denmark}
\affiliation[b]{Vienna University of Technology, Wiedner Hauptstr., 8-10/136, A-1040 Vienna, Austria}
\emailAdd{xander@nbi.dk, jcmyers@nbi.dk, peter.pedersen@nbi.dk}
\emailAdd{rosseelj@hep.itp.tuwien.ac.at}
\abstract{We develop a lattice diagrammatic technique for calculating the chiral condensate of QCD at infinite coupling inspired by recent work of Tomboulis and earlier work from the 80's. The technique involves calculating the contribution of gauge link diagrams formed from all possible combinations of a number of sub-diagram types. This is achieved by performing a resummation, using a truncated number of sub-diagram types. We show how to calculate the relevant sub-diagrams, including a new technique for evaluating group integrals with arbitrary number of gauge link elements, using Young Projectors. Including up to four different diagram types we calculate the chiral condensate as a function of $N_f$, and show that two real solutions result, which are non-zero for all integer $N_f$. We analyse these solutions and find signs of convergence of the expansion at small $N_f$. We discuss sources of error associated with this approach in detail and implement a technique to reduce over-counting of diagrams.}

\notoc
\begin{document}

\maketitle

\newpage


\section{Introduction}

Until recently, it was thought that the chiral condensate of QCD at infinite coupling would remain non-zero for any number of fundamental fermion flavours $N_f$. This is in contrast to the restoration of chiral symmetry which is observed at some critical $N_f$ for more moderate couplings, resulting in the appearance of a conformal window (see for example \cite{Deuzeman:2012ee,Cheng:2013xha,Fodor:2011tu,Lin:2012iw,Itou:2013faa,Bursa:2010xn} for a selection of lattice simulation results with fundamental representation fermions). The belief that the chiral symmetry remains broken for $g = \infty$ is based on the results of a few studies in the $80$'s. Among these is the work of \cite{KlubergStern:1982bs}, in which the authors calculate the normalized chiral condensate $\frac{1}{N_f N_c} \langle {\bar \psi} \psi \rangle$ from a $1/d$ expansion. They obtain a non-zero result which is independent of $N_f$ for the first two orders in the expansion.

The approach in \cite{KlubergStern:1982bs} is considered to be reliable. In the limit $N_f \rightarrow 0$ the normalized chiral condensate approaches the result in \cite{Blairon:1980pk}, which employed a quite different analytic lattice diagrammatic approach, up to ${\cal O}(1/d)$ corrections. Subsequently, the diagrammatic lattice approach of \cite{Blairon:1980pk} was extended in \cite{Martin:1982tb} by systematically removing certain diagrams which lead to over-counting. In this way the authors in \cite{Martin:1982tb} obtain a result for $\frac{1}{N_f N_c}\langle {\bar \psi} \psi \rangle$ as $N_f \rightarrow 0$, which is equivalent to that in \cite{KlubergStern:1982bs}, including the ${\cal O}(1/d)$ corrections.

More recently, lattice simulations have been performed with $g = \infty$ and the chiral condensate was obtained as a function of $N_f$ \cite{deForcrand:2012vh}. Surprisingly these simulations on $4^4$ and $6^4$ lattices indicate that the chiral condensate drops discontinuously to a value close to zero at a critical value of $N_f \sim 13$ staggered flavours. These results are clearly in contrast with the results in \cite{KlubergStern:1982bs} from the $1/d$ expansion. Moreover, the authors of \cite{deForcrand:2012vh} also show that in contrast to their simulation results, a mean field calculation \cite{Damgaard:1985bn} of the critical temperature $T_c$, above which chiral symmetry is expected to be restored, gives a non-zero result for all $N_f$.

Shortly after the simulation results in \cite{deForcrand:2012vh} appeared, the presence of a possible transition in the chiral condensate at some critical $N_f$ at infinite coupling was also indicated using a lattice diagrammatic approach in \cite{Tomboulis:2012nr}. The approach used in \cite{Tomboulis:2012nr} is an extension of the earlier works \cite{Blairon:1980pk,Martin:1982tb}, to the case of $N_f \ne 0$, by including in the resummation a second type of ``mesonic" graph (each bond in the diagram contains one gauge link $U$ and one gauge link $U^{\dag}$), which contains a closed loop, contributing an $N_f$-dependence. The result is that the normalized chiral condensate is non-zero up to a critical value of $N_f \sim 10.7$ staggered flavours, beyond which only complex-valued solutions exist.

The motivation of this work is to examine the effect on $\frac{1}{N_f N_c}\langle {\bar \psi} \psi \rangle$ of including various different types of diagrams in an approach which is inspired by \cite{Tomboulis:2012nr}, and is also an extension of \cite{Blairon:1980pk,Martin:1982tb}. Focussing specifically on generalising \cite{Martin:1982tb}, the different types of base diagrams are resummed in a hopping expansion, to form all possible diagrams made out of these building blocks, and from these obtain the chiral condensate. Our results indicate that, up to the order at which we work, there are multiple solutions for the normalized chiral condensate as a function of $N_f$. Only one of these solutions has a sensible $N_f \rightarrow 0$ limit, matching onto the results of \cite{KlubergStern:1982bs,Martin:1982tb}. This solution for the chiral condensate approaches zero extremely slowly as a function of $N_f$, and there is no sign of any discontinuity, or of chiral symmetry restoration at any finite $N_f$. However, one can show that there is a second solution for the chiral condensate which is much larger at small $N_f$, and decreases more rapidly towards zero as $N_f$ increases. There is also no discontinuity or chiral symmetry restoration at any $N_f$ for the second solution. However, it cannot be ruled out that the chiral condensate jumps from one of these solutions to the other at some critical $N_f$.

As a technical by-product of this work, we will present a technique for evaluating group integrals, using Young projectors. Indeed, in order to calculate higher order diagrams with multiple overlapping gauge links $U$ and $U^\dag$, it becomes necessary to evaluate SU$(N_c)$ group integrals of the form
\begin{equation}
\int_{\mathrm{SU}(N_c)} d U \, U_{a_1}^{\ b_1} \cdots U_{a_m}^{\ b_m} (U^\dag)_{c_1}^{\ d_1} \cdots (U^\dag)_{c_n}^{\ d_n} \, ,
\end{equation}
for some number of $U_{a}^{\ b}$, $(U^\dag)_{c}^{\ d}$. We propose a simplified technique for evaluating this type of integral, using Young projectors. We comment on how this technique is related to previous approaches that appeared in \cite{Bars:1979xb,Creutz:1984mg,Cvitanovic:2008zz,Wilson:1975id}.

The outline of this paper is as follows. In Section \ref{tomboulis}, we will review how the chiral condensate at infinite coupling can be obtained from a lattice diagrammatic expansion \cite{Tomboulis:2012nr}. In Section \ref{martin-siu}, we will explain how the diagrammatic expansion can be resummed in a hopping expansion, that allows one to calculate the normalized chiral condensate from irreducible diagrams. Here, we generalize the analysis of \cite{Martin:1982tb}, that only included $N_f$-independent tree graph contributions (that enclose zero area), to include irreducible diagrams that are built out of $N_f$-dependent base sub-diagrams that no longer lead to tree graphs. The relevant fundamental base sub-diagrams are given and calculated in Section \ref{diags}. In Section \ref{groupsection} we comment on various techniques to calculate SU$(N_c)$ group integrals and explain a technique to evaluate these integrals in terms of Young projectors. In Section \ref{errorsection}, we discuss sources of error that are associated with our techniques and we show how over-counting of diagrams can be reduced. Our results are contained in Section \ref{results}, where we also compare our methods with the ones used in \cite{Tomboulis:2012nr}. We conclude in Section \ref{concl}.

\section{Expansion of $\langle {\bar \psi} \psi \rangle$ at $g = \infty$}
\label{tomboulis}

Our objective is to investigate the behaviour of the chiral condensate as a function of the number of fermion flavours $N_f$. To extend the procedure of obtaining $\frac{1}{N_f N_c}\langle {\bar \psi} \psi \rangle$ in \cite{Blairon:1980pk,Martin:1982tb}, for $N_f  \rightarrow 0$, and in \cite{Tomboulis:2012nr} for $N_f \ne 0$, to systematically account for the contributions which dominate in a diagrammatic expansion, order by order, it is necessary to understand how the diagrams contribute mathematically.  Using the notation in \cite{Tomboulis:2012nr}, the chiral condensate $\langle {\bar \psi} \psi \rangle$ is obtained from
\EQ{
\langle {\bar \psi}(x) \psi(x) \rangle = - \lim_{m\rightarrow 0}  \partial_m \log Z \, ,
}
where the partition function $Z$ (after integrating out the fermion fields) is given by
\EQ{
Z = \int dU\,  \det\left[ 1 + K^{-1} M(U) \right] \, ,
\label{defZ}
}
with
\EQ{
M_{xy} \equiv \frac{1}{2} \sum_\mu \left[ \gamma_{\mu} U_{\mu}(x) \delta_{y,x+{\hat \mu}} - \gamma_{\mu} U_{\mu}^{\dag}(x-{\hat \mu}) \delta_{y,x-{\hat \mu}} \right] \, ,
}
\EQ{
K_{xy} = m {\field I}_{N_f} {\field I}_{N_c} \delta_{x y} \, ,
}
for $\mu = 1, ..., d$, including $N_f$ fermion flavours, and $N_c$ colours. The chiral condensate is thus given by \cite{Tomboulis:2012nr}
\EQ{
\langle {\bar \psi}(x) \psi(x) \rangle = - \lim_{m\rightarrow 0} \tr \left[ G(x,x) \right]\, ,
\label{2ptcorr}
}
where
\EQ{
G(x,x) = \frac{\int dU \, \det\left[ 1 + K^{-1} M(U) \right] \left[ \left[ 1 + K^{-1} M(U) \right]^{-1} K^{-1} \right]_{xx}}{\int dU \, \det\left[ 1 + K^{-1} M(U) \right]} \, .
\label{Gxx}
}

Expanding in powers of $K^{-1} M(U)$ one obtains
\EQ{
\det\left[ 1 + K^{-1} M \right] = \exp \tr \left[ \sum_{n=1}^{\infty} \frac{(-1)^{n+1}}{n} (K^{-1} M)^n \right] \, ,
\label{detD}
}
\EQ{
\left[ \left[ 1 + K^{-1} M \right]^{-1} K^{-1} \right]_{xx} = \frac{1}{m} \left[ \sum_{n=0}^{\infty} (-1)^n (K^{-1}M)^n \right]_{xx} \, .
}
Note that $\tr \left[ \text{odd \# of} ~\gamma_{\mu}\text{'s} \right] = 0$ implies that only contributions from $(K^{-1}M)^n$ with $n$ even contribute to the integrals in (\ref{Gxx}). The trace in (\ref{detD}) (and (\ref{2ptcorr})) extends over colour, flavour, and spinor degrees of freedom. For example,
\SP{
\left[ (K^{-1} M)^2\right]_{xx} = &\frac{1}{(2 m)^2} \sum_{\mu, \nu} \sum_{y} \left[ \gamma_{\mu} \gamma_{\nu} \right] \\
&\times \left[ U_{\mu}(x) \delta_{y,x+{\hat \mu}} - U_{\mu}^{\dag}(x-{\hat \mu}) \delta_{y,x-{\hat \mu}} \right] \left[ U_{\nu}(y) \delta_{x,y+{\hat \nu}} - U_{\nu}^{\dag}(y-{\hat \nu}) \delta_{x,y-{\hat \nu}} \right] \, ,
}
and so on. In general, the trace in \eqref{detD} leads to a closed loop of link variables, because the first and last lattice site are identified. Each loop also comes with a factor of $N_f$. The traces over the gamma matrices can be determined from
\EQ{
\{\gamma_{\mu}, \gamma_{\nu}\} = 2 \delta_{\mu \nu} {\field I}_{N_s} \, ,
}
where $\gamma_{\mu}$ are the Euclidean gamma matrices and $N_s$ denotes the number of spinor degrees of freedom.

It is also useful to notice that certain types of contributions will lead to cancellations with the denominator in \eqref{Gxx}. Since all diagrams resulting from the determinant are closed loops, the contributions to $\langle \bar{\psi} \psi \rangle$ which cancel are closed loop diagrams which can be disconnected from the path of gauge links beginning and ending at $x$. For example, in the diagram

\noindent
\begin{equation}
\begin{tikzpicture}[baseline=(current  bounding  box.center),scale=0.7]

\def \xoff {0}

\draw [directed] (0.0+\xoff,0.0) -- (0.0+\xoff,2.0) -- (2.0+\xoff,2.0) -- (2.0,0.0) -- (0.2,0.0);
\draw [reverse directed] (0.2+\xoff,0.2) rectangle (1.8+\xoff,1.8);

\def \xoff {4.5}

\draw [directed] (0.0+\xoff,0.0) rectangle (2.0+\xoff,2.0);
\draw [reverse directed] (0.2+\xoff,0.2) rectangle (1.8+\xoff,1.8);

\end{tikzpicture}\ \ \ ,
\end{equation}
the closed loops on the right cancel with a contribution from the denominator. Note that this would even be true when there is partial overlap with links coming from $\left[ \left[ 1 + K^{-1} M \right]^{-1} K^{-1} \right]_{xx}$, as in 

\noindent
\begin{equation}
\begin{tikzpicture}[baseline=(current  bounding  box.center),scale=0.7]

\def \xoff {0}

\draw [directed] (0.0+\xoff,0.0) -- (0.0+\xoff,2.0) -- (2.0+\xoff,2.0) -- (2.0,0.0) -- (0.2,0.0);
\draw [reverse directed] (0.2+\xoff,0.2) rectangle (1.8+\xoff,1.8);

\def \xoff {2.2}

\draw [directed] (0.0+\xoff,0.0) rectangle (2.0+\xoff,2.0);
\draw [reverse directed] (0.2+\xoff,0.2) rectangle (1.8+\xoff,1.8);

\node[anchor=west] at (2.3+\xoff,1.0) {\Large{
$= \frac{1}{N_c}$
}};

\def \xoff {7.4}

\draw [directed] (0.0+\xoff,0.0) -- (0.0+\xoff,2.0) -- (2.0+\xoff,2.0) -- (2.0+\xoff,0.0) -- (0.2+\xoff,0.0);
\draw [reverse directed] (0.2+\xoff,0.2) rectangle (1.8+\xoff,1.8);

\def \xoff {9.6}

\draw [directed] (0.0+\xoff,0.0) -- (0.0+\xoff,2.0) -- (0.2+\xoff,2.0) -- (0.2+\xoff,0.0) -- (0.0+\xoff,0.0);

\node[anchor=west] at (0.3+\xoff,1.0) {\large{
$=$
}};

\def \xoff {11.8}

\draw [directed] (0.0+\xoff,0.0) -- (0.0+\xoff,2.0) -- (2.0+\xoff,2.0) -- (2.0+\xoff,0.0) -- (0.2+\xoff,0.0);
\draw [reverse directed] (0.2+\xoff,0.2) rectangle (1.8+\xoff,1.8);

\end{tikzpicture}\ \ \ ,
\end{equation}
\noindent where the second equality is obtained by using
\EQ{
U_{a}^{\ b} (U^{\dag})_{b}^{\ c} = \delta_a^c \,,
}
due to the unitarity of the $U$'s. So one sees that, even in this case, where there is partial overlap, the integrations can be separated.

\section{Building $\langle {\bar \psi} \psi \rangle$ from irreducible diagrams}
\label{martin-siu}

To generalise the diagram building procedure of \cite{Martin:1982tb} we calculate the chiral condensate obtained (from $\langle {\bar \psi} \psi \rangle = - \lim_{m\rightarrow0} \tr[G(x,x)]$) by performing a hopping expansion, summing over gauge links order by order in the number of links
\EQ{
\frac{\tr [G(x,x)]}{N_s N_f N_c} = \frac{1}{m} \sum_{L=0}^{\infty} (-1)^L \frac{A(L)}{(2 m)^{2L}} \, ,
}
where $A(L)$ is the contribution from all graphs with $2 L$ links which start and end at some site $x$. A general graph can be obtained by combining irreducible graphs $I(l)$ of $2 l$ links which start and end at $x$, where an irreducible graph is defined as one that cannot be separated into smaller segments which start and end at $x$.

\begin{minipage}{0.5\textwidth}

\begin{tikzpicture}[scale=0.5]

\def \xoff {8.0}
\def \yoff {0.0}

\draw [directed] (0.0+\xoff,0.0+\yoff) -- (0.0+\xoff,1.8+\yoff);
\draw [directed] (0.0+\xoff,1.8+\yoff) -- (-1.0+\xoff,3.2+\yoff) -- (-0.8+\xoff,3.4+\yoff) -- (0.2+\xoff,1.9+\yoff) -- (0.4+\xoff,3.8+\yoff) -- (0.6+\xoff,3.7+\yoff) -- (0.4+\xoff,2.0+\yoff) -- (2.1+\xoff,2.8+\yoff) -- (2.3+\xoff,2.6+\yoff) -- (0.2+\xoff,1.7+\yoff);
\draw [directed] (0.2+\xoff,1.7+\yoff) -- (0.2+\xoff,0.0+\yoff);

\node[anchor=west] at (\xoff-0.5,-0.5) {\Large{
$x$
}};

\node[anchor=west] at (\xoff-2.0,5.0) {\Large{
Irreducible
}};

\end{tikzpicture}\\
\end{minipage}
\begin{minipage}{0.5\textwidth}

\begin{tikzpicture}[scale=0.5]

\def \xoff {14.0}

\draw (0.0+\xoff,0.0) -- (-1.0+\xoff,1.4) -- (-0.8+\xoff,1.6) -- (0.2+\xoff,0.25);
\draw [directed] (0.2+\xoff,0.25) -- (2.1+\xoff,1.0) -- (2.3+\xoff,2.9) -- (2.5+\xoff,2.8) -- (2.3+\xoff,0.8) -- (0.2+\xoff,0.0);

\node[anchor=west] at (\xoff-0.5,-0.5) {\Large{
$x$
}};

\node[anchor=west] at (\xoff-1.7,4.0) {\Large{
Reducible
}};

\end{tikzpicture}
\end{minipage}

\noindent The contribution $A(L)$ obeys the recursion relation 
\EQ{
A(L) = \sum_{l=1}^{L} I(l) A(L-l) \, , \hspace{1cm} L \ge 1 \, ; \hspace{1cm} A(0) = 1 \, ,
}
where the irreducible graphs are built iteratively out of all possible combinations of smaller segments
\SP{ \label{IL}
I(L) = &2 d F_0(L-1) - 4 d(d-1) \frac{N_f}{N_c} F_1(L-4)^7 + ... \,,
}
with $I(0) = 0$, and the quantity $F_n(L)$ represents all possible graphs of length $2 L$ which start and end on a site on a sub-diagram of area $n$. It is given by
\EQ{ \label{Fn}
F_n(L) = \sum_{\substack{l_i = 1, 2, ... ,\\ k_j = 4, 8, ...,\\ \sum l_i + k_j = L - 1}} I_a(l_1) I_a(l_2) ... I_a(l_p) I_b(k_1) I_b(k_2) ... I_b(k_q) ... ~ \widehat{a}_n^p\, \widehat{b}_n^q ... \, ,
} 
with $F_n(0) = 1$. In this formula, $I_a$ refers to irreducible graphs which begin with an `$a$-type' sub-diagram,
\begin{tikzpicture}[scale=0.3]

\draw [directed] (0.0,0.0) -- (0.0,2.0);
\draw (0.0,2.0) -- (0.2,2.0);
\draw [directed] (0.2,2.0) -- (0.2,0.0);

\end{tikzpicture}
,  and $I_b$ refers to irreducible graphs which begin with a $L=4$ box, that is a `$b$-type' sub-diagram,
\begin{tikzpicture}[scale=0.3]

\def \xoff {0.0}
\def \yoff {0.0}

\draw [directed] (0.0+\xoff,0.0+\yoff) -- (0.0+\xoff,2.0+\yoff) -- (2.0+\xoff,2.0+\yoff) -- (2.0+\xoff,0.0+\yoff) -- (0.2+\xoff,0.0+\yoff);
\draw [reverse directed] (0.2+\xoff,0.2+\yoff) rectangle (1.8+\xoff,1.8+\yoff);

\end{tikzpicture}
. Further types of sub-diagrams that can appear at larger $L$ will be denoted by `$c$-type', `$d$-type', ... and will be defined later on in Section \ref{diags}. In \eqref{Fn}, we have also introduced the notation ${\widehat x}_n \equiv \frac{x_n}{d_x}$, where $x_n$ is the dimensionality of an attachment of type $x$ to an area $n$ diagram, and $d_x$ is the total dimensionality of a type $x$ diagram. These are catalogued in Appendix \ref{A}.  For example,
\EQ{
{\widehat a}_0 = \frac{2d-1}{2d} \, ,
}
\EQ{
{\widehat b}_0 = \frac{4(d-1)^2}{4d(d-1)} = \frac{d-1}{d} \, .
}
In particular, an $a$-type sub-diagram,
\begin{tikzpicture}[scale=0.3]

\draw [directed] (0.0,0.0) -- (0.0,2.0);
\draw (0.0,2.0) -- (0.2,2.0);
\draw [directed] (0.2,2.0) -- (0.2,0.0);

\end{tikzpicture}
attaches with dimensionality $2 d {\widehat a}_n$, to a graph of area $n$. All ``tree" graphs are of this type (tree graphs don't include internal plaquettes). A $b$-type sub-diagram,
\begin{tikzpicture}[scale=0.3]

\def \xoff {0.0}
\def \yoff {0.0}

\draw [directed] (0.0+\xoff,0.0+\yoff) -- (0.0+\xoff,2.0+\yoff) -- (2.0+\xoff,2.0+\yoff) -- (2.0+\xoff,0.0+\yoff) -- (0.2+\xoff,0.0+\yoff);
\draw [reverse directed] (0.2+\xoff,0.2+\yoff) rectangle (1.8+\xoff,1.8+\yoff);

\end{tikzpicture}
attaches with dimensionality $4d(d-1) {\widehat b}_n$, to a graph of area $n$, such as $b$-type diagrams attached to $a$-type diagrams or other area $1$ diagrams. The specific forms of ${\widehat a}_n, {\widehat b}_n, ...$ have been determined to avoid over-counting of graphs \footnote{Regardless, there is some over-counting of attachments to certain winding diagrams, which will be discussed later.}.

As an illustration of \eqref{IL} and \eqref{Fn}, we note that the irreducible graphs $I(L)$ have the following form\\
\begin{flalign}
\begin{tikzpicture}[baseline=(current  bounding  box.center),scale=0.7]
\node[anchor=east] at (0.2,1.0) {\large{
$I(1) = ~$
}};
\draw [directed] (0.0,0.0) -- (0.0,2.0);
\draw (0.0,2.0) -- (0.2,2.0);
\draw [directed] (0.2,2.0) -- (0.2,0.0);
\node[anchor=west] at (0.2,1.0) {\large{
$= I_a(1) = 2 d$\ ,
}};
\end{tikzpicture}&&
\end{flalign}

\noindent
\begin{flalign}
\begin{tikzpicture}[baseline=(current  bounding  box.center),scale=0.7]
\node[anchor=east] at (0.2,1.0) {\large{
$I(2) = ~$
}};
\draw [directed] (0.0,0.0) -- (0.0,2.0);
\draw [directed] (0.0,2.0) -- (2.0,2.7) -- (2.1,2.5) -- (0.2,1.8);
\draw [directed] (0.2,1.8) -- (0.2,0.0);
\node[anchor=west] at (0.2,1.0) {\large{
$= I_a(2) = 2 d ~  \left[ I_a(1) ~ {\widehat a}_0 \right]$\ ,
}}; 
\end{tikzpicture}&&
\end{flalign}

\noindent
\begin{flalign}
\begin{tikzpicture}[baseline=(current  bounding  box.center),scale=0.7]
\def \xoff {6.0}
\node[anchor=east] at (0.2,1.0) {\large{
$I(3) = ~$
}};
\draw [directed] (0.0,0.0) -- (0.0,2.0);
\draw [directed] (0.0,2.0) -- (1.9,2.7) -- (1.5,4.5) -- (1.7,4.5) -- (2.2,2.5) -- (0.2,1.8);
\draw [directed] (0.2,1.8) -- (0.2,0.0);
\node[anchor=west] at (0.2,1.0) {\large{
$\hspace{20mm}+$
}};
\draw [directed] (0.0+\xoff,0.0) -- (0.0+\xoff,1.8);
\draw [directed] (0.0+\xoff,1.8) -- (-1.5+\xoff,3.2) -- (-1.4+\xoff,3.4) -- (0.1+\xoff,2.0) -- (1.9+\xoff,2.7) -- (2.1+\xoff,2.5) -- (0.2+\xoff,1.8);
\draw [directed] (0.2+\xoff,1.8) -- (0.2+\xoff,0.0);
\node[anchor=west] at (0.2+\xoff,1.0) {\large{
$~~= I_a(3) = 2 d \left[ I_a(2) {\widehat a}_0 + I_a(1)^2 {\widehat a}_0^2 \right]$\ ,
}}; 
\end{tikzpicture}&&
\end{flalign}

\noindent
\begin{flalign}
\begin{tikzpicture}[baseline=(current  bounding  box.center),scale=0.7]
\def \xoff {6.0}
\node[anchor=east] at (0.2,1.0) {\large{
$I(4) = ~$
}};
\draw [directed] (0.0,0.0) -- (0.0,2.0);
\draw [directed] (0.0,2.0) -- (1.9,2.7) -- (1.5,4.5) --(3.0,6.2) -- (3.2,6.0) -- (1.8,4.5) -- (2.2,2.5) -- (0.2,1.8);
\draw [directed] (0.2,1.8) -- (0.2,0.0);
\node[anchor=west] at (0.2,1.0) {\large{
$\hspace{20mm}+$
}};
\draw [directed] (0.0+\xoff,0.0) -- (0.0+\xoff,1.8);
\draw [directed] (0.0+\xoff,1.8) -- (1.9+\xoff,2.7) -- (0.5+\xoff,4.2) -- (0.6+\xoff,4.4) -- (2.1+\xoff,2.8) -- (3.0+\xoff,4.7) -- (3.2+\xoff,4.6) -- (2.2+\xoff,2.6) -- (0.2+\xoff,1.7);
\draw [directed] (0.2+\xoff,1.7) -- (0.2+\xoff,0.0);
\def \xoff {12.0}
\node[anchor=west] at (0.2+6.0,1.0) {\large{
$\hspace{20mm}+ ~~2$
}};
\draw [directed] (0.0+\xoff,0.0) -- (0.0+\xoff,1.8);
\draw [directed] (0.0+\xoff,1.8) -- (-1.0+\xoff,3.2) -- (-0.8+\xoff,3.4) -- (0.2+\xoff,1.9) -- (2.1+\xoff,2.8) -- (2.3+\xoff,4.7) -- (2.5+\xoff,4.6) -- (2.3+\xoff,2.6) -- (0.2+\xoff,1.7);
\draw [directed] (0.2+\xoff,1.7) -- (0.2+\xoff,0.0);
\def \xoff {2.0}
\def \yoff {-4.0}
\node[anchor=west] at (0.2-3.0,1.0+\yoff) {\large{
$\hspace{20mm}+$
}};
\draw [directed] (0.0+\xoff,0.0+\yoff) -- (0.0+\xoff,1.8+\yoff);
\draw [directed] (0.0+\xoff,1.8+\yoff) -- (-1.0+\xoff,3.2+\yoff) -- (-0.8+\xoff,3.4+\yoff) -- (0.2+\xoff,1.9+\yoff) -- (0.4+\xoff,3.8+\yoff) -- (0.6+\xoff,3.7+\yoff) -- (0.4+\xoff,2.0+\yoff) -- (2.1+\xoff,2.8+\yoff) -- (2.3+\xoff,2.6+\yoff) -- (0.2+\xoff,1.7+\yoff);
\draw [directed] (0.2+\xoff,1.7+\yoff) -- (0.2+\xoff,0.0+\yoff);
\def \xoff {7.0}
\def \yoff {-4.0}
\node[anchor=west] at (0.2+1.5,1.0+\yoff) {\large{
$\hspace{20mm}+$
}};
\draw [directed] (0.0+\xoff,0.0+\yoff) -- (0.0+\xoff,2.0+\yoff) -- (2.0+\xoff,2.0+\yoff) -- (2.0+\xoff,0.0+\yoff) -- (0.2+\xoff,0.0+\yoff);
\draw [reverse directed] (0.2+\xoff,0.2+\yoff) rectangle (1.8+\xoff,1.8+\yoff);
\node[anchor=west] at (0.2-2.5,-6.0) {\large{
$~~= I_a(4)+ I_b(4)$
}};
\node[anchor=west] at (0.2-2.5,-9.0) {\large{
$~~= 2 d \left[ I_a(3) {\widehat a}_0 + 2 I_a(1) I_a(2) {\widehat a}_0^2 + I_a(1)^3 {\widehat a}_0^3 \right] - 4d(d-1) \frac{N_f}{N_c}$\ ,
}};
\end{tikzpicture}&&
\end{flalign}

\begin{flalign}
\begin{tikzpicture}[baseline=(current  bounding  box.center),scale=0.7]
\def \xoff {6.0}
\node[anchor=east] at (0.2,1.0) {\large{
$...$\ .
}};
\end{tikzpicture}&&
\end{flalign}

The generating function, which gives the total contribution of all irreducible graphs including the mass dependence, is
\EQ{
W_I = \sum_{l=0}^{\infty} \left( -\frac{1}{4 m^2} \right)^l I(l) \, .
}
Using \eqref{IL} for the $I(l)$ and defining $x = - \frac{1}{4 m^2}$ results in
\EQ{
W_I = W_a + W_b + ... \, ,
}
where $W_a$ is all irreducible graphs starting with an $a$-type base diagram
\begin{tikzpicture}[scale=0.3]

\draw [directed] (0.0,0.0) -- (0.0,2.0);
\draw (0.0,2.0) -- (0.2,2.0);
\draw [directed] (0.2,2.0) -- (0.2,0.0);

\end{tikzpicture}
. $W_b$ is all irreducible graphs starting with a $b$-type base diagram
\begin{tikzpicture}[scale=0.3]

\def \xoff {0.0}
\def \yoff {0.0}

\draw [directed] (0.0+\xoff,0.0+\yoff) -- (0.0+\xoff,2.0+\yoff) -- (2.0+\xoff,2.0+\yoff) -- (2.0+\xoff,0.0+\yoff) -- (0.2+\xoff,0.0+\yoff);
\draw [reverse directed] (0.2+\xoff,0.2+\yoff) rectangle (1.8+\xoff,1.8+\yoff);

\end{tikzpicture}
, etc. These take the form
\EQ{
W_a = 2 d x \sum_{n=0}^{\infty} \left[ {\widehat a}_0 W_a + {\widehat b}_0 W_b + ... \right]^n = \frac{2 d x}{1 - {\widehat a}_0 W_a - {\widehat b}_0 W_b - ...} \, ,
}
\EQ{
W_b = -4d(d-1) \frac{N_f}{N_c} x^4 \left[ \sum_{n=0}^{\infty} \left[ {\widehat a}_1 W_a + {\widehat b}_1 W_b + ...\right]^n \right]^7 = \frac{-4d(d-1)\frac{N_f}{N_c} x^4}{(1 - {\widehat a}_1 W_a - {\widehat b}_1 W_b - ... )^7} \, ,
}
\EQ{
... \, .
}
where the ``$...$" include higher order (in $x$) base diagrams. The normalized chiral condensate is obtained by adding all possible combinations of irreducible graphs, such that
\EQ{
\frac{\langle \bar{\psi} \psi \rangle}{N_s N_f N_c} = \lim_{m\rightarrow0}\frac{\tr [G(x,x)]}{N_s N_f N_c} = \lim_{m \rightarrow 0} \frac{1}{m} \left( \frac{1}{1-W_I} \right) \, .
}
In order to take the massless limit it is convenient to introduce the variables $g_x \equiv - \frac{2 m W_x}{d_x}$, for dimensional pre-factors $d_a = 2 d$, $d_b = 4d(d-1)$, $d_c = 12d(d-1)(2d-3)$, ..., such that the chiral condensate can be obtained from
\EQ{
g \equiv d_a g_a + d_b g_b + ... \, ,
\label{gser}
}
with, taking $m \rightarrow 0$,
\EQ{
g_a = \frac{1}{a_0 g_a + b_0 g_b + ...} \, ,
\label{ga}
}
\EQ{
g_b = \frac{\frac{N_f}{N_c}}{(a_1 g_a + b_1 g_b + ...)^7} \, ,
\label{gb}
}
\EQ{
g_c = \frac{\frac{N_f}{N_c}}{(a_2 g_a + b_2 g_b + ...)^{11}} \, ,
}
\EQ{
... \, ,
\label{gall}
}
using
\EQ{
\lim_{m\rightarrow 0}\frac{\tr [G(x,x)]}{N_s N_f N_c} = \frac{2}{g} \, .
\label{chicong}
}
We derive the pre-factors $x_n$ in (\ref{ga}) - (\ref{gall}) in Section \ref{over-counting}. What we find is that the contributions to $g$ from the $g_x$ in general decrease in magnitude with increasing number of links in the base diagram (See Figure \ref{g-figs} in Section \ref{results}). Thus it appears that the series in (\ref{gser}) tends towards convergence.

A few comments are in order. First, it is useful to notice that for all $N_f$, diagram contributions with unit area will dominate over contributions with higher areas $n$. Since at leading order in $d$, the $x_n$ are the same for all $n$ and equivalent to $d_x$, then at this order the quantity $a_n g_a + b_n g_b + ...$ is independent of $n$ and equivalent to $d_a g_a + d_b g_b + ...$. In general the results in Section \ref{results} indicate that\footnote{In general we find in Section \ref{results} that $g > 1$ except at very small $N_f$ for solution $2$ when working only to order $L = 4$.}
\EQ{
g = d_a g_a + d_b g_b + ... > 1 \, .
}
This is already true at $N_f \rightarrow 0$, and the magnitude of $d_a g_a + d_b g_b + ... $ grows as a function of $N_f$, causing the magnitude of the chiral condensate to decrease. This implies that diagrams with a higher power of $(a_n g_a + b_n g_b + ...)^{-1}$ are suppressed at a fixed order in $N_f$. However, for sufficiently large $N_f$, diagrams which are higher order in $N_f$ will dominate regardless of whether they have higher powers of $(a_n g_a + b_n g_b + ...)^{-1}$. Therefore since larger areas result in more powers of $(a_n g_a + b_n g_b + ...)^{-1}$, at each order in $N_f$, the diagrams with the smallest area dominate.

In addition, the prefactors $x_n$ in the system of equations in \eqref{ga} - \eqref{gall} can be adjusted to reduce over-counting resulting from certain types of diagram attachments. The prefactors $x_n$ are derived in Section \ref{over-counting}, and tabulated in Appendix \ref{A}. These considerations are taken into account in the results for the normalized chiral condensate in Section \ref{results}. 

\section{Fundamental base diagrams}
\label{diags}

In this section we calculate the leading order fundamental base diagrams, from which irreducible graphs can be built. The contributions can be categorised based on the information in Sections \ref{tomboulis}, \ref{martin-siu}. The calculations include the following components:
\begin{itemize}
\item A factor $\frac{1}{i!} (-N_f N_s)^i$, for a number $i$, of overlapping closed internal loops,
\item A mass factor $\left( - \frac{1}{4m^2} \right)^n$, for $n$ pairs of links,
\item $(-1)^k$ for $k$ permutations of $\gamma$ matrices,
\item $\left[ ... \right]$, containing the result obtained by performing the group integrations,
\item $\{ ... \}$, containing the dimensionality of the graph. 
\end{itemize}

The group integrations can be performed using the techniques described in the next section (based on e.g.  \cite{Bars:1979xb,Creutz:1984mg,Cvitanovic:2008zz,Wilson:1975id}). For this section, we will in particular need the expressions \eqref{I1} and \eqref{I2}, that we repeat here for convenience:
\EQ{
\int_{SU(N_c)} d U \, U_{a}^{\ b} (U^\dag)_{c}^{\ d} = \frac{1}{N_c} \delta_{a}^{d} \delta_{c}^{b} \, ,
}
\begin{align}
& \int_{SU(N_c)} d U \, \Useries{U}{2}{a}{b} \Useries{(U^\dag)}{2}{c}{d} = \frac{1}{2 N_c(N_c+1)} \left( \delta_{a_1}^{d_1} \delta_{a_2}^{d_2} + \delta_{a_1}^{d_2} \delta_{a_2}^{d_1} \right) \left( \delta_{c_1}^{b_1} \delta_{c_2}^{b_2} + \delta_{c_1}^{b_2} \delta_{c_2}^{b_1} \right) \nonumber \\
&\qquad \qquad + \frac{1}{2 N_c(N_c-1)} \left(  \delta_{a_1}^{d_1} \delta_{a_2}^{d_2} - \delta_{a_1}^{d_2} \delta_{a_2}^{d_1}  \right) \left(\delta_{c_1}^{b_1} \delta_{c_2}^{b_2} - \delta_{c_1}^{b_2} \delta_{c_2}^{b_1}  \right) \, .
\end{align}
These integrals are sufficient to calculate diagrams with up to $4$ overlapping links. In the next section, we will explain in more generality how group integrals can be calculated. The techniques explained there will enable us to also calculate diagrams that contain more than $4$ overlapping links.

In the case of finite $N_c$, it is necessary to include additional `baryonic' contributions, arising from integrals \eqref{baryonint}
\begin{equation}
\int_{SU(N_c)} d U\, U_{a_1}^{\ b_1} \cdots U_{a_{N_c}}^{\ b_{N_c}} = \frac{1}{N_c!}\epsilon_{a_1 \cdots  a_{N_c}} \epsilon^{b_1\cdots b_{N_c}} \,.
\end{equation}
In the following, we will list such contributions explicitly for the case $N_c = 3$. We will moreover also restrict ourselves to the case of staggered fermions, for which $N_s = 1$ and for which backtracking of the gauge links results in non-zero contributions.

The base diagrams up to order $L = 9$ are as follows, where we also indicate the type the diagram belongs to.
\subsection{$L=1$ : `$a$-type'}

\begin{flalign}
\begin{tikzpicture}[baseline=(current  bounding  box.center),scale=0.7]
\draw [directed] (0.0,0.0) -- (0.0,2.0);
\draw (0.0,2.0) -- (0.2,2.0);
\draw [directed] (0.2,2.0) -- (0.2,0.0);
\node[anchor=west] at (0.2,1.0) {\large{$= - \frac{1}{4m^2} \{ 2d \}$}};
\end{tikzpicture}&&
\label{type-a}
\end{flalign}

\subsection{$L=4$ : `$b$-type'}

\begin{flalign} 
\begin{tikzpicture}[baseline=(current  bounding  box.center),scale=0.7]
\def \yoff {0}
\draw [directed] (0.0,0.0) -- (0.0,2.0) -- (2.0,2.0) -- (2.0,0.0) -- (0.2,0.0);
\draw [reverse directed] (0.2,0.2) rectangle (1.8,1.8);
\node[anchor=west] at (2.0,1.0) {\large{
$= \left( - \frac{1}{4m^2} \right)^4 (-1)^2 (-N_f) \left[ \frac{1}{N_c} \right] \{ 4d(d-1) \}$
}};
\end{tikzpicture}&&
\label{type-b}
\end{flalign}

\subsection{$L=6$}

\subsubsection{`$c$-type'}
\begin{flalign}
\begin{tikzpicture}[baseline=(current  bounding  box.center),scale=0.7]
\def \yoff {0}
\draw [directed] (0.0,0.0) -- (0.0,2.0) -- (4.0,2.0) -- (4.0,0.0) -- (0.2,0.0);
\draw [reverse directed] (0.2,0.2) rectangle (3.8,1.8);
\node[anchor=west] at (4.0,1.0) {\large{
$= \left( - \frac{1}{4m^2} \right)^6 (-N_f) \left[ \frac{1}{N_c} \right] \{ 12d(d-1)(2d-3) \}$
}};
\end{tikzpicture}&&
\end{flalign}

\subsubsection{$N_c = 3$ : `$d$-type'}

\begin{flalign}
\begin{tikzpicture}[baseline=(current  bounding  box.center),scale=0.7]
\def \yoff {0}
\draw [directed] (0.0,0.0) -- (0.0,2.0) -- (2.0,2.0) -- (2.0,0.0) -- (0.2,0.0);
\draw [directed] (0.2,0.2) rectangle (1.8,1.8);
\draw [directed] (0.4,0.4) rectangle (1.6,1.6);
\node[anchor=west] at (2.0,1.0) {\large{
$= \frac{1}{2!} \left( - \frac{1}{4m^2} \right)^6 (-1)^3 (-N_f)^2 \left[ \frac{1}{3} \right] \{ 4d(d-1) \}$
}};
\end{tikzpicture}&&
\label{type-d31}
\end{flalign}

\noindent
\begin{flalign}
\begin{tikzpicture}[baseline=(current  bounding  box.center),scale=0.7]
\def \yoff {0}
\draw [directed] (0.0,0.0) -- (0.0,2.0) -- (2.0,2.0) -- (2.0,0.0) -- (0.2,0.0) -- (0.2,1.8) -- (1.8,1.8) -- (1.8,0.2) -- (-0.2,0.2);
\draw [directed] (0.4,0.4) rectangle (1.6,1.6);
\node[anchor=west] at (2.0,1.0) {\large{
$= \left( - \frac{1}{4m^2} \right)^6 (-1)^3 (-N_f) \left[ -\frac{1}{3} \right] \{ 4d(d-1) \}$
}};
\end{tikzpicture}&&
\end{flalign}

\noindent
\begin{flalign}
\begin{tikzpicture}[baseline=(current  bounding  box.center),scale=0.7]
\def \yoff {0}
\draw [directed] (0.0,0.0) -- (0.0,2.0) -- (2.0,2.0) -- (2.0,0.0) -- (0.2,0.0)
                             -- (0.2,1.8) -- (1.8,1.8) -- (1.8,0.2) -- (0.4,0.2)
                             -- (0.4,1.6) -- (1.6,1.6) -- (1.6,0.4) -- (-0.4,0.4);
%
\node[anchor=west] at (2.0,1.0) {\large{
$= \left( - \frac{1}{4m^2} \right)^6 (-1)^3 \left[ \frac{1}{3} \right] \{ 4d(d-1) \}$
}};
\end{tikzpicture}&&
\end{flalign}

\noindent
\begin{flalign}
\begin{tikzpicture}[baseline=(current  bounding  box.center),scale=0.7]
\def \yoff {0}
\draw [directed] (0.0,0.0) -- (0.0,2.0) -- (2.0,2.0) -- (2.0,0.0) -- (0.2,0.0);
\draw [directed] (0.2,0.4) -- (0.2,1.8) -- (1.8,1.8) -- (1.8,0.2) -- (0.4,0.2) -- (0.4,1.6) -- (1.6,1.6) -- (1.6,0.4) -- (0.2,0.4);
%
\node[anchor=west] at (2.0,1.0) {\large{
$= \left( - \frac{1}{4m^2} \right)^6 (-1)^3 (-N_f) \left[ - \frac{1}{3} \right] \{ 4d(d-1) \}$
}};
\end{tikzpicture}&&
\label{type-d34}
\end{flalign}

\subsection{$L=7$ : `$e$-type'}

\begin{flalign}
\begin{tikzpicture}[baseline=(current  bounding  box.center),scale=0.7]
\def \yoff {0}
\draw [directed] (0.0,0.0) -- (0.0,2.0) -- (4.0,2.0) -- (4.0,0.0) -- (0.2,0.0);
\draw [reverse directed] (0.2,0.2) rectangle (1.9,1.8);
\draw [reverse directed] (2.1,0.2) rectangle (3.8,1.8);
\node[anchor=west] at (4.0,1.0) {\large{
$= \frac{1}{2!} \left( - \frac{1}{4m^2} \right)^7 (-1)^2 (-N_f)^2 \left[ \frac{1}{N_c^2} \right] \{ 12d(d-1)(2d-3) \}$
}};
\end{tikzpicture}&&
\end{flalign}

\subsection{$L=8$ }

\subsubsection{`$f$-type'}
\begin{flalign}
\begin{tikzpicture}[baseline=(current  bounding  box.center),scale=0.7]
\def \yoff {0}
\draw [directed] (0.0,0.0) -- (0.0,2.0) -- (6.0,2.0) -- (6.0,0.0) -- (0.2,0.0);
\draw [reverse directed] (0.2,0.2) rectangle (5.8,1.8);
\node[anchor=west] at (6.0,1.0) {\large{
$= \left( - \frac{1}{4m^2} \right)^8 (-1)^2 (-N_f) \left[ \frac{1}{N_c} \right] \{ 48 d (d-1) (2d-3)^2 \}$
}};
\end{tikzpicture}&&
\end{flalign}
\subsubsection{`$g$-type'}
\noindent
\begin{flalign}
\begin{tikzpicture}[baseline=(current  bounding  box.center),scale=0.7]
\def \yoff {0}
\draw [directed] (0.0,0.0) -- (0.0,2.0) -- (2.0,2.0) -- (2.0,0.0) -- (0.2,0.0);
\draw [reverse directed] (0.2,0.2) rectangle (1.8,1.8);
\draw [directed] (0.4,0.4) rectangle (1.6,1.6);
\draw [reverse directed] (0.6,0.6) rectangle (1.4,1.4);
\node[anchor=west] at (2.0,1.0) {\large{
$= \frac{3}{3!} \left( - \frac{1}{4m^2} \right)^8 (-1)^4 (-N_f)^3 \left[ \frac{2}{N_c} \right] \{ 4d(d-1) \}$
}};
\end{tikzpicture}&&
\label{type-g1}
\end{flalign}

\noindent
\begin{flalign}
\begin{tikzpicture}[baseline=(current  bounding  box.center),scale=0.7]
\def \yoff {0}
\draw [directed] (0.0,0.0) -- (0.0,2.0) -- (2.0,2.0) -- (2.0,0.0) -- (0.2,0.0);
\draw [reverse directed] (0.2,0.4) -- (0.2,1.8) -- (1.8,1.8) -- (1.8,0.2) -- (0.4,0.2) -- (0.4,1.6) -- (1.6,1.6) -- (1.6,0.4) -- (0.2,0.4);
\draw [directed] (0.6,0.6) rectangle (1.4,1.4);
\node[anchor=west] at (2.0,1.0) {\large{
$= \frac{2}{2!} \left( - \frac{1}{4m^2} \right)^8 (-1)^4 (-N_f)^2 \left[ 0 \right] \{ 4d(d-1) \}$
}};
\end{tikzpicture}&&
\label{type-g2}
\end{flalign}

\noindent
\begin{flalign}
\begin{tikzpicture}[baseline=(current  bounding  box.center),scale=0.7]
\def \yoff {0}
\draw [directed] (0.0,0.0) -- (0.0,2.0) -- (2.0,2.0) -- (2.0,0.0) -- (0.2,0.0) -- (0.2,1.8) -- (1.8,1.8) -- (1.8,0.2) -- (-0.2,0.2);
\draw [reverse directed] (0.4,0.6) -- (0.4,1.6) -- (1.6,1.6) -- (1.6,0.4) -- (0.6,0.4) -- (0.6,1.4) -- (1.4,1.4) -- (1.4,0.6) -- (0.4,0.6);
\node[anchor=west] at (2.0,1.0) {\large{
$= \left( - \frac{1}{4m^2} \right)^8 (-1)^4 (-N_f) \left[ \frac{2}{N_c} \right] \{ 4d(d-1) \}$
}};
\end{tikzpicture}&&
\label{type-g3}
\end{flalign}

%
%
%
%

\subsection{$L=9$}

\begin{flalign}
\begin{tikzpicture}[baseline=(current  bounding  box.center),scale=0.6]
\def \yoff {0}
\draw [directed] (0.0,0.0) -- (0.0,2.0) -- (6.0,2.0) -- (6.0,0.0) -- (0.2,0.0);
\draw [reverse directed] (0.2,0.2) rectangle (3.9,1.8);
\draw [reverse directed] (4.1,0.2) rectangle (5.8,1.8);
\node[anchor=west] at (6.0,1.0) {\large{
$= 2 \left( - \frac{1}{4m^2} \right)^9 (-1)^2 (-N_f)^2 \left[ \frac{1}{N_c^2}\right] \{ {48 d(d-1)(2d-3)^2} \}$
}};
\end{tikzpicture}&&
\end{flalign}

\subsubsection{$N_c=3$}

\begin{flalign}
\begin{tikzpicture}[baseline=(current  bounding  box.center),scale=0.7]
\def \yoff {0}
\draw [directed] (0.0,0.0) -- (0.0,2.0) -- (4.0,2.0) -- (4.0,0.0) -- (0.2,0.0);
\draw [reverse directed] (0.2,0.2) rectangle (1.9,1.8);
\draw [directed] (2.1,0.2) rectangle (3.8,1.8);
\draw [directed] (2.3,0.4) rectangle (3.6,1.6);
\node[anchor=west] at (4.0,1.0) {\large{
$= \frac{2}{2!} \left( - \frac{1}{4m^2} \right)^9 (-1)^3 (-N_f)^3 \left[ \frac{1}{9} \right] \{ 12d(d-1)(2d-3) \}$
}};
\end{tikzpicture}&&
\end{flalign}

\noindent
\begin{flalign}
\begin{tikzpicture}[baseline=(current  bounding  box.center),scale=0.7]
\def \yoff {0}
\draw [directed] (0.0,0.0) -- (0.0,2.0) -- (4.0,2.0) -- (4.0,0.0) -- (0.2,0.0);
\draw [directed] (0.2,0.2) rectangle (3.8,1.8);
\draw [directed] (0.4,0.4) rectangle (3.6,1.6);
\node[anchor=west] at (4.0,1.0) {\large{
$= \frac{1}{2!} \left( - \frac{1}{4m^2} \right)^9 (-N_f)^2 \left[ \frac{1}{3} \right] \{ 12d(d-1)(2d-3) \}$
}};
\end{tikzpicture}&&
\end{flalign}
\section{Calculating SU($N_c$) group integrals} \label{groupsection}

To obtain diagrams up to ${\cal O}\left(\left( \frac{1}{m^2} \right)^{16}\right)$, we need the following additional group integrals for general number of colours $N_c$
\begin{align} \label{3and4int}
& \int_{SU(N_c)} d U \, \Useries{U}{3}{a}{b} \Useries{(U^\dag)}{3}{c}{d} \,, \nonumber \\ &  \int_{SU(N_c)} d U \, \Useries{U}{4}{a}{b} \Useries{(U^\dag)}{4}{c}{d} \,.
\end{align}
Moreover, since we are interested in the case $N_c=3$, the following integrals also give a non-zero contribution at this order
\begin{align} \label{Ncis3ints}
& \int_{SU(3)} d U \, \Useries{U}{4}{a}{b} (U^\dag)_{c_1}^{\ d_1} \,, \quad \int_{SU(3)} d U\, \Useries{U}{6}{a}{b} \,, \nonumber \\ 
& \int_{SU(3)} d U \, \Useries{U}{5}{a}{b} \Useries{(U^\dag)}{2}{c}{d} \,.
\end{align}
In this section, we will explain how integrals of this type can be calculated in full generality. Methods to calculate integrals of this type have appeared in the literature at various occasions (see e.g. \cite{Bars:1979xb,Creutz:1984mg,Cvitanovic:2008zz,Wilson:1975id}). In this section, we will employ a method that is loosely based on techniques that appeared in \cite{Cvitanovic:2008zz} and that, to our knowledge, has not yet appeared in the literature. It uses tensor product decompositions to write the required integrals in terms of Young projectors. It has the advantage that it can easily be implemented using a symbolic computer algebra system. This method, that we will explain in Section \ref{sec:sec61} can be used to perform the group integrations associated to general diagrams. Diagrammatic methods to do these group integrations  are given in \cite{Creutz:1984mg}. For more complicated diagrams, these can quickly become cumbersome. For relatively simple diagrams, they can however be quick and useful, so we will give a brief summary of these techniques in Section \ref{sec:sec62}.

\subsection{General procedure} \label{sec:sec61}
 
In order to calculate the diagrams considered in this work, we need to evaluate various  integrals of products of matrix elements of SU($N_c$) group elements. Let us first focus on integrals of the form
\begin{equation} \label{mUmUdag}
I_m = \int_{\mathrm{SU}(N_c)} d U \, U_{a_1}^{\ b_1} \cdots U_{a_m}^{\ b_m} (U^\dag)_{c_1}^{\ d_1} \cdots (U^\dag)_{c_m}^{\ d_m} \,,
\end{equation}
where $U$ represents a SU($N_c$) group element in the fundamental representation. Integrals of this form were calculated in an implicit manner in \cite{Bars:1979xb}, where an
iterative way of calculating the quantities
\begin{equation}
 F_m(A) = \int_{\mathrm{SU}(N_c)} d U \, \left(\mathrm{tr} A U \right)^m \left(\mathrm{tr} A^\dag U^\dag \right)^m \,,
\end{equation}
for an arbitrary, constant matrix $A$, was given. In particular, it was argued that $F_{m}(A)$ is a linear combination of $(\mathrm{tr}(A A^\dag))^k\mathrm{tr}(A A^\dag)^{m-k}$
(for $k = 0, \cdots, m$) and that the coefficients of the linear combination can be obtained from knowledge of $F_{1}(A)$, $\cdots$, $F_{m-1}(A)$. Once such an 
expression for $F_{m}(A)$ is obtained, it can be used to extract the integral \eqref{mUmUdag}, by writing out all traces explicitly in terms of matrix elements and 
Kronecker delta symbols. The integral \eqref{mUmUdag} can then be found in terms of Kronecker delta symbols as the coefficient of $A_{b_1}^{\ a_1}\cdots A_{b_m}^{\ a_m} (A^\dag)_{d_1}^{\ c_1}\cdots (A^\dag)_{d_m}^{\ c_m}$, 
as can be seen by writing
\begin{align} 
& \int_{\mathrm{SU}(N_c)} d U \, \left(\mathrm{tr} A U \right)^m \left(\mathrm{tr} A^\dag U^\dag \right)^m = \nonumber \\ 
& \sum_{a_i, b_i, c_i,d_i} A_{b_1}^{\  a_1}\cdots A_{b_m}^{\ a_m} (A^\dag)_{d_1}^{\ c_1}\cdots (A^\dag)_{d_m}^{\ c_m} \int_{\mathrm{SU}(N_c)} d U \, U_{a_1}^{\ b_1} \cdots U_{a_m}^{\ b_m} (U^\dag)_{c_1}^{\ d_1} \cdots (U^\dag)_{c_m}^{\ d_m}\,.
\end{align}
Note that in extracting the integral \eqref{mUmUdag} in this way, care has to be taken of making sure that the result has the correct symmetry properties for the indices.
In particular, various symmetrizations have to be performed by hand. While in principle this gives a straightforward way to calculate the integrals \eqref{mUmUdag}, calculating
the $F_m(A)$ and extracting the wanted integrals from it can be cumbersome, especially as $m$ gets larger. For the purpose of this paper, we will therefore use a different
method, that allows one to directly and explicitly construct the integrals $I_m$, in a way that can be easily implemented using a symbolic computer program. We have
explicitly checked that the results we get for $I_m$ agree with the results one can get from the formulas of \cite{Bars:1979xb} for $m=1,\cdots,4$. We will now
outline our method and illustrate it in two examples.

The general procedure to evaluate $I_m$ consists of the following steps:
\begin{enumerate}
\item First, one writes the decomposition of $m$ fundamental representations. This decomposition is given by the sum of all standard Young tableaux with $m$ entries.
\item Next, one constructs the Young projectors associated with the standard Young tableaux that appear in this decomposition. These Young projectors can be constructed by symmetrizing the expression $\delta_{a_1}^{b_1} \cdots \delta_{a_m}^{b_m}$ in the $a_i$-indices of the first row of the Young tableau. The resulting expression is then symmetrized  in the $a_i$-indices appearing in the second row of the Young tableau and one continues this symmetrization procedure for all rows (from top to bottom). The result of this symmetrization is then antisymmetrized in the $a_i$-indices that appear in the first column of the tableau and similarly for all columns (from left to right). The Young projector is  given by the result of these consecutive symmetrizations and antisymmetrizations, multiplied by a factor that is the inverse of the product of all hook lengths of the tableau. This factor guarantees that the Young projector squares to itself.
\item Using the decomposition of step 1, the integral \eqref{mUmUdag} can be turned into a sum of integrals that are schematically of the form \cite{Cvitanovic:2008zz}
\begin{equation} \label{rulecvit}
\int_{\mathrm{SU}(N_c)} d U\, R_\alpha{}^\beta (S^\dag)_\gamma{}^\delta = \frac{1}{d_R} \, \mathbb{P}^R_\alpha{}^{\bar{\delta}}\, \mathbb{P}^S_\gamma{}^{\bar{\beta}}\, \delta_{R,S} \,.
\end{equation}
In this formula $R$ and $S$ are irreducible representations, that correspond to standard Young tableaux in the tensor product of $m$ fundamental representations. The dimension of $R$ has been denoted by $d_R$, while $\mathbb{P}^R_\alpha{}^\beta$ corresponds to the Young projector that picks out the representation $R$ in the tensor product. The $\delta_{R,S}$ indicates that the above integral is only non-zero when $R$, $S$ correspond to representations with the same Young tableau shape. Note that we have used a schematic notation for the indices $\alpha$, $\beta$, $\gamma$, $\delta$ of the matrix elements of $R$ and $S$. These indices are composite and consist of $m$ indices in the fundamental representation, with symmetry properties indicated by the standard Young tableau that corresponds to $R$ or $S$. Note that in \eqref{rulecvit}, the composite index $\delta$ has symmetry properties indicated by the Young tableau corresponding to $S$, whereas it has to appear in the Young projector corresponding to $R$. In 
case $R$ and $S$ correspond to different standard Young tableaux, one must reorder the indices that make up the composite index $\delta$ in such a way that the reordered collection, indicated by $\bar{\delta}$ in \eqref{rulecvit}, has symmetry properties of the Young tableau that corresponds to $R$. Such a reordering is possible for Young tableaux with the same shape. An analogous remark holds for the composite index $\beta$.
\end{enumerate}

All integrals $I_m$ can be calculated along the lines described above. The simplest integral is of course $I_1$, which by directly applying \eqref{rulecvit} is given by
\begin{equation} \label{I1}
I_1 = \int_{\mathrm{SU(N_c)}} d U \, U_a^{\ b} (U^\dag)_c^{\ d} = \frac{1}{N_c} \delta_a^d \delta_c^b \,.
\end{equation}
Let us now illustrate the above procedure via the calculation of $I_2$ and $I_3$.

Consider first the integral $I_2$
\begin{equation} \label{2U2Udag}
I_2 = \int_{\mathrm{SU}(N_c)} d U \, \Useries{U}{2}{a}{b} \Useries{(U^\dag)}{2}{c}{d} \,.
\end{equation}
Since $\Useries{U}{2}{a}{b}$ acts in the tensor product of two fundamental representations ($\ytableausetup{aligntableaux=center}\ytableaushort[a_]{1} \otimes \ytableaushort[a_]{2}$) and since
\begin{equation} \label{decomp21}
\ytableausetup{aligntableaux=center}
\ytableaushort[a_]{1} \otimes \ytableaushort[a_]{2} = \ytableaushort[a_]{12} \oplus \ytableaushort[a_]{1,2}\,,
\end{equation}
we can write
\begin{equation} \label{decomp2}
\Useries{U}{2}{a}{b} = \tens{S}{2}{a}{b} +\tens{A}{2}{a}{b}  \,, 
\end{equation}
where $\tens{S}{2}{a}{b}$ acts in the representation $\ytableaushort[a_]{12}$ and $\tens{A}{2}{a}{b}$ acts in the representation $\ytableaushort[a_]{1,2}$.
The symmetric and antisymmetric representation matrices $\tens{S}{2}{a}{b}$, $\tens{A}{2}{a}{b}$ can be obtained explicitly via
\begin{align}
\tens{S}{2}{a}{b} & = \proj{S}{2}{a}{c}\left(U_{c_1}^{\ d_1} U_{c_2}^{\ d_2} \right)  \proj{S}{2}{d}{b}\,, \nonumber \\
\tens{A}{2}{a}{b} & = \proj{A}{2}{a}{c}\left(U_{c_1}^{\ d_1} U_{c_2}^{\ d_2} \right)  \proj{A}{2}{d}{b}\,,
\end{align}
where the Young projectors $\proj{S}{2}{a}{b}$, $\proj{A}{2}{a}{b}$ on the symmetric and anti-symmetric representations are given by
\begin{align}
\proj{S}{2}{a}{b} & =  \frac12 \left(\delta_{a_1}^{b_1} \delta_{a_2}^{b_2} + \delta_{a_1}^{b_2} \delta_{a_2}^{b_1}   \right) \,, \nonumber \\
\proj{A}{2}{a}{b} & =  \frac12 \left(\delta_{a_1}^{b_1} \delta_{a_2}^{b_2} - \delta_{a_1}^{b_2} \delta_{a_2}^{b_1}   \right) \,.
\end{align}
Using the decomposition \eqref{decomp2}, the integral \eqref{2U2Udag} can be written as a sum of four terms
\begin{align}
I_2 & =  \int_{\mathrm{SU}(N_c)} dU \, \tens{S}{2}{a}{b} \tens{(S^\dag)}{2}{c}{d} + \int_{\mathrm{SU}(N_c)} dU \, \tens{A}{2}{a}{b} \tens{(A^\dag)}{2}{c}{d} \nonumber \\ &  \quad + \int_{\mathrm{SU}(N_c)} dU \, \tens{S}{2}{a}{b} \tens{(A^\dag)}{2}{c}{d}  + \int_{\mathrm{SU}(N_c)} dU \, \tens{A}{2}{a}{b} \tens{(S^\dag)}{2}{c}{d} \,.
\end{align}
The last two terms involve an integral of a product of two representations with different Young tableau shape and are therefore zero according to \eqref{rulecvit}. The first two terms can be evaluated using the same rule, resulting in
\begin{equation} \label{I2}
I_2  =  \frac{2}{N_c(N_c+1)} \proj{S}{2}{a}{d}\  \proj{S}{2}{c}{b} + \frac{2}{N_c(N_c-1)} \proj{A}{2}{a}{d}\  \proj{A}{2}{c}{b} \,.
\end{equation}
As a slightly more involved example, let us also consider the integral
\begin{equation} \label{3U3Udag}
I_3 = \int_{\mathrm{SU}(N_c)} d U \, \Useries{U}{3}{a}{b} \Useries{(U^\dag)}{3}{c}{d}\,.
\end{equation}
In this case, we can use the decomposition
\begin{equation} \label{decomp31}
\ytableausetup{aligntableaux=center}
\ytableaushort[a_]{1} \otimes \ytableaushort[a_]{2} \otimes  \ytableaushort[a_]{3}  = \ytableaushort[a_]{123}\ (S)  \oplus \ytableaushort[a_]{12,3}\ (M)\oplus \ytableaushort[a_]{13,2}\ (\tilde{M})  \oplus \ytableaushort[a_]{1,2,3}\ (A) \,,
\end{equation}
where in brackets we have given a shorthand notation to denote the corresponding tableaux, to write
\begin{equation} \label{decomp32}
U_{a_1}^{\ b_1} U_{a_2}^{\ b_2}U_{a_3}^{\ b_3} = \tens{S}{3}{a}{b} + \tens{M}{3}{a}{b} +\tens{\tilde{M}}{3}{a}{b}
 + \tens{A}{3}{a}{b} \,,
\end{equation}
where $\tens{S}{3}{a}{b}$, $\tens{M}{3}{a}{b}$, $\tens{\tilde{M}}{3}{a}{b}$, $\tens{A}{3}{a}{b}$ act in the representations indicated by the Young tableaux on the right-hand-side of eq. \eqref{decomp31}. They are explicitly obtained by acting with the appropriate Young projectors
\begin{align} \label{defSMA3}
\tens{S}{3}{a}{b} & =  \proj{S}{3}{a}{c}\left(U_{c_1}^{\ d_1} U_{c_2}^{\ d_2} U_{c_3}^{\ d_3} \right) \proj{S}{3}{d}{b}\,, \nonumber \\
\tens{M}{3}{a}{b} & =  \proj{M}{3}{a}{c}\left(U_{c_1}^{\ d_1} U_{c_2}^{\ d_2} U_{c_3}^{\ d_3} \right) \proj{M}{3}{d}{b}\,, \nonumber \\
\tens{\tilde{M}}{3}{a}{b} & = \proj{\tilde{M}}{3}{a}{c}\left(U_{c_1}^{\ d_1} U_{c_2}^{\ d_2} U_{c_3}^{\ d_3} \right) \proj{\tilde{M}}{3}{d}{b}\,, \nonumber \\
\tens{A}{3}{a}{b} & =  \proj{A}{3}{a}{c}\left(U_{c_1}^{\ d_1} U_{c_2}^{\ d_2} U_{c_3}^{\ d_3} \right) \proj{A}{3}{d}{b}\,,
\end{align}
where the Young projectors are given by
\begin{align} \label{youngproj}
\proj{S}{3}{a}{b} &= \frac16 \Big( \delta_{a_1}^{b_1} \delta_{a_2}^{b_2} \delta_{a_3}^{b_3} + \delta_{a_1}^{b_1} \delta_{a_2}^{b_3} \delta_{a_3}^{b_2} + \delta_{a_1}^{b_3} \delta_{a_2}^{b_1} \delta_{a_3}^{b_2} + \delta_{a_1}^{b_3} \delta_{a_2}^{b_2} \delta_{a_3}^{b_1} \nonumber \\ & \qquad + \delta_{a_1}^{b_2} \delta_{a_2}^{b_3} \delta_{a_3}^{b_1} + \delta_{a_1}^{b_2} \delta_{a_2}^{b_1} \delta_{a_3}^{b_3} \Big) \,, \nonumber \\
\proj{M}{3}{a}{b} &=\frac13 \Big(\delta_{a_1}^{b_1} \delta_{a_2}^{b_2} \delta_{a_3}^{b_3} +  \delta_{a_1}^{b_2} \delta_{a_2}^{b_1} \delta_{a_3}^{b_3} - \delta_{a_1}^{b_3} \delta_{a_2}^{b_2} \delta_{a_3}^{b_1} - \delta_{a_1}^{b_3} \delta_{a_2}^{b_1} \delta_{a_3}^{b_2}\Big)\,, \nonumber \\
\proj{\tilde{M}}{3}{a}{b} &=\frac13 \Big(\delta_{a_1}^{b_1} \delta_{a_2}^{b_2} \delta_{a_3}^{b_3} +  \delta_{a_1}^{b_3} \delta_{a_2}^{b_2} \delta_{a_3}^{b_1} - \delta_{a_1}^{b_2} \delta_{a_2}^{b_1} \delta_{a_3}^{b_3} - \delta_{a_1}^{b_2} \delta_{a_2}^{b_3} \delta_{a_3}^{b_1}\Big)\,, \nonumber \\
\proj{A}{3}{a}{b} &= \frac16 \Big( \delta_{a_1}^{b_1} \delta_{a_2}^{b_2} \delta_{a_3}^{b_3} - \delta_{a_1}^{b_1} \delta_{a_2}^{b_3} \delta_{a_3}^{b_2} + \delta_{a_1}^{b_3} \delta_{a_2}^{b_1} \delta_{a_3}^{b_2} - \delta_{a_1}^{b_3} \delta_{a_2}^{b_2} \delta_{a_3}^{b_1} \nonumber \\ & \qquad + \delta_{a_1}^{b_2} \delta_{a_2}^{b_3} \delta_{a_3}^{b_1} - \delta_{a_1}^{b_2} \delta_{a_2}^{b_1} \delta_{a_3}^{b_3} \Big) \,.
\end{align}
Using the decomposition \eqref{decomp32}, the integral \eqref{3U3Udag} can be written as a sum of integrals of the form \eqref{rulecvit}
\begin{align} \label{I3inbetween}
I_3 &= \int_{\mathrm{SU}(N_c)} dU \, \tens{S}{3}{a}{b} \tens{(S^\dag)}{3}{c}{d} + \int_{\mathrm{SU}(N_c)} dU \, \tens{A}{3}{a}{b} \tens{(A^\dag)}{3}{c}{d} \nonumber \\ & \quad + \int_{\mathrm{SU}(N_c)} dU \, \tens{M}{3}{a}{b} \tens{(M^\dag)}{3}{c}{d}  + \int_{\mathrm{SU}(N_c)} dU \, \tens{\tilde{M}}{3}{a}{b} \tens{(\tilde{M}^\dag)}{3}{c}{d} \nonumber \\ & \quad + \int_{\mathrm{SU}(N_c)} dU \, \tens{M}{3}{a}{b} \tens{(\tilde{M}^\dag)}{3}{c}{d}  + \int_{\mathrm{SU}(N_c)} dU \, \tens{\tilde{M}}{3}{a}{b} \tens{(M^\dag)}{3}{c}{d} \,,
\end{align}
where we have not written down the integrals involving representations with different Young tableau shape, as they are zero.
The above integrals can be evaluated using the rule \eqref{rulecvit}, with the understanding that for the two integrals on the last line, proper care should be taken of the correct placement of the indices. Specifically, in the integral
\begin{equation}
 \int_{\mathrm{SU}(N_c)} dU \, \tens{M}{3}{a}{b} \tens{(\tilde{M}^\dag)}{3}{c}{d} \,,
\end{equation}
the indices $d_1$, $d_2$, $d_3$ have the symmetry property indicated by the Young tableau $\ytableaushort[d_]{13,2}$ of $\tilde{M}$. According to \eqref{rulecvit}, they should be distributed on the Young projector corresponding to $M$, i.e. they should be re-ordered such that they have the symmetry property indicated by $\ytableaushort[d_]{12,3}$. This is done by interchanging $d_2$ and $d_3$. 
A similar remark holds for the indices $b_1$, $b_2$, $b_3$, so that 
\begin{equation}
\int_{\mathrm{SU}(N_c)} dU \, \tens{M}{3}{a}{b} \tens{(\tilde{M}^\dag)}{3}{c}{d} = \frac{3}{N_c(N_c^2-1)} \projnc{M}{3}{a}{d}{{1,3,2}}\ \projnc{\tilde{M}}{3}{c}{b}{{1,3,2}} \,.
\end{equation}
The last term of \eqref{I3inbetween} can be evaluated from analogous considerations. One then finds the following results for the integral \eqref{3U3Udag}
\begin{align} \label{I3}
I_3 & = \frac{6}{N_c(N_c+1)(N_c+2)}\proj{S}{3}{a}{d}\ \proj{S}{3}{c}{b}  +  \frac{3}{N_c(N_c^2-1)} \proj{M}{3}{a}{d} \ \proj{M}{3}{c}{b} \nonumber \\ & \quad +  \frac{3}{N_c(N_c^2-1)}\proj{\tilde{M}}{3}{a}{d} \ \proj{\tilde{M}}{3}{c}{b}  +\frac{3}{N_c(N_c^2-1)}  \projnc{M}{3}{a}{d}{{1,3,2}}\ \projnc{\tilde{M}}{3}{c}{b}{{1,3,2}}  \nonumber \\ & \quad + \frac{3}{N_c(N_c^2-1)} \projnc{\tilde{M}}{3}{a}{d}{{1,3,2}}\ \projnc{M}{3}{c}{b}{{1,3,2}}+  \frac{6}{N_c(N_c-1)(N_c-2)} \proj{A}{3}{a}{d} \ \proj{A}{3}{c}{b} \,.
\end{align}
The other $I_m$ can be calculated in a similar manner. We have given the result for $I_4$ in Appendix \ref{B}. 

Using the above results, other non-zero integrals can be derived by making use of the $\mathrm{SU}(N_c)$ identities
\begin{align}\label{suncid}
U_{a_1}^{\ b_1} & = \frac{1}{(N_c-1)!} \epsilon_{a_1 a_2 \cdots a_N} \epsilon^{b_1 b_2 \cdots b_N} (U^\dag)_{b_2}^{\ a_2} \cdots (U^\dag)_{b_N}^{\ a_N} \,, \nonumber \\ 
(U^\dag)_{a_1}^{\ b_1} & = \frac{1}{(N_c-1)!} \epsilon_{a_1 a_2 \cdots a_N} \epsilon^{b_1 b_2 \cdots b_N} U_{b_2}^{\ a_2} \cdots U_{b_N}^{\ a_N} \,.
\end{align}
These identities often allow one to reduce group integrals to integrals of the form of $I_m$, that can be calculated according to the method outlined above. In this way, one can for instance calculate the baryonic integral
\begin{equation} \label{baryonint}
\int_{\mathrm{SU}(N_c)} d U \, U_{a_1}^{\ b_1} \cdots U_{a_{N_c}}^{\ b_{N_c}} = \frac{1}{N_c !} \epsilon_{a_1 \cdots a_{N_c}} \epsilon^{b_1 \cdots b_{N_c}} \,.
\end{equation}
Moreover, the calculation of \eqref{Ncis3ints} can now be reduced to the calculation of \eqref{3and4int}
\begin{align}
& \int_{\mathrm{SU}(3)} d U \, \Useries{U}{4}{a}{b} (U^\dag)_{c_1}^{\ d_1}  = \nonumber \\ & \qquad \qquad  \frac12 \epsilon_{a_4 d_2 d_3} \epsilon^{b_4 c_2 c_3} \int_{\mathrm{SU}(3)} d U \, \Useries{U}{3}{a}{b} \Useries{(U^\dag)}{3}{c}{d} \,, \nonumber  \\
& \int_{\mathrm{SU}(3)} d U \,  \Useries{U}{6}{a}{b} = \nonumber \\ & 
\quad \frac14 \epsilon_{a_5 d_1 d_2 } \epsilon^{b_5 c_1c_2}\epsilon_{a_6 d_3 d_4 } \epsilon^{b_6 c_3 c_4}\int_{\mathrm{SU}(3)} d U \, \Useries{U}{4}{a}{b} \Useries{(U^\dag)}{4}{c}{d} \,, \nonumber \\ 
&  \int_{\mathrm{SU}(3)} d U \,  \Useries{U}{5}{a}{b} \Useries{(U^\dag)}{2}{c}{d} = \nonumber \\ 
& \qquad \frac12 \epsilon_{a_5 d_3 d_4} \epsilon^{b_5 c_3 c_4} \int_{\mathrm{SU}(3)} d U \, \Useries{U}{4}{a}{b}  \Useries{(U^\dag)}{4}{c}{d} \,.
\end{align}
Finally, let us note for the sake of completeness that an expression for integrals of the type
\begin{equation}
\int_{\mathrm{SU}(N_c)} d U \, U_{a_1}^{\ b_1} \cdots U_{a_{N_c}}^{\ b_{N_c}} U_{a_{N_c+1}}^{\ b_{N_c+1}} \cdots U_{a_{2 N_c}}^{\ b_{2 N_c}} \cdots U_{a_{(p-1)N_c + 1}}^{\ b_{(p-1)N_c +1}} \cdots U_{a_{p N_c}}^{\ b_{p N_c}}\,,
\end{equation}
is known in terms of $\epsilon$-symbols (see e.g. \cite{Creutz:1984mg} for a derivation). In particular, the result is given by
\begin{align} \label{creutzform}
& \frac{2! \cdot 3! \cdots (N_c - 1)!}{(p+1)! \cdots (p+N_c-1)!} \epsilon_{a_1 \cdots a_{N_c}}\epsilon^{b_1 \cdots b_{N_c}} \cdots \epsilon_{a_{(p-1)N_c + 1} \cdots a_{p N_c}}\epsilon^{b_{(p-1)N_c + 1} \cdots b_{p N_c}} \nonumber \\ & \qquad + \mathrm{permutations}\,,
\end{align}
where `$+$ permutations' indicates that one has to add similar terms as the first, where however the indices of the first term are permuted in such a way as to render the resulting expression symmetric under the interchange of all $(a_i,b_i)$ index pairs. In principle, one can use this result along with the SU$(N_c)$ identities \eqref{suncid} to calculate the integrals \eqref{mUmUdag}. One can then rewrite the result in terms of Kronecker-deltas by contracting the various $\epsilon$-symbols and using the identity
\begin{equation}
\epsilon_{a_1 \cdots a_{N_c}} \epsilon^{b_1 \cdots b_{N_c}} = N_c!\  \delta_{[a_1}^{[b_1} \cdots \delta_{a_{N_c}]}^{b_{N_c}]} \,.
\end{equation}
Given the number of permutations one has to add by hand in \eqref{creutzform}, extracting the integrals \eqref{mUmUdag} in this way can however be rather cumbersome.

\subsection{Diagrammatic techniques} \label{sec:sec62}

The technique described in the above section is general and can be used to calculate any type of non-zero $\mathrm{SU}(N_c)$ integral. Since a diagram consists of a number of links attached to each other, the group integrals associated to a diagram can be obtained by multiplying the integrals corresponding to the links and by properly contracting their group indices. These contractions can easily be carried out by a symbolic computer program. For simple diagrams, the contractions can also be easily done using diagrammatic techniques explained in reference \cite{Wilson:1975id,Creutz:1984mg}, to which we refer for diagrammatic notations and conventions. For example the result for the integral $\int_{\mathrm{SU}(N_c)} dU \, U_a^{\ b} (U^\dag)_c^{\ d}$ (given in \eqref{I1}) can be written diagrammatically as

\noindent
\begin{equation}
\begin{tikzpicture}[baseline=(current  bounding  box.center),scale=0.7]

\def \yoff {0}
\def \xoff {0.0}


\node[anchor=west] at (0.0+\xoff,0.5) {\Large{
$\frac{1}{N_c}$
}};

\draw (1.5+\xoff,0) arc (-90:90:5mm);
\draw (3.2+\xoff,1) arc (90:270:5mm);


\end{tikzpicture}\ \ \ .
\end{equation}

\noindent Carefully identifying the links which are connected it is possible to calculate any of the diagrams in Section \ref{diags} diagrammatically using the appropriate integral equations in Section \ref{groupsection}. As a simple example consider the diagram 
\begin{tikzpicture}[scale=0.3]

\def \xoff {0.0}
\def \yoff {0.0}

\draw [directed] (0.0+\xoff,0.0+\yoff) -- (0.0+\xoff,2.0+\yoff) -- (2.0+\xoff,2.0+\yoff) -- (2.0+\xoff,0.0+\yoff) -- (0.2+\xoff,0.0+\yoff);
\draw [reverse directed] (0.2+\xoff,0.2+\yoff) rectangle (1.8+\xoff,1.8+\yoff);

\end{tikzpicture}
in (\ref{type-b}). This can be evaluated as

\noindent
\begin{equation*}
\begin{tikzpicture}[baseline=(current  bounding  box.center),scale=1.0]

\def \yoff {0}
\def \xoff {0.0}


\node[anchor=west] at (-0.5+\xoff,0.5) {\Large{
$\bigg{[} \frac{1}{N_c}$
}};

\draw [dashed] (0.9+\xoff,0.0) -- (1.5+\xoff,0.0);

\draw (1.5+\xoff,0) arc (-90:90:5mm);
\draw (3.2+\xoff,1) arc (90:270:5mm);

\node[anchor=west] at (3.0+\xoff,0.5) {\Large{
$\bigg{]}$
}};

\def \xoff {3.5}

\node[anchor=west] at (0.0+\xoff,0.5) {\Large{
$\bigg{[} \frac{1}{N_c}$
}};

\draw (1.5+\xoff,0) arc (-90:90:5mm);
\draw (3.2+\xoff,1) arc (90:270:5mm);

\node[anchor=west] at (3.0+\xoff,0.5) {\Large{
$\bigg{]}$
}};

\def \xoff {7.0}

\node[anchor=west] at (0.0+\xoff,0.5) {\Large{
$\bigg{[} \frac{1}{N_c}$
}};

\draw (1.5+\xoff,0) arc (-90:90:5mm);
\draw (3.2+\xoff,1) arc (90:270:5mm);

\node[anchor=west] at (3.0+\xoff,0.5) {\Large{
$\bigg{]}$
}};

\def \xoff {10.5}

\node[anchor=west] at (0.0+\xoff,0.5) {\Large{
$\bigg{[} \frac{1}{N_c}$
}};

\draw (1.5+\xoff,0) arc (-90:90:5mm);
\draw (3.2+\xoff,1) arc (90:270:5mm);

\draw [dashed] (3.2+\xoff,0.0) -- (3.8+\xoff,0.0);

\node[anchor=west] at (3.5+\xoff,0.5) {\Large{
$\bigg{]}$
}};


\end{tikzpicture}
\end{equation*}

\noindent
\begin{equation*}
\begin{tikzpicture}[baseline=(current  bounding  box.center),scale=1.0]

\def \yoff {0}
\def \xoff {1.5}


\node[anchor=west] at (-0.5,0.5) {\Large{
$= \bigg{[} \frac{1}{N_c} \bigg{]}^4 \bigg{[}$
}};

\draw [dashed] (0.9+\xoff,0.0) -- (1.5+\xoff,0.0);

\draw (1.5+\xoff,0) arc (-90:90:5mm);
\draw (3.2+\xoff,1) arc (90:270:5mm);


\def \xoff {3.2}


\draw (1.5+\xoff,0) arc (-90:90:5mm);
\draw (3.2+\xoff,1) arc (90:270:5mm);


\def \xoff {4.9}


\draw (1.5+\xoff,0) arc (-90:90:5mm);
\draw (3.2+\xoff,1) arc (90:270:5mm);


\def \xoff {6.6}


\draw (1.5+\xoff,0) arc (-90:90:5mm);
\draw (3.2+\xoff,1) arc (90:270:5mm);

\draw [dashed] (3.2+\xoff,0.0) -- (3.8+\xoff,0.0);

\node[anchor=west] at (3.5+\xoff,0.5) {\Large{
$\bigg{]}$
}};


\end{tikzpicture}
\end{equation*}

\noindent
\begin{equation}
\begin{tikzpicture}[baseline=(current  bounding  box.center),scale=1.0]

\def \yoff {0}
\def \xoff {1.5}


\node[anchor=west] at (0.0,0.5) {\Large{
$= \bigg{[} \frac{1}{N_c} \bigg{]} \bigg{[}$
}};

\draw (1.3+\xoff,0.5) arc (180:360:5mm);




\def \xoff {1.3}




\node[anchor=west] at (2.5+\xoff,0.5) {\Large{
$\bigg{]}$
}};


\end{tikzpicture} \ \ \ .
\end{equation}

\noindent Similarly the result for $\int_{\mathrm{SU}(N_c)} dU \, \Useries{U}{2}{a}{b} \Useries{(U^\dag)}{2}{c}{d}$ (given in \eqref{I2}) can be written as

\noindent
\begin{equation}
\begin{tikzpicture}[baseline=(current  bounding  box.center),scale=1.0]

\def \yoff {0}
\def \xoff {0.0}


\node[anchor=west] at (0.0+\xoff,0.6) {\Large{
$\bigg{[}$
}};

\draw (1.0+\xoff,0.75) arc (-90:90:2.2mm);
\draw (1.0+\xoff,0) arc (-90:90:2.2mm);
\draw (2.0+\xoff,1.2) arc (90:270:2.2mm);
\draw (2.0+\xoff,0.45) arc (90:270:2.2mm);

\node[anchor=west] at (2.1+\xoff,0.6) {\Large{
$+$
}};

\def \xoff {2.3}

\draw (1.0+\xoff,0.0) arc (-90:90:6mm);
\draw (1.0+\xoff,0.4) arc (-90:90:2.2mm);
\draw (2.5+\xoff,1.2) arc (90:270:6mm);
\draw (2.5+\xoff,0.8) arc (90:270:2.2mm);

\node[anchor=west] at (2.6+\xoff,0.6) {\Large{
$\bigg{]} \alpha_+$
}};

\node[anchor=west] at (4.5+\xoff,0.6) {\Large{
$+$
}};

\def \xoff {8.0}

\node[anchor=west] at (0.0+\xoff,0.6) {\Large{
$\bigg{[}$
}};

\draw (1.0+\xoff,0.75) arc (-90:90:2.2mm);
\draw (1.0+\xoff,0) arc (-90:90:2.2mm);
\draw (2.2+\xoff,1.2) arc (90:270:6mm);
\draw (2.2+\xoff,0.8) arc (90:270:2.2mm);

\node[anchor=west] at (2.3+\xoff,0.6) {\Large{
$+$
}};

\def \xoff {10.5}

\draw (1.0+\xoff,0.0) arc (-90:90:6mm);
\draw (1.0+\xoff,0.4) arc (-90:90:2.2mm);
\draw (2.2+\xoff,1.2) arc (90:270:2.2mm);
\draw (2.2+\xoff,0.45) arc (90:270:2.2mm);

\node[anchor=west] at (2.2+\xoff,0.6) {\Large{
$\bigg{]} \alpha_-$
}};

\end{tikzpicture} \ \ \ ,
\end{equation}

\noindent with
\EQ{
\alpha_{\pm} \equiv \frac{1}{2} \left[ \frac{1}{N_c(N_c+1)} \pm \frac{1}{N_c(N_c-1)} \right] \, ,
}
which can be used to calculate diagrams with four overlapping links, and so on.

Diagrams of one-tile area, that are open in one corner, can also be easily integrated. Since such diagrams have only two free indices, the final result must be given by a constant $C$ times a Kronecker delta for the two indices
\begin{equation}
\begin{tikzpicture}[baseline=(current  bounding  box.center),scale=1.0, every node/.style={transform shape}]
\draw [directed] (0,0) -- (0,1);
\draw [directed] (0,1) -- (1,1);'
\draw [directed] (1,1) -- (1,0);
\draw [directed] (1,0) -- (0.2,0);

\draw [reverse directed] (0.4,0.3) -- (0.4,0.6) -- (0.6,0.6) -- (0.6,0.4) -- (0.3,0.4);
\draw [reverse directed] (0.3,0.4) -- (0.3,0.7) -- (0.7,0.7) -- (0.7,0.3) -- (0.4,0.3);

\node at (0.5,0.1) {\tiny$.$};
\node at (0.5,0.15) {\tiny$.$};
\node at (0.5,0.2) {\tiny$.$};

\node at (0.5,0.8) {\tiny$.$};
\node at (0.5,0.85) {\tiny$.$};
\node at (0.5,0.9) {\tiny$.$};

\node at (0.1,0.5) {\tiny$.$};
\node at (0.15,0.5) {\tiny$.$};
\node at (0.2,0.5) {\tiny$.$};

\node at (0.8,0.5) {\tiny$.$};
\node at (0.85,0.5) {\tiny$.$};
\node at (0.9,0.5) {\tiny$.$};

\node [below] at (0,-0.06) {\scriptsize$a$};
\node [below] at (0.2,0) {\scriptsize$b$};

\node [right] at (1.1,0.5) {$=C$};
\draw [black] (2.1,0.4) to[out=90,in=90, distance=6pt] (2.3,0.4);
\node [below] at (2.1,0.39)  {\scriptsize$a$};
\node [below] at (2.3,0.45) {\scriptsize$b$};
\end{tikzpicture} \, .
\label{OpenIntDelta}
\end{equation}
In order to determine the constant $C$ we multiply by $\delta_{ab}$. This leads to a closed diagram. By calculating this closed diagram in two different ways, the constant $C$ can be calculated, as illustrated in the following diagrammatic equation:
\begin{equation}
\begin{tikzpicture}[baseline=(current  bounding  box.center),scale=1.0, every node/.style={transform shape}]
\hspace{-11.5mm}
\draw [directed] (0,0) -- (0,1);
\draw [directed] (0,1) -- (1,1);'
\draw [directed] (1,1) -- (1,0);
\draw [directed] (1,0) -- (0.2,0);

\draw [reverse directed] (0.4,0.3) -- (0.4,0.6) -- (0.6,0.6) -- (0.6,0.4) -- (0.3,0.4);
\draw [reverse directed] (0.3,0.4) -- (0.3,0.7) -- (0.7,0.7) -- (0.7,0.3) -- (0.4,0.3);

\node at (0.5,0.1) {\tiny$.$};
\node at (0.5,0.15) {\tiny$.$};
\node at (0.5,0.2) {\tiny$.$};

\node at (0.5,0.8) {\tiny$.$};
\node at (0.5,0.85) {\tiny$.$};
\node at (0.5,0.9) {\tiny$.$};

\node at (0.1,0.5) {\tiny$.$};
\node at (0.15,0.5) {\tiny$.$};
\node at (0.2,0.5) {\tiny$.$};

\node at (0.8,0.5) {\tiny$.$};
\node at (0.85,0.5) {\tiny$.$};
\node at (0.9,0.5) {\tiny$.$};

\draw [black] (0,0) to[out=270,in=270, distance=6pt] (0.2,0);

\node at (1.9,0.5) {$\displaystyle = \left\{ \begin{array}{c} \\ \\ \\ \\ \\ \\\end{array} \right.$};

\hspace{23mm}

\draw [directed] (0,0.75) -- (0,1.75);
\draw [directed] (0,1.75) -- (1,1.75);
\draw [directed] (1,1.75) -- (1,0.75);
\draw [directed] (1,0.75) -- (0,0.75);

\draw [reverse directed] (0.4,1.05) -- (0.4,1.35) -- (0.6,1.35) -- (0.6,1.15) -- (0.3,1.15);
\draw [reverse directed] (0.3,1.15) -- (0.3,1.45) -- (0.7,1.45) -- (0.7,1.05) -- (0.4,1.05);

\node at (0.5,0.84) {\tiny$.$};
\node at (0.5,0.9) {\tiny$.$};
\node at (0.5,0.95) {\tiny$.$};

\node at (0.5,1.55) {\tiny$.$};
\node at (0.5,1.60) {\tiny$.$};
\node at (0.5,1.65) {\tiny$.$};

\node at (0.1,1.25) {\tiny$.$};
\node at (0.15,1.25) {\tiny$.$};
\node at (0.2,1.25) {\tiny$.$};

\node at (0.8,1.25) {\tiny$.$};
\node at (0.85,1.25) {\tiny$.$};
\node at (0.9,1.25) {\tiny$.$};

\node [right] at (1.2,1.25) {$= \, I$};

\node [right] at (0,-0.25) {$C \hspace{5mm} = \, N_c C$};

\draw [black] (0.6,-0.23) to[out=270,in=270, distance=6pt] (0.8,-0.23);
\draw [black] (0.6,-0.23) to[out=90,in=90, distance=6pt] (0.8,-0.23);
\end{tikzpicture} \hspace{11.5mm}\,.
\end{equation}
The $I$ in the upper equation on the r.h.s. is the value of the integrated closed diagram, while the lower equation is obtained from \eqref{OpenIntDelta}, and using that $\delta_{aa} = N_c$ for the fundamental representation. Equating these two different ways of calculating the same diagram, we can thus write
\SP{
C = \frac{1}{N_c} I = \frac{1}{N_c}\int_{{\rm SU}(N_c)} \hspace{-8mm} d U \, \cdots \, ,
\label{CByIntegral}
}
where the integrand indicated by $\cdots$ depends on the diagram under consideration. Since $I$ corresponds to a closed one-tile diagram, there can be no free indices and the integrand indicated by $\cdots$ is given entirely in terms of traces of powers of $U$ and $U^\dag$. As a rule of thumb that can be used to write this integrand down, one can use that for every loop in the diagram that winds around $n$ times in one direction, one should include a factor of $\tr\,U^n$ in the integrand. Likewise a factor of $\tr\,{U^\dag}^n$ should be included in the integrand for every loop that winds around $n$ times in the other direction. These one-tile closed diagram integrals can then be evaluated very easily using the Young projector formulas of the previous section, or using the diagrammatic techniques of  \cite{Creutz:1984mg}. In this way, the calculation of this type of diagrams can be reduced to calculating a single group integral, instead of calculating four group integrals (one for every link) and multiplying and contracting the results.

As an illustrative example we calculate the value of the diagram
\begin{equation} \label{opendiagramex}
\begin{tikzpicture}[baseline=(current  bounding  box.center),scale=1.0, every node/.style={transform shape}]
\draw [directed] (0,0) -- (0,1);
\draw [directed] (0,1) -- (1,1);
\draw [directed] (1,1) -- (1,0);
\draw [directed] (1,0) -- (0.2,0);

\draw [directed] (0.1,0.1) -- (0.1,0.9);
\draw [directed] (0.1,0.9) -- (0.9,0.9);
\draw [directed] (0.9,0.9) -- (0.9,0.1);
\draw [directed] (0.9,0.1) -- (0.1,0.1);

\draw [reverse directed] (0.2,0.2) -- (0.2,0.8) -- (0.8,0.8) -- (0.8,0.3) -- (0.3,0.3) -- (0.3,0.7) -- (0.7,0.7) -- (0.7,0.2) -- (0.2,0.2);

\node [below] at (0,-0.08) {\scriptsize$a$};
\node [below] at (0.2,0) {\scriptsize$b$};
\end{tikzpicture} \, ,
\end{equation}
where the corresponding closed diagram is
\begin{equation} \label{closeddiagram}
\begin{tikzpicture}[baseline=(current  bounding  box.center),scale=1.0, every node/.style={transform shape}]
\draw [directed] (0,0) -- (0,1);
\draw [directed] (0,1) -- (1,1);
\draw [directed] (1,1) -- (1,0);
\draw [directed] (1,0) -- (0,0);

\draw [directed] (0.1,0.1) -- (0.1,0.9);
\draw [directed] (0.1,0.9) -- (0.9,0.9);
\draw [directed] (0.9,0.9) -- (0.9,0.1);
\draw [directed] (0.9,0.1) -- (0.1,0.1);

\draw [reverse directed] (0.2,0.2) -- (0.2,0.8) -- (0.8,0.8) -- (0.8,0.3) -- (0.3,0.3) -- (0.3,0.7) -- (0.7,0.7) -- (0.7,0.2) -- (0.2,0.2);
\end{tikzpicture} \, .
\end{equation}
Using equations \eqref{OpenIntDelta} and \eqref{CByIntegral} the open diagram evaluates to
\begin{equation} 
\begin{tikzpicture}[baseline=(current  bounding  box.center),scale=1.0, every node/.style={transform shape}]
\draw [directed] (0,0) -- (0,1);
\draw [directed] (0,1) -- (1,1);
\draw [directed] (1,1) -- (1,0);
\draw [directed] (1,0) -- (0.2,0);

\draw [directed] (0.1,0.1) -- (0.1,0.9);
\draw [directed] (0.1,0.9) -- (0.9,0.9);
\draw [directed] (0.9,0.9) -- (0.9,0.1);
\draw [directed] (0.9,0.1) -- (0.1,0.1);

\draw [reverse directed] (0.2,0.2) -- (0.2,0.8) -- (0.8,0.8) -- (0.8,0.3) -- (0.3,0.3) -- (0.3,0.7) -- (0.7,0.7) -- (0.7,0.2) -- (0.2,0.2);

\node [below] at (0,-0.07) {\scriptsize$a$};
\node [below] at (0.2,0) {\scriptsize$b$};
\hspace{11.5 mm}
\node [right] at (0,0.5) {$\displaystyle = \frac{\delta_{ab}}{N_c}\int_{{\rm SU}(N_c)} \hspace{-8mm} d U \,\tr^2U^\dag \, \tr\,U^2$};
\end{tikzpicture}  \hspace{11.5mm} \, ,
\end{equation}
where the integral corresponds to the value of the closed diagram \eqref{closeddiagram}. The integrand is determined using the above stated rule of thumb, by noting that the closed diagram consists of three loops : the outer two winding one time in one direction, while the inner loop winds two times in the other direction. This integral can be very easily evaluated using e.g. the Young projector formula \eqref{I2} as
\begin{align}
\int_{{\rm SU}(N_c)} \hspace{-8mm} d U \, \tr^2U^\dag \, \tr\,U^2 & = \int_{{\rm SU}(N_c)} \hspace{-8mm} d U \,  U_{a_1}^{\ b_1} U_{b_1}^{\ a_1} \Useries{(U^\dag)}{2}{c}{c}  = 0 \,,
\end{align}
where the last equation is obtained by plugging the indices in in \eqref{I2} and evaluating the resulting formula explicitly. We thus find that the diagram \eqref{opendiagramex} evaluates to zero.

\section{Sources of error} \label{errorsection}

\subsection{Mis-counting of overlapping graphs} \label{overlap}

One of the potentially problematic aspects of our approach is that since each diagram type can be placed at a site $x$, any number of times, in any possible direction, over-counting will result from contributions with overlapping diagrams\footnote{We note that overlapping diagrams are not mis-counted when including only $a$-type contributions, as in the $N_f = 0$ calculations \cite{Blairon:1980pk,Martin:1982tb}.}. This is a problem which arises at $N_f \ne 0$ due to the link integrations. It is in principle possible to systematically account for mis-counted graphs order by order by adding the appropriate counter term. However, practically speaking, it is difficult to do this within the formulation we are using. Here are some examples of mis-counted overlapping graphs.

\subsubsection{$L=8$}
\begin{flalign}
\begin{tikzpicture}[baseline=(current  bounding  box.center),scale=0.7]
\def \yoff {0}
\draw [directed] (0.0,0.0) -- (0.0,2.0) -- (2.0,2.0) -- (2.0,0.0) -- (0.2,0.0) -- (0.2,1.8) -- (1.8,1.8) -- (1.8,0.2) -- (-0.2,0.2);
\draw [reverse directed] (0.4,0.4) rectangle (1.6,1.6);
\draw [reverse directed] (0.6,0.6) rectangle (1.4,1.4);
\node[anchor=west] at (2.0,1.0) {\large{
$= \frac{1}{2!} \left( - \frac{1}{4m^2} \right)^{8} (-1)^4 (-N_f)^2 \left[ 0 \right] \, ,$
}};
\end{tikzpicture}&&
\end{flalign}

\noindent however, it gets counted as
\EQ{
\left( - \frac{1}{4m^2} \right)^{8} (-1)^4 (-N_f)^2 \left[ \frac{1}{N_c^2} \right] \, .
}
To account for the above mis-counting, it is necessary to add a counter term at $L = 8$ of the form

\begin{flalign}
\begin{tikzpicture}[baseline=(current  bounding  box.center),scale=0.7]
\def \yoff {0}
\draw [directed] (0.0,0.0) -- (0.0,2.0) -- (2.0,2.0) -- (2.0,0.0) -- (0.2,0.0) -- (0.2,1.8) -- (1.8,1.8) -- (1.8,0.2) -- (-0.2,0.2);
\draw [reverse directed] (0.4,0.4) rectangle (1.6,1.6);
\draw [reverse directed] (0.6,0.6) rectangle (1.4,1.4);
\node[anchor=west] at (2.0,0.0){\large{c.t.}};
\node[anchor=west] at (2.0,1.0) {\large{
$= - \left( - \frac{1}{4m^2} \right)^{8} (-1)^4 (-N_f)^2 \left[ \frac{1}{N_c^2} \right] \{ 4d(d-1) \} \, .$
}};
\end{tikzpicture}&&
\label{ctL8}
\end{flalign}
\subsubsection{$L=12$}

\noindent
\begin{tikzpicture}[scale=0.7]

\def \off {1.0}

\draw [directed] (0.0,0.0) -- (0.0,2.0+\off) -- (2.0+\off,2.0+\off) -- (2.0+\off,0.0) -- (0.2,0.0) -- (0.2,1.8+\off) -- (1.8+\off,1.8+\off) -- (1.8+\off,0.2) -- (0.4,0.2) -- (0.4,1.6+\off) -- (1.6+\off,1.6+\off) -- (1.6+\off,0.4) -- (-0.2,0.4);
\draw [reverse directed] (0.6,0.6) rectangle (1.4+\off,1.4+\off);
\draw [reverse directed] (0.8,0.8) rectangle (1.2+\off,1.2+\off);
\draw [reverse directed] (1.0,1.0) rectangle (1.0+\off,1.0+\off);

\node[anchor=west] at (2.0+\off,1.0+0.5) {\large{
$= \left( - \frac{1}{4m^2} \right)^{12} (-1)^6 (-N_f)^3 \left[ 0 \right] \, ,$
}};

\end{tikzpicture}\\

\noindent for $N_c \ge 3$. For $N_c = 2$ the result is $\left( - \frac{1}{4m^2} \right)^{12} (-1)^6 (-N_f)^3 \left[ -\frac{1}{2} \right]$. However, in either case it gets counted as
\EQ{
\left( - \frac{1}{4m^2} \right)^{12} (-1)^6 (-N_f)^3 \left[ \frac{1}{N_c^3} \right] \, .
}

The difficulties in adding counter terms are 1.) it is difficult to determine where exactly to add them within our formulation, and 2.) the counter terms lead to mis-counting at higher orders, requiring the addition of even more counter terms. Since the second issue can be resolved order by order, the first issue is the most critical. If one naively adds the counter term (\ref{ctL8}) as a base diagram at order $L=8$, then indeed the wrong contributions obtained with two overlapping $b$-type diagrams can be cancelled off. However, in addition, new diagrams would be created with both contributions from overlapping $b$-type diagrams, and counter terms of the form (\ref{ctL8}). These mixed diagrams should not be included and would introduce a different, difficult to quantify source of error. Therefore, at this point, we don't attempt to correct for errors resulting from overlapping diagrams. A proper treatment of the issue of overlapping diagrams is left for future research.

\subsection{Avoiding over-counting of graphs}
\label{over-counting}

Another source of error results from over-counting or under-counting of graphs. This happens, for example, when attaching a trunk,
\begin{tikzpicture}[scale=0.3]

\draw [directed] (0.0,0.0) -- (0.0,2.0);
\draw (0.0,2.0) -- (0.2,2.0);
\draw [directed] (0.2,2.0) -- (0.2,0.0);

\end{tikzpicture}
 ($a$-type), to either of two adjacent corners of a box,
\begin{tikzpicture}[scale=0.3]

\def \xoff {0.0}
\def \yoff {0.0}

\draw [directed] (0.0+\xoff,0.0+\yoff) -- (0.0+\xoff,2.0+\yoff) -- (2.0+\xoff,2.0+\yoff) -- (2.0+\xoff,0.0+\yoff) -- (0.2+\xoff,0.0+\yoff);
\draw [reverse directed] (0.2+\xoff,0.2+\yoff) rectangle (1.8+\xoff,1.8+\yoff);

\end{tikzpicture}
 ($b$-type) diagram. This results in graphs of the form \cite{Tomboulis:2012nr} 

\begin{equation}
\begin{tikzpicture}[baseline=(current  bounding  box.center),scale=0.7]
\def \yoff {0}
\def \xoff {4.0}
\draw [directed] (0.0,0.0) -- (0.0,2.4) -- (2.0,2.4) -- (2.0,2.2) -- (0.2,2.2) -- (0.2,2.0) -- (2.0,2.0) -- (2.0,0.0) -- (0.2,0.0);
\draw [reverse directed] (0.2,0.2) rectangle (1.8,1.8);
\draw [directed] (0.0+\xoff,0.0) -- (0.0+\xoff,2.0) -- (1.8+\xoff,2.0) -- (1.8+\xoff,2.2) -- (0.0+\xoff,2.2) -- (0.0+\xoff,2.4) -- (2.0+\xoff,2.4) -- (2.0+\xoff,0.0) -- (0.2+\xoff,0.0);
\draw [reverse directed] (0.2+\xoff,0.2) rectangle (1.8+\xoff,1.8);
\end{tikzpicture}\ \ \ ,
\end{equation}

\noindent which are identical since the same sequence of links, $U_{\nu}(x) U_{\mu}(x+\hat{\nu}) U_{\mu}^{\dagger}(x+\hat{\nu}) U_{\mu}(x+\hat{\nu}) U_{\nu}^{\dagger}(x+\hat{\mu}) U_{\mu}^{\dagger}(x)$ (outside, plus inside plaquette), appears in both diagrams. To deal with this issue we follow \cite{Tomboulis:2012nr} and subtract off one possible direction when attaching a trunk ($a$-type) to a box ($b$-type) diagram. At one corner it is necessary to subtract off two directions to avoid over-counting either of

\begin{equation}
\begin{tikzpicture}[baseline=(current  bounding  box.center),scale=0.7]

\def \yoff {0}
\def \xoff {4.0}

\draw [directed] (0.0,0.0) -- (0.0,1.8) -- (-0.2,1.8) -- (-0.2,0.0) -- (-0.4,0.0) -- (-0.4,2.0) -- (2.0,2.0) -- (2.0,0.0) -- (0.2,0.0);
\draw [reverse directed] (0.2,0.2) rectangle (1.8,1.8);

\draw [directed] (0.0+\xoff,0.0) -- (0.0+\xoff,2.0) -- (2.0+\xoff,2.0) -- (2.0+\xoff,-0.4) -- (0.2+\xoff,-0.4) -- (0.2+\xoff,-0.2) -- (1.8+\xoff,-0.2) -- (1.8+\xoff,0.0) -- (0.2+\xoff,0.0);
\draw [reverse directed] (0.2+\xoff,0.2) rectangle (1.8+\xoff,1.8);


\end{tikzpicture} \ \ \ ,
\end{equation}

\noindent which also appear by attaching both an $a$-type and $b$-type diagram directly at $x$ (when irreducible diagrams are combined). That is, they correspond to

\begin{equation}
\begin{tikzpicture}[baseline=(current  bounding  box.center),scale=0.7]

\def \yoff {0}
\def \xoff {4.0}

\draw [directed] (-0.4,0.0) -- (-0.4,2.0) -- (-0.2,2.0) -- (-0.2,0.2) -- (0.0,0.2) -- (0.0,2.0) -- (2.0,2.0) -- (2.0,0.0) -- (0.2,0.0);
\draw [reverse directed] (0.2,0.2) rectangle (1.8,1.8);

\draw [directed] (0.0+\xoff,0.0) -- (0.0+\xoff,2.0) -- (2.0+\xoff,2.0) -- (2.0+\xoff,0.0) -- (0.2+\xoff,0.0) -- (0.2+\xoff,-0.2) -- (2.0+\xoff,-0.2) -- (2.0+\xoff,-0.4) -- (0.0+\xoff,-0.4);
\draw [reverse directed] (0.2+\xoff,0.2) rectangle (1.8+\xoff,1.8);


\end{tikzpicture}\ \ \ ,
\end{equation}

\noindent respectively. This result can be generalised for attachment of an $a$-type diagram to any area $1$-type diagram. Therefore, the dimensionalities are $a_1 = 2d-1$, $a_1' = 2(d-1)$, where $a_1'$ corresponds to attachment at one (outer) corner of an area $1$ diagram, and $a_1$ corresponds to attachment at any of the other $6$ possible locations. For example, one can choose the outer corner farthest from $x$,

\begin{equation}
\begin{tikzpicture}[baseline=(current  bounding  box.center),scale=0.7]

\def \yoff {0}
\def \xoff {4.0}

\draw [directed] (0.0,0.0) -- (0.0,2.0) -- (2.0,2.0) -- (2.0,0.0) -- (0.2,0.0);
\draw [reverse directed] (0.2,0.2) rectangle (1.8,1.8);

\def \xoff {0.2}
\def \yoff {0.2}


\fill [green][rotate around={120:(0.0-\xoff,2.0+\yoff)}] (0.0-\xoff,2.0+\yoff) ellipse (0.3 and 0.18);
\fill [blue][rotate around={60:(2.0 + \xoff,2.0+\yoff)}] (2.0 + \xoff,2.0+\yoff) ellipse (0.3 and 0.18);
\fill [green][rotate around={-60:(2.0 + \xoff,0.0-\yoff)}] (2.0 + \xoff,0.0-\yoff) ellipse (0.3 and 0.18);

\def \xoff {0.24}
\def \yoff {0.24}

\fill [green][rotate around={60:(0.2+\xoff,0.2+\yoff)}] (0.2+\xoff,0.2+\yoff) ellipse (0.3 and 0.18);
\fill [green][rotate around={-60:(0.2+\xoff,1.8-\yoff)}] (0.2+\xoff,1.8-\yoff) ellipse (0.3 and 0.18);
\fill [green][rotate around={-120:(1.8-\xoff,1.8-\yoff)}] (1.8-\xoff,1.8-\yoff) ellipse (0.3 and 0.18);
\fill [green][rotate around={120:(1.8-\xoff,0.2+\yoff)}] (1.8-\xoff,0.2+\yoff) ellipse (0.3 and 0.18);


\end{tikzpicture}\ \ \ ,
\end{equation}

\noindent where the blue leaf corresponds to an $a_1'$ attachment site and the green leaves correspond to an $a_1$ attachment site.

\subsubsection{Overlapping of $b$-type graphs}

In the calculation of the dimensionality for attaching $b$-type graphs one can also make improvements by removing contributions which lead to over-counting. One example results from allowing $b$-type diagrams to overlap. For example,

\begin{equation}
\begin{tikzpicture}[baseline=(current  bounding  box.center),scale=0.7]

\def \yoff {0}
\def \xoff {4.0}

\draw [directed] (0.0,0.0) -- (0.0,2.0) -- (2.0,2.0) -- (2.0,0.0) -- (0.2,0.0);
\draw [reverse directed] (0.2,0.2) -- (0.2,1.8) -- (1.4,1.8) -- (1.4,1.6) -- (0.4,1.6) -- (0.4,0.4) -- (1.6,0.4) -- (1.6,1.8) -- (1.8,1.8) -- (1.8,0.2) -- (0.2,0.2);
\draw [reverse directed] (0.6,0.6) rectangle (1.4,1.4);

\draw [directed] (0.0+\xoff,0.0) -- (0.0+\xoff,2.0) -- (2.0+\xoff,2.0) -- (2.0+\xoff,0.0) -- (0.2+\xoff,0.0);
\draw [reverse directed] (0.2+\xoff,0.2) -- (0.2+\xoff,1.8) -- (1.6+\xoff,1.8) -- (1.6+\xoff,0.4) -- (0.4+\xoff,0.4) -- (0.4+\xoff,1.6) -- (1.8+\xoff,1.6) -- (1.8+\xoff,0.2) -- (0.2+\xoff,0.2);
\draw [directed] (0.6+\xoff,0.6) rectangle (1.4+\xoff,1.4); 

\end{tikzpicture}\ \ \ .
\end{equation} 

\noindent The first graph is already counted as it corresponds to

\begin{equation}
\begin{tikzpicture}[baseline=(current  bounding  box.center),scale=0.7]

\def \yoff {0}
\def \xoff {4.0}

\draw [directed] (0.0,0.0) -- (0.0,2.0) -- (2.0,2.0) -- (2.0,0.0) -- (0.2,0.0);
\draw [reverse directed] (0.2,0.2) rectangle (1.8,1.8);

\node[anchor=west] at (2.4,1.0) {\Large{
$\times$
}};

\draw [reverse directed] (0.2+\xoff,0.2) -- (0.2+\xoff,1.8) -- (1.4+\xoff,1.8) -- (1.4+\xoff,1.6) -- (0.4+\xoff,1.6) -- (0.4+\xoff,0.4) -- (1.6+\xoff,0.4) -- (1.6+\xoff,1.8) -- (1.8+\xoff,1.8) -- (1.8+\xoff,0.2) -- (0.2+\xoff,0.2);

\end{tikzpicture} \ \ \ .
\end{equation}

\noindent Since it factorises into a separately integrable contribution from the correlator (left) and a contribution from the determinant (right), the contribution from the determinant cancels against the denominator, resulting in a contribution already contained in

\begin{equation}
\begin{tikzpicture}[baseline=(current  bounding  box.center),scale=0.7]

\def \yoff {0}
\def \xoff {4.0}

\draw [directed] (0.0,0.0) -- (0.0,2.0) -- (2.0,2.0) -- (2.0,0.0) -- (0.2,0.0);
\draw [reverse directed] (0.2,0.2) rectangle (1.8,1.8);

\end{tikzpicture} \ \ \ .
\end{equation}

\noindent The second graph is not already included so one could allow for it. However, performing the group integrations, the contribution from this graph is

\begin{equation}
\begin{tikzpicture}[baseline=(current  bounding  box.center),scale=0.7]

\def \yoff {0}
\def \xoff {0}

\draw [directed] (0.0+\xoff,0.0) -- (0.0+\xoff,2.0) -- (2.0+\xoff,2.0) -- (2.0+\xoff,0.0) -- (0.2+\xoff,0.0);
\draw [reverse directed] (0.2+\xoff,0.2) -- (0.2+\xoff,1.8) -- (1.6+\xoff,1.8) -- (1.6+\xoff,0.4) -- (0.4+\xoff,0.4) -- (0.4+\xoff,1.6) -- (1.8+\xoff,1.6) -- (1.8+\xoff,0.2) -- (0.2+\xoff,0.2);
\draw [directed] (0.6+\xoff,0.6) rectangle (1.4+\xoff,1.4);

\node[anchor=west] at (2.0,1.0) {\Large{
$= 0$
}};

\end{tikzpicture} \ \ \ .
\end{equation}

\noindent Since this graph would be counted incorrectly by multiplying the separate contributions of the two $b$-type graphs we should disallow it as well. The same arguments can be used to justify disallowing overlapping $b$-type diagrams of the form

\begin{equation}
\begin{tikzpicture}[baseline=(current  bounding  box.center),scale=0.7]

\def \yoff {0}
\def \xoff {4.0}

\draw [directed] (0.0,0.0) -- (0.0,2.0) -- (1.6,2.0) -- (1.6,1.8) -- (0.2,1.8) -- (0.2,0.2) -- (1.8,0.2) -- (1.8,2.0) -- (2.0,2.0) -- (2.0,0.0) -- (0.2,0.0);
\draw [directed] (0.4,0.4) rectangle (1.6,1.6);
\draw [reverse directed] (0.6,0.6) rectangle (1.4,1.4);

\draw [directed] (0.0+\xoff,0.0) -- (0.0+\xoff,2.0) -- (1.8+\xoff,2.0) -- (1.8+\xoff,0.2) -- (0.2+\xoff,0.2) -- (0.2+\xoff,1.8) -- (2.0+\xoff,1.8) -- (2.0+\xoff,0.0) -- (0.2+\xoff,0.0);
\draw [reverse directed] (0.4+\xoff,0.4) rectangle (1.6+\xoff,1.6);
\draw [reverse directed] (0.6+\xoff,0.6) rectangle (1.4+\xoff,1.4);

\end{tikzpicture} \ \ \ .
\end{equation}

\noindent Allowing $b$-type graphs to overlap as in the first diagram would result in over-counting due to factorisation. Allowing them to overlap as in the second diagram would also result in mis-counting, since the diagram evaluates to zero.

\subsubsection{Avoiding over-counting of $b$-type graphs}

To improve the dimensionality $b_1$, for attaching $b$-type graphs to area $1$ type graphs, it is useful to subtract off dimensions which lead to over-counting. For example, attaching a $b$-type graph to the leaf in

\begin{equation}
\begin{tikzpicture}[baseline=(current  bounding  box.center),scale=0.7]

\def \yoff {0}
\def \xoff {4.0}

\draw [directed] (0.0,0.0) -- (0.0,2.0) -- (2.0,2.0) -- (2.0,0.0) -- (0.2,0.0);
\draw [reverse directed] (0.2,0.2) rectangle (1.8,1.8);

\def \xoff {0.2}
\def \yoff {0.2}


\fill [green][rotate around={120:(0.0-\xoff,2.0+\yoff)}] (0.0-\xoff,2.0+\yoff) ellipse (0.3 and 0.18);

\end{tikzpicture}\ \ \ ,
\end{equation}

\noindent could result (among others) in diagrams of the form

\begin{equation}
\begin{tikzpicture}[baseline=(current  bounding  box.center),scale=0.7]

\def \yoff {0}
\def \xoff {8.0}

\draw [directed] (0.0,0.0) -- (0.0,1.8) -- (-0.2,1.8) -- (-0.2,0.0) -- (-2.2,0.0) -- (-2.2,2.0) -- (2.0,2.0) -- (2.0,0.0) -- (0.2,0.0);
\draw [reverse directed] (-2.0,0.2) rectangle (-0.4,1.8);
\draw [reverse directed] (0.2,0.2) rectangle (1.8,1.8);

\draw [directed] (0.0+\xoff,0.0) -- (0.0+\xoff,1.8) -- (-2.2+\xoff,1.8) -- (-2.2+\xoff,0.0) -- (-0.2+\xoff,0.0) -- (-0.2+\xoff,2.0) -- (2.0+\xoff,2.0) -- (2.0+\xoff,0.0) -- (0.2+\xoff,0.0);
\draw [directed] (-2.0+\xoff,0.2) rectangle (-0.4+\xoff,1.6);
\draw [reverse directed] (0.2+\xoff,0.2) rectangle (1.8+\xoff,1.8);


\end{tikzpicture}\ \ \ ,
\end{equation}

\noindent which would lead to over-counting. The first diagram corresponds to attaching

\begin{equation}
\begin{tikzpicture}[baseline=(current  bounding  box.center),scale=0.7]

\def \yoff {0}
\def \xoff {0.0}

\draw [reverse directed] (0.0+\xoff,0.0) -- (0.0+\xoff,1.8) -- (0.2+\xoff,1.8) -- (0.2+\xoff,0.0);

\node[anchor=west] at (1.0,1.0) {\Large{
$\times$
}};

\def \xoff {5.0}

\draw [directed] (0.0+\xoff,0.0) -- (-1.8+\xoff,0.0) -- (-1.8+\xoff,2.0) -- (2.0+\xoff,2.0) -- (2.0+\xoff,0.0) -- (0.2+\xoff,0.0);
\draw [reverse directed] (-1.6+\xoff,0.2) rectangle (0.0+\xoff,1.8);
\draw [reverse directed] (0.2+\xoff,0.2) rectangle (1.8+\xoff,1.8);


\end{tikzpicture}\ \ \ ,
\end{equation}

\noindent at $x$. The second corresponds to
\begin{equation}
\begin{tikzpicture}[baseline=(current  bounding  box.center),scale=0.7]

\def \yoff {0}
\def \xoff {0.0}

\draw [directed] (-0.2+\xoff,0.0) -- (-0.2+\xoff,1.8) -- (-2.2+\xoff,1.8) -- (-2.2+\xoff,0.2) -- (0.0+\xoff,0.2) -- (0.0+\xoff,2.0) -- (2.0+\xoff,2.0) -- (2.0+\xoff,0.0) -- (0.2+\xoff,0.0);
\draw [directed] (-2.0+\xoff,0.4) rectangle (-0.4+\xoff,1.6);
\draw [reverse directed] (0.2+\xoff,0.2) rectangle (1.8+\xoff,1.8);


\end{tikzpicture}\ \ \ ,
\end{equation}

\noindent which is formed by combining two $b$-type diagrams at $x$. Avoiding also direct overlap of $b$-type diagrams discussed in the previous subsection, the dimensionality at the external corners neighbouring $x$ is $b_1 = 4(d-1)^2$.

Consider the addition of a $b$-type diagram at one of the internal corners

\begin{equation}
\begin{tikzpicture}[baseline=(current  bounding  box.center),scale=0.7]

\def \yoff {0}
\def \xoff {4.0}

\draw [directed] (0.0,0.0) -- (0.0,2.0) -- (2.0,2.0) -- (2.0,0.0) -- (0.2,0.0);
\draw [reverse directed] (0.2,0.2) rectangle (1.8,1.8);

\def \xoff {0.24}
\def \yoff {0.24}

\fill [blue][rotate around={-120:(1.8-\xoff,1.8-\yoff)}] (1.8-\xoff,1.8-\yoff) ellipse (0.3 and 0.18);


\end{tikzpicture}\ \ \ .
\end{equation}

\noindent One possible attachment would look like

\begin{equation}
\begin{tikzpicture}[baseline=(current  bounding  box.center),scale=0.7]

\def \yoff {0}
\def \xoff {4.0}

\draw [directed] (0.0,0.0) -- (0.0,2.0) -- (2.0,2.0) -- (2.0,0.0) -- (0.2,0.0);
\draw [reverse directed] (0.2,0.2) -- (0.2,1.8) -- (2.2,1.8) -- (2.2,0.0) -- (4.2,0.0) -- (4.2,1.6) -- (1.8,1.6) -- (1.8,0.2) -- (0.2,0.2);
\draw [reverse directed] (2.4,0.2) rectangle (4.0,1.4);

\end{tikzpicture}\ \ \ ,
\end{equation}

\noindent however, this one is equivalent to

\begin{equation}
\begin{tikzpicture}[baseline=(current  bounding  box.center),scale=0.7]

\def \yoff {0}
\def \xoff {4.0}

\draw [directed] (0.0,0.0) -- (0.0,2.0) -- (2.0,2.0) -- (2.0,0.0) -- (0.2,0.0);
\draw [reverse directed] (0.2,0.2) -- (0.2,1.8) -- (1.8,1.8) -- (1.8,0.4) -- (4.2,0.4) -- (4.2,2.0) -- (2.2,2.0) -- (2.2,0.2) -- (0.2,0.2);
\draw [reverse directed] (2.4,0.6) rectangle (4.0,1.8);

\end{tikzpicture}\ \ \ ,
\end{equation}

\noindent where the attachment is at the lower right internal corner. It is therefore important, when attaching a neighbouring area $1$-diagram, to remove the contributions to the dimensionality from re-tracing along the internal plaquette. There are $2(d-1)$ ways to attach in this way from one of the internal corners, and we need to remove an additional contribution from direct overlap of area $1$-diagrams by backtracking along a link. The remaining contribution is $b_1' = 4d(d-1)-[2(d-1)+1]$.

Finally consider attachment to the far external corner

\begin{equation}
\begin{tikzpicture}[baseline=(current  bounding  box.center),scale=0.7]

\def \yoff {0}
\def \xoff {4.0}

\draw [directed] (0.0,0.0) -- (0.0,2.0) -- (2.0,2.0) -- (2.0,0.0) -- (0.2,0.0);
\draw [reverse directed] (0.2,0.2) rectangle (1.8,1.8);

\def \xoff {0.2}
\def \yoff {0.2}


\fill [green][rotate around={60:(2.0 + \xoff,2.0+\yoff)}] (2.0 + \xoff,2.0+\yoff) ellipse (0.3 and 0.18);

\end{tikzpicture}\ \ \ .
\end{equation}

\noindent One possible attachment is

\begin{equation}
\begin{tikzpicture}[baseline=(current  bounding  box.center),scale=0.7]

\def \yoff {0}
\def \xoff {4.0}

\draw [directed] (0.0,0.0) -- (0.0,2.0) -- (2.2,2.0) -- (2.2,0.0) -- (4.2,0.0) -- (4.2,1.8) -- (2.0,1.8) -- (2.0,0.0) -- (0.2,0.0);
\draw [reverse directed] (0.2,0.2) rectangle (1.8,1.8);
\draw [directed] (2.4,0.2) rectangle (4.0,1.6);

\end{tikzpicture}\ \ \ ,
\end{equation}

\noindent which is equivalent to

\begin{equation}
\begin{tikzpicture}[baseline=(current  bounding  box.center),scale=0.7]

\def \yoff {0}
\def \xoff {4.0}

\draw [directed] (0.0,0.2) -- (0.0,2.0) -- (2.0,2.0) -- (2.0,0.0) -- (4.2,0.0) -- (4.2,1.8) -- (2.2,1.8) -- (2.2,0.2) -- (0.2,0.2);
\draw [reverse directed] (0.2,0.4) rectangle (1.8,1.8);
\draw [directed] (2.4,0.2) rectangle (4.0,1.6);

\end{tikzpicture}\ \ \ .
\end{equation}

\noindent Including all possible ways of folding the diagram which would lead to double counting, the contribution to subtract off the dimensionality is $2(d-1)$. Since an area-$1$ diagram can also neighbour the top link in the same way this amount needs to be subtracted twice. The total dimensionality at the external corner is therefore $b_1 = 4(d-1)^2$.

\subsection{Over-counting resulting from symmetries}
\label{winding}

In this section we examine diagrams with symmetries. In the first case, this symmetry leads to over-counting, and in the second case it does not.

Consider a graph of the form (\ref{type-d34}),

\begin{equation}
\begin{tikzpicture}[baseline=(current  bounding  box.center),scale=0.7]
\def \yoff {0}
\draw [directed] (0.0,0.0) -- (0.0,2.0) -- (2.0,2.0) -- (2.0,0.0) -- (0.2,0.0);
\draw [directed] (0.2,0.4) -- (0.2,1.8) -- (1.8,1.8) -- (1.8,0.2) -- (0.4,0.2) -- (0.4,1.6) -- (1.6,1.6) -- (1.6,0.4) -- (0.2,0.4);
%
%
\end{tikzpicture}\ \ \ ,
\end{equation}

\noindent which contains a gauge field loop that winds twice before closing on itself. The graphs in (\ref{type-g2}), (\ref{type-g3}) also belong to this category. One source of over-counting comes about when asymmetric attachments are made to the multiply-wound loop. In this case, the over-counting results due to symmetry under rotations by $4$ lattice sites of the internal loop. For example, consider two different attachments, represented by the green and blue leaves in

\begin{equation}
\begin{tikzpicture}[baseline=(current  bounding  box.center),scale=0.7]
\def \yoff {0}
\def \xoff {0}
\draw [directed] (0.0,0.0) -- (0.0,2.0) -- (2.0,2.0) -- (2.0,0.0) -- (0.2,0.0);
\draw [directed] (0.2,0.4) -- (0.2,1.8) -- (1.8,1.8) -- (1.8,0.2) -- (0.4,0.2) -- (0.4,1.6) -- (1.6,1.6) -- (1.6,0.4) -- (0.2,0.4);
\fill [green][rotate around={120:(1.4-\xoff,0.6+\yoff)}] (1.4-\xoff,0.6+\yoff) ellipse (0.3 and 0.18);
\fill [blue][rotate around={-60:(1.95+\xoff,0.05-\yoff)}] (1.95+\xoff,0.05-\yoff) ellipse (0.3 and 0.18);
\end{tikzpicture}\ \ \ .
\end{equation}

\noindent Since these are both attached to the same loop in the same corner it makes no difference if one attaches at the green leaf or the blue one. Such attachments result in identical diagrams which can be transformed into each other under rotations by $4$ lattice sites. Therefore if attachment at both sites is allowed with the same dimensionality then there will be over-counting. Over-counting also results when the attachments are made in two different corners on the same loop, since the attachments can be shifted $4$ sites along the loop to give an identical diagram, which gets counted separately. Notice that, if identical attachments are made at both the green and blue attachment sites simultaneously then there is no over-counting. We have not yet accounted for this effect in our calculations, so it is a source of error.

It is important to note that not all symmetries lead to over-counting. There also exists a symmetry in diagrams of the form

\begin{equation}
\begin{tikzpicture}[baseline=(current  bounding  box.center),scale=0.7]
\def \yoff {0}
\draw [directed] (0.0,0.0) -- (0.0,2.0) -- (2.0,2.0) -- (2.0,0.0) -- (0.2,0.0);
\draw [directed] (0.2,0.2) rectangle (1.8,1.8);
\draw [directed] (0.4,0.4) rectangle (1.6,1.6);
%
%
\end{tikzpicture}\ \ \ ,
\end{equation}

\noindent with respect to interchange of the two internal loops (also true in $L=8$ diagrams of the form (\ref{type-g1})). In this case there is no over-counting when making asymmetric attachments to the internal loops. The contributions from the internal loops come down from the exponential in (\ref{detD}), so if one of the loops takes a different shape, then it is necessary to count it twice.

\section{Results}
\label{results}

Using the procedure outlined in Section \ref{martin-siu}, and the considerations outlined in the previous section for reducing over-counting, it is possible to obtain the chiral condensate to some order by solving the appropriate truncated system of equations. In what follows we present results including area $0$ and $1$ diagrams up to order $L = 8$.


\subsection{Asymptotic solutions for large $N_f/N_c$}

First consider the contributions up to $L = 4$, that is all possible diagrams formed from type $a$ (\ref{type-a}) and type $b$ (\ref{type-b}) sub-diagrams. The system of equations, using (\ref{ga}), (\ref{gb}), and the considerations in Section \ref{over-counting}, is
\EQ{
g_a = \frac{1}{a_0 g_a + b_0 g_b} \, ,
\label{ga-1}
}
\EQ{
g_b = \frac{\frac{N_f}{N_c}}{(a_1 g_a + b_1' g_b)^4 (a_1 g_a + b_1 g_b)^2 (a_1' g_a + b_1 g_b)} \, ,
\label{gb-1}
}
where the dimensionalities $x_n$ are given in Appendix \ref{A}. The chiral condensate as a function of $N_f$ can be obtained from (\ref{gser}) and (\ref{chicong}).

We are interested in finding real roots of the set of self-consistent equations for large $N_f/N_c$, where we will take $d=4$ in what follows. Solving \eqref{ga-1} for $g_b$ and plugging this solution in \eqref{gb-1}, we find that solutions for $g_a$ are determined by the roots of the polynomial equation
\begin{align} \label{polyga}
& \frac{2825761}{10504375}-\frac{32255028}{10504375} g_a^2 +\frac{15618171}{1500625}g_a^4-\frac{707824}{42875}g_a^6+\left(\frac{2403}{175}-\frac{60466176}{10504375}\frac{N_f}{N_c} \right) g_a^8 \nonumber \\ & \qquad -\frac{204}{35}g_a^{10} +g_a^{12}=0 \,.
\end{align}
Once real solutions for $g_a$ of the above polynomial have been found, the corresponding real solutions for $g_b$ are found from \eqref{ga-1}
\begin{equation} \label{gbsol}
g_b = \frac{1}{b_0 g_a} - \frac{a_0}{b_0} g_a \,.
\end{equation}
The number of real roots of \eqref{polyga} in a certain interval can be found by applying Sturm's theorem. For generic\footnote{For very small values of $N_f/N_c$ ($N_f/N_c < 2 \cdot 10^{-6}$), the polynomial \eqref{polyga} has four real roots.} values of $N_f/N_c$, one finds that the number of real roots in the interval $(0,+\infty)$ is given by 2. Since the polynomial \eqref{polyga} is even in $g_a$, the negatives of these roots are also roots and hence there are four real roots in total.

\begin{figure}[h]
\centering
\includegraphics[scale=0.54]{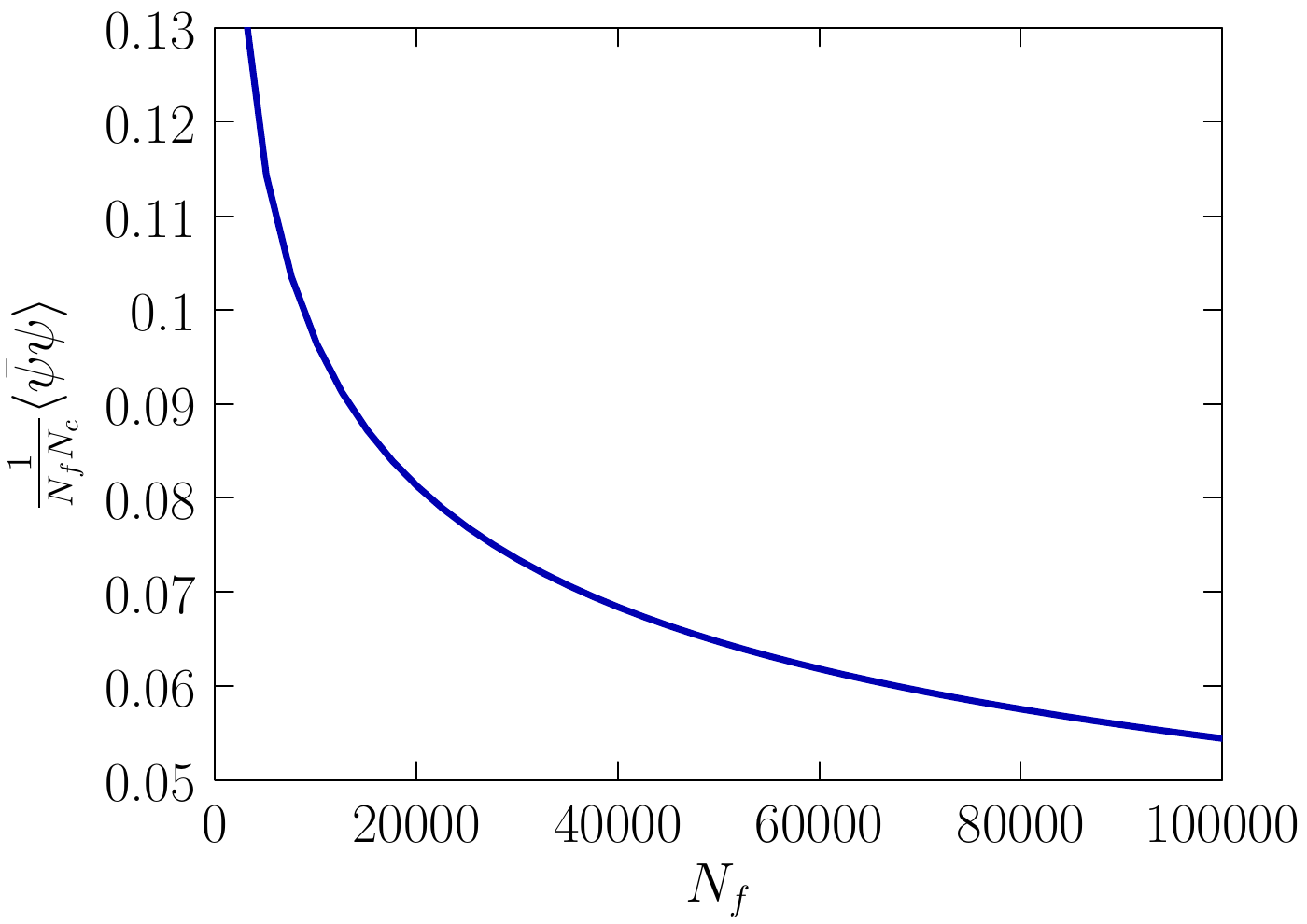}
\includegraphics[scale=0.54]{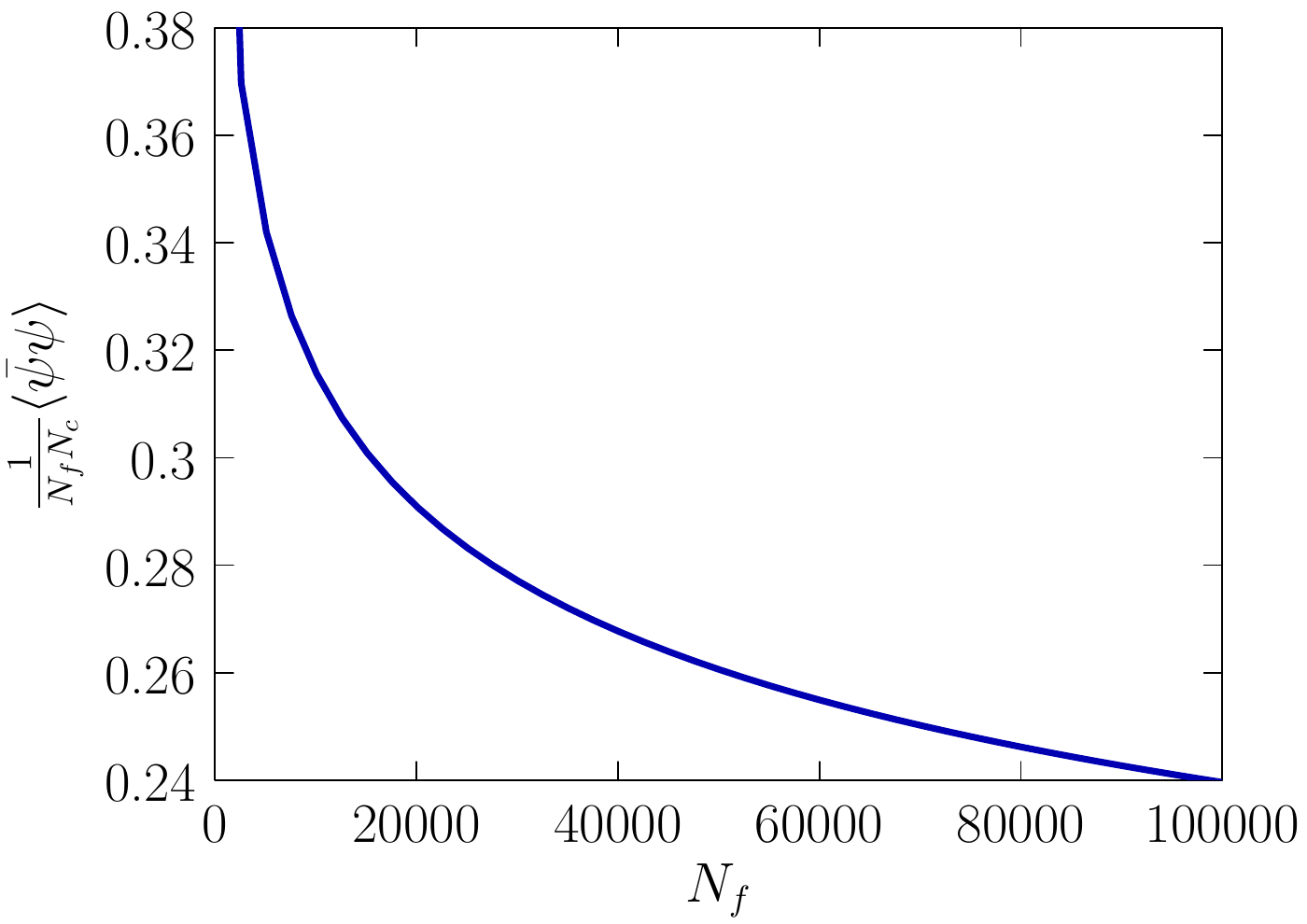}
\caption{Plots of the two (positive) approximate solutions for the chiral condensate, for large $N_f/N_c$. The left figure represents the solution of \eqref{cc1}, the right figure the solution of \eqref{cc2}.} \label{figcc}
\end{figure}

Here, we are interested in finding asymptotic expansions for the roots of \eqref{polyga}, for large $N_f/N_c \gg 1$. Multiplying \eqref{polyga} by $\epsilon = N_c/N_f$, we wish to apply perturbation theory to obtain real solutions of 
\begin{align} \label{polyga2}
& \epsilon \frac{2825761}{10504375}- \epsilon \frac{32255028}{10504375} g_a^2 +\epsilon \frac{15618171}{1500625}g_a^4-\epsilon \frac{707824}{42875}g_a^6+\left(\epsilon \frac{2403}{175}-\frac{60466176}{10504375}\right) g_a^8 \nonumber \\ & \qquad -\epsilon \frac{204}{35}g_a^{10} +\epsilon g_a^{12}=0 \,,
\end{align}
for $\epsilon \ll 1$. Asymptotic expansions in $\epsilon$ for the roots of this polynomial can then be found via singular perturbation theory \cite{SimmondsMann,BenderBook}. In particular, one looks for roots of the form
\begin{equation}
g_a = \epsilon^P w(\epsilon) \,,
\end{equation}
where $w(\epsilon)$ is regular in the limit $\epsilon \rightarrow 0$ and $\lim_{\epsilon \rightarrow 0} w(\epsilon)$ is assumed to be non-zero. The exponent $P$ can be determined via singular perturbation theory to be either $-1/4$ or $1/8$. Let us focus on solutions with $P=-1/4$ first. Plugging $g_a = \epsilon^{-1/4} w(\epsilon)$ in \eqref{polyga2}, one obtains
\begin{align}
& \frac{2825761}{10504375}\epsilon^3-\frac{32255028}{10504375}\epsilon^{5/2} w(\epsilon)^2+\frac{15618171}{1500625}\epsilon^2 w(\epsilon)^4-\frac{707824}{42875}\epsilon^{3/2} w(\epsilon)^6 \nonumber \\ & \qquad -\frac{60466176}{10504375}w(\epsilon)^8 +\frac{2403}{175} \epsilon  w(\epsilon)^8-\frac{204}{35} \sqrt{\epsilon} w(\epsilon)^{10}+w(\epsilon)^{12} = 0\,.
\end{align}
Upon renaming $\epsilon = \beta^2$, one obtains an expression that only involves integer powers of $\beta$
\begin{align} \label{polygabeta}
& \frac{2825761}{10504375}\beta^6-\frac{32255028}{10504375}\beta^{5} w(\beta)^2+\frac{15618171}{1500625}\beta^4 w(\beta)^4-\frac{707824}{42875}\beta^{3} w(\beta)^6 \nonumber \\ & \qquad -\frac{60466176}{10504375}w(\beta)^8 +\frac{2403}{175} \beta^2  w(\beta)^8-\frac{204}{35} \beta w(\beta)^{10}+w(\beta)^{12} = 0\,.
\end{align}
One can then propose an ordinary series solution for $w(\beta)$
\begin{equation} \label{seriesw}
w(\beta) = \sum_{i=0}^\infty \omega_i \beta^i\,.
\end{equation}
The coefficients $\omega_i$ can be solved for by plugging \eqref{seriesw} in \eqref{polygabeta} and requiring that the result is zero at every order in $\beta$. This leads to a set of equations for $\omega_i$, that can be solved in an iterative manner. Restricting ourselves to sixth order in $\beta$, we thus obtain asymptotic expansions for two solutions, that are each others negatives. Expressed again in terms of $\epsilon$, these are given by
\begin{align}
g_a & =  \pm \frac{1}{\epsilon^{1/4}}\Bigg(\frac{36 \sqrt{6}}{35 7^{1/4}}+\frac{17 7^{1/4} \sqrt{\epsilon }}{12 \sqrt{6}}-\frac{45185 7^{3/4} \epsilon }{124416 \sqrt{6}}+\frac{11150869975 7^{1/4} \epsilon ^{3/2}}{8707129344 \sqrt{6}} \nonumber \\ &\quad \quad -\frac{145105138793125 7^{3/4} \epsilon ^2}{180551034077184 \sqrt{6}}  +\frac{616080011402463125 7^{1/4} \epsilon ^{5/2}}{155996093442686976 \sqrt{6}}\nonumber \\ &\quad \quad -\frac{43215120505930210709375 7^{3/4} \epsilon ^3}{14556307471324007104512 \sqrt{6}}\Bigg) \,.
\end{align}
Similarly, for $P = 1/8$, one obtains asymptotic expansions for two solutions, that are each others negatives and are given by
\begin{align}
g_a & =  \pm \epsilon ^{1/8} \Bigg(\frac{\sqrt{41}}{6 6^{1/4}}-\frac{13 \sqrt{41} \epsilon ^{1/4}}{48 6^{3/4}}+\frac{3995 \sqrt{41} \sqrt{\epsilon }}{62208 6^{1/4}}-\frac{2827435 \sqrt{41} \epsilon ^{3/4}}{13436928 6^{3/4}} +\frac{85021433 \sqrt{41} \epsilon }{859963392 6^{1/4}}\nonumber \\ & \quad \quad -\frac{141672440399 \sqrt{41} \epsilon ^{5/4}}{557256278016 6^{3/4}}+\frac{13186932605159 \sqrt{41} \epsilon ^{3/2}}{80244904034304 6^{1/4}}\Bigg) \,.
\end{align}
Asymptotic expansions for $g_b$ can then be found by using these expansions for $g_a$ in \eqref{gbsol}. The expansions for $g_a$ and $g_b$ can be used to obtain approximate solutions for the chiral condensate, that are valid for $N_f/N_c \gg 1$. In particular, one obtains two positive solutions for the chiral condensate $\lim_{m\rightarrow 0}\mathrm{tr}\left[G(x,x) \right]/(N_s N_f N_c) = 2/g$, given by
\begin{align} \label{cc1}
\frac{2}{g} & =  \frac{35 \cdot 7^{1/4}}{24 \sqrt{6}} \left(\frac{N_c}{N_f}\right)^{1/4}-\frac{1225\cdot 7^{3/4}}{11664 \sqrt{6}} \left(\frac{N_c}{N_f}\right)^{3/4} +\frac{1651587875 \cdot7^{1/4}}{5804752896 \sqrt{6}} \left(\frac{N_c}{N_f}\right)^{5/4} \nonumber \\ & \quad-\frac{1810166421875\cdot 7^{3/4}}{11284439629824 \sqrt{6}}\left(\frac{N_c}{N_f}\right)^{7/4}+\frac{2087791584389809375\cdot 7^{1/4}}{2807929681968365568 \sqrt{6}}\left(\frac{N_c}{N_f}\right)^{9/4}\nonumber \\ & \quad -\frac{163362753019994171875 \cdot 7^{3/4} }{303256405652583481344 \sqrt{6}} \left(\frac{N_c}{N_f}\right)^{11/4} + {\cal O} \left((N_c/N_f)^{13/4} \right)\,,
\end{align}
and 
\begin{align} \label{cc2}
\frac{2}{g} & = \frac{\sqrt{41}}{4\cdot 6^{1/4}}  \left(\frac{N_c}{N_f}\right)^{1/8}-\frac{35 \sqrt{41}}{288 \cdot 6^{3/4}} \left(\frac{N_c}{N_f}\right)^{3/8}-\frac{10073 \sqrt{41}}{124416\cdot 6^{1/4}} \left(\frac{N_c}{N_f}\right)^{5/8}\nonumber \\ & \quad+\frac{35399 \sqrt{41}}{2985984\cdot 6^{3/4}} \left(\frac{N_c}{N_f}\right)^{7/8} -\frac{103461197 \sqrt{41}}{15479341056\cdot 6^{1/4}}\left(\frac{N_c}{N_f}\right)^{9/8}\nonumber \\ & \quad+\frac{171638444453\cdot \sqrt{41}}{1114512556032\cdot 6^{3/4}}  \left(\frac{N_c}{N_f}\right)^{11/8} + {\cal O}\left((N_c/N_f)^{13/8} \right)\,.
\end{align}
These two solutions are plotted (for large $N_f/N_c$) in Figure \ref{figcc}.

\subsection{Numerical results for $N_c = 3$}

Consider again the contributions up to $L = 4$, formed from type $a$ (\ref{type-a}) and type $b$ (\ref{type-b}) sub-diagrams. The system of equations for $g_a$ and $g_b$ is as in the previous subsection given by (\ref{ga-1}), (\ref{gb-1}), including the considerations in Section \ref{over-counting}, such that the chiral condensate as a function of $N_f$ can be obtained from (\ref{gser}) and (\ref{chicong}) by solving the system of equations numerically.

\begin{figure}[h!]
\begin{minipage}{0.5\textwidth}
\includegraphics[scale=0.55]{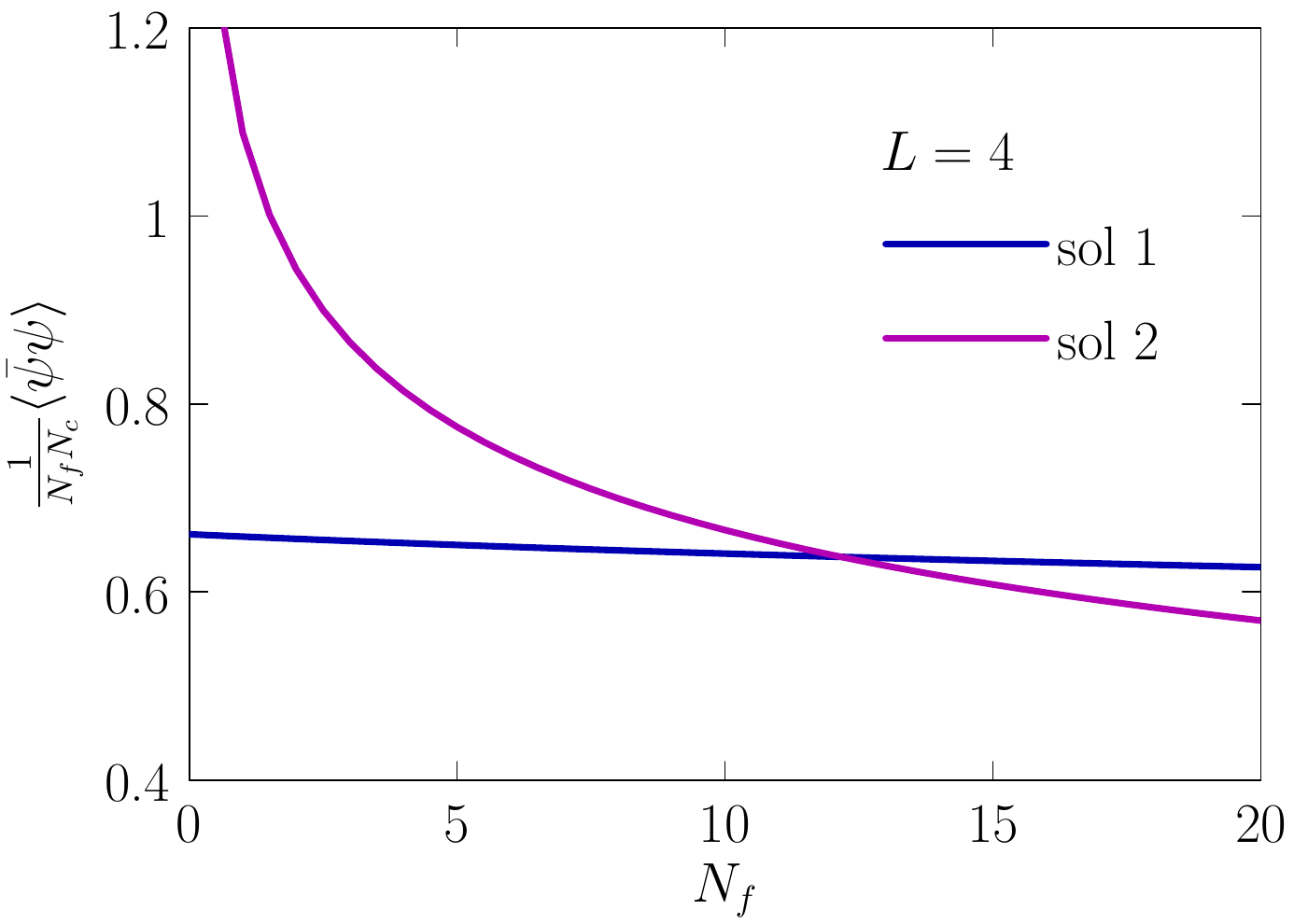}
\end{minipage}
\begin{minipage}{0.5\textwidth}
\includegraphics[scale=0.55]{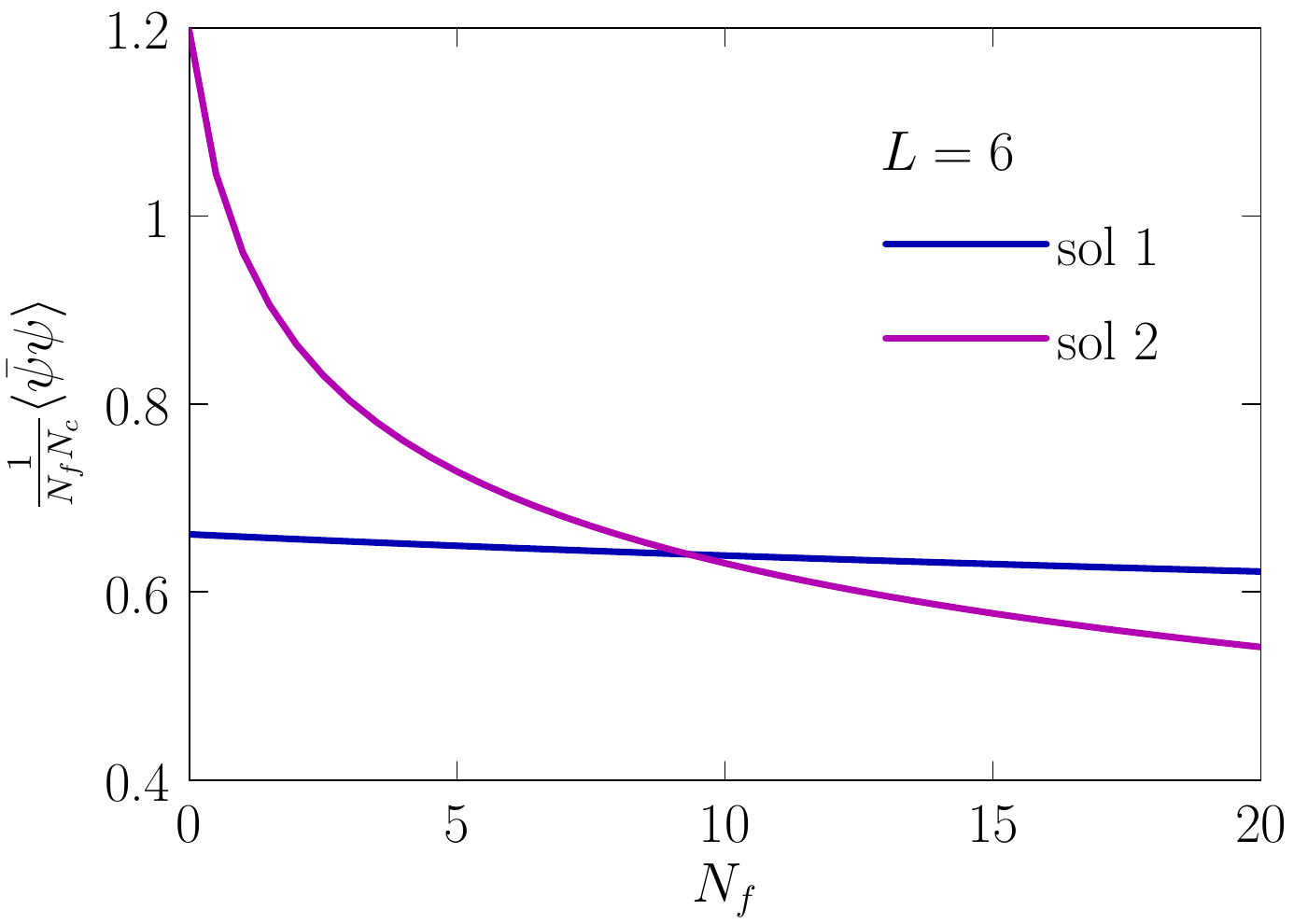}
\end{minipage}
\caption{$\frac{1}{N_f N_c} \langle {\bar \psi} \psi \rangle$ vs. $N_f$ including area $1$ diagrams up to order $L = 4$ (left), and $L = 6$ (right).}
\label{chi-bd}
\end{figure}

Results for $\frac{1}{N_f N_c} \langle \bar{\psi} \psi \rangle$, including base diagrams up to $L=4$, with $N_c = 3$ and $d = 4$ are shown in Figure \ref{chi-bd} (left). As in the previous section, solving the system of equations results in two solutions. One of these, solution $1$, approaches the result of \cite{KlubergStern:1982bs,Martin:1982tb} as $N_f \rightarrow 0$. For the other, solution $2$, $\frac{1}{N_f N_c} \langle {\bar \psi} \psi \rangle \rightarrow \infty$ as $N_f \rightarrow 0$. In the limit $N_f \rightarrow \infty$, both solutions approach zero, solution $2$ falling off more quickly. There is no sign of a discontinuity, at any $N_f$, for either of the solutions.

To determine the effect of including higher order diagrams consider the contributions of area $1$ diagrams up to $L = 6$, formed from type $a$, $b$, and $d$ (\ref{type-d31}) - (\ref{type-d34}) sub-diagrams. The system of equations is
\EQ{
g_a = \frac{1}{a_0 g_a + b_0 g_b + b_0 g_d} \, ,
\label{ga-2}
}
\EQ{
g_b = \frac{\frac{N_f}{N_c}}{(a_1 g_a + b_1' g_b + b_1' g_d)^4 (a_1 g_a + b_1 g_b + b_1 g_d)^2 (a_1' g_a + b_1 g_b + b_1 g_d)} \, ,
}
\EQ{
g_d = \frac{\frac{1}{3} \left( \frac{1}{2} N_f^2 + 2 N_f + 1 \right)}{(a_1 g_a + b_1' g_b + b_1' g_d)^8 (a_1 g_a + b_1 g_b + b_1 g_d)^2 (a_1' g_a + b_1 g_b + b_1 g_d)} \, .
\label{gd-2}
}
In \eqref{gd-2}, we have explicitly set $N_c = 3$, as the contribution from $d$-type diagrams is otherwise zero. 
The results for $\frac{1}{N_f N_c} \langle {\bar \psi} \psi \rangle$ from \eqref{ga-2} - \eqref{gd-2} as a function of $N_f$ (and with $N_c=3$) are shown in Figure \ref{chi-bd} (right). The results are quite similar to the case of $L = 4$, suggesting that the solutions are converging, however, solution $2$ now approaches a finite value around $1.2$ in the limit $N_f \rightarrow 0$. For all $N_f$, the values of $\frac{1}{N_f N_c} \langle {\bar \psi} \psi \rangle$ have decreased. In the limit $N_f \rightarrow \infty$, the differences from the $L=4$ truncation become more apparent but both solutions still approach zero, without exhibiting any discontinuities.

\begin{figure}[h!]
\begin{minipage}{0.5\textwidth}
\includegraphics[scale=0.55]{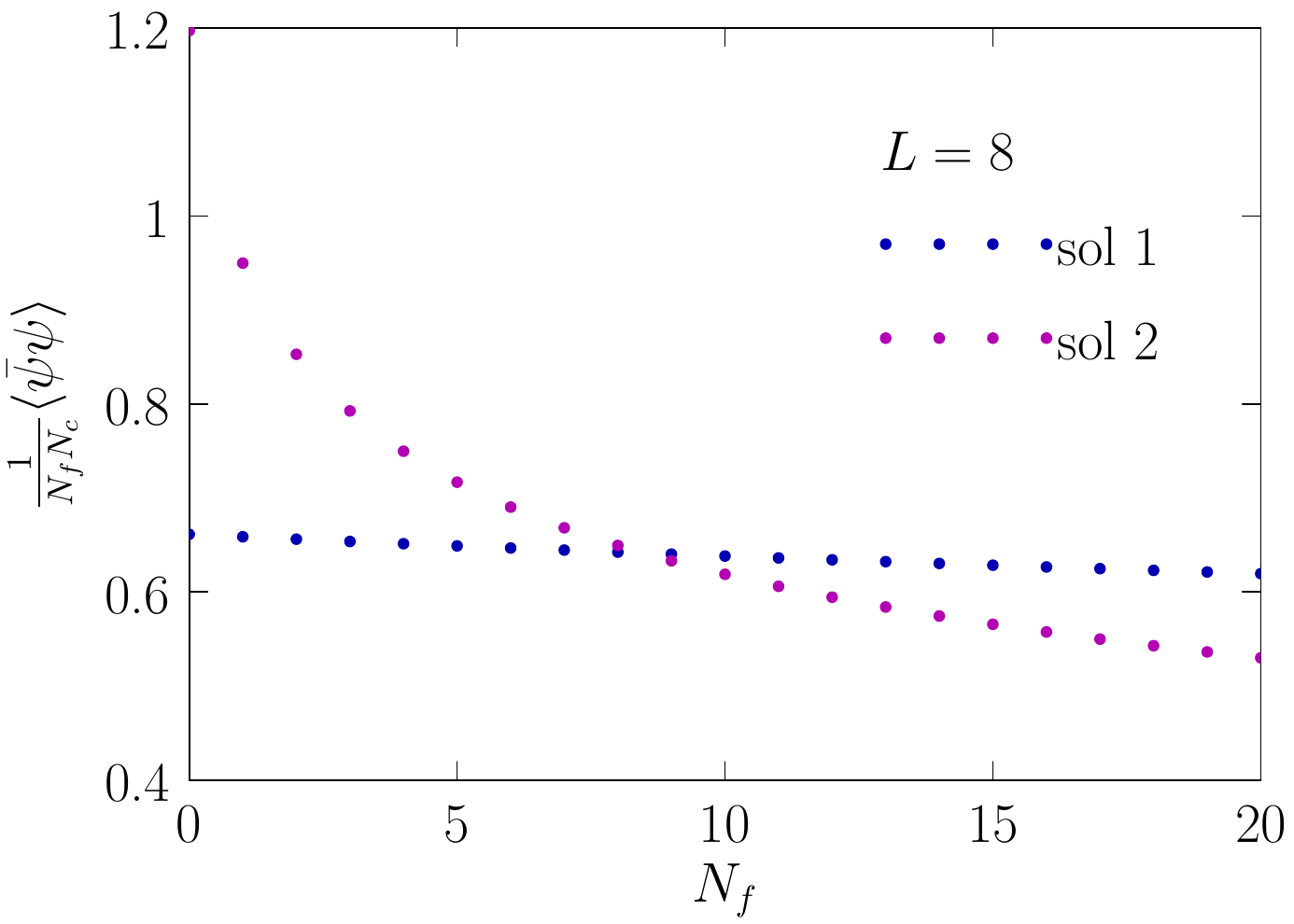}
\end{minipage}
\begin{minipage}{0.5\textwidth}

\end{minipage}
\caption{$\frac{1}{N_f N_c} \langle {\bar \psi} \psi \rangle$ vs. $N_f$ including area $1$ diagrams up to order $L = 8$ (left).}
\label{chi-g}
\end{figure}

Finally, consider the effect of including contributions of area $1$ diagrams up to $L = 8$, formed from type $a$, $b$, $d$, and $g$ (\ref{type-g1}) - (\ref{type-g3}) sub-diagrams. The system of equations is
\EQ{
g_a = \frac{1}{a_0 g_a + b_0 g_b + b_0 g_d + b_0 g_g} \, ,
\label{ga-all}
}
\EQ{
g_b = \frac{\frac{N_f}{N_c}}{(a_1 g_a + b_1' g_b + b_1' g_d + b_1' g_g)^4 (a_1 g_a + b_1 g_b + b_1 g_d + b_1 g_g)^2 (a_1' g_a + b_1 g_b + b_1 g_d + b_1 g_g)} \, ,
}
\EQ{
g_d = \frac{\frac{1}{3} \left( \frac{1}{2} N_f^2 + 2 N_f + 1 \right)}{(a_1 g_a + b_1' g_b + b_1' g_d + b_1' g_g)^8 (a_1 g_a + b_1 g_b + b_1 g_d + b_1 g_g)^2 (a_1' g_a + b_1 g_b + b_1 g_d + b_1 g_g)} \, ,
\label{gd-all}
}
\EQ{
g_g = \frac{\frac{1}{N_c} \left( N_f^3 + 2 N_f \right)}{(a_1 g_a + b_1' g_b + b_1' g_d + b_1' g_g)^{12} (a_1 g_a + b_1 g_b + b_1 g_d + b_1 g_g)^2 (a_1' g_a + b_1 g_b + b_1 g_d + b_1 g_g)} \, .
\label{gg-all}
}
The results for $\frac{1}{N_f N_c}\langle {\bar \psi} \psi \rangle$ as a function of $N_f$ are given in Figure \ref{chi-g}. While the data points haven't shifted so much from the $L = 6$ results, one notable difference is the absence of real solutions for $\frac{1}{N_f N_c}\langle {\bar \psi} \psi \rangle$ for non-integer values of $N_f$. At lower orders, the two real solutions were continuous solutions for all $N_f$.

\begin{figure}[h!]
\begin{minipage}{0.5\textwidth}
\includegraphics[scale=0.55]{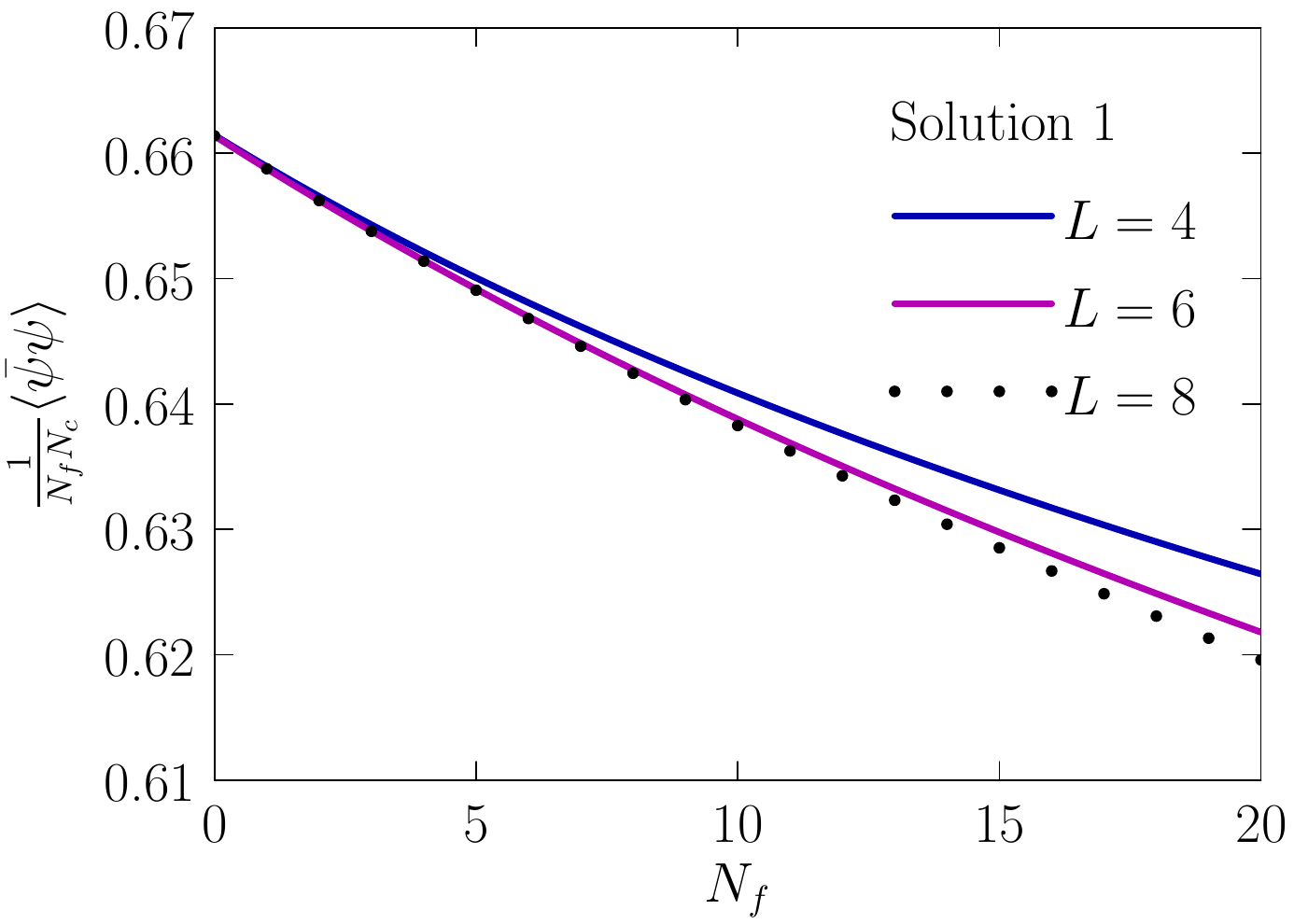}
\end{minipage}
\begin{minipage}{0.5\textwidth}
\includegraphics[scale=0.55]{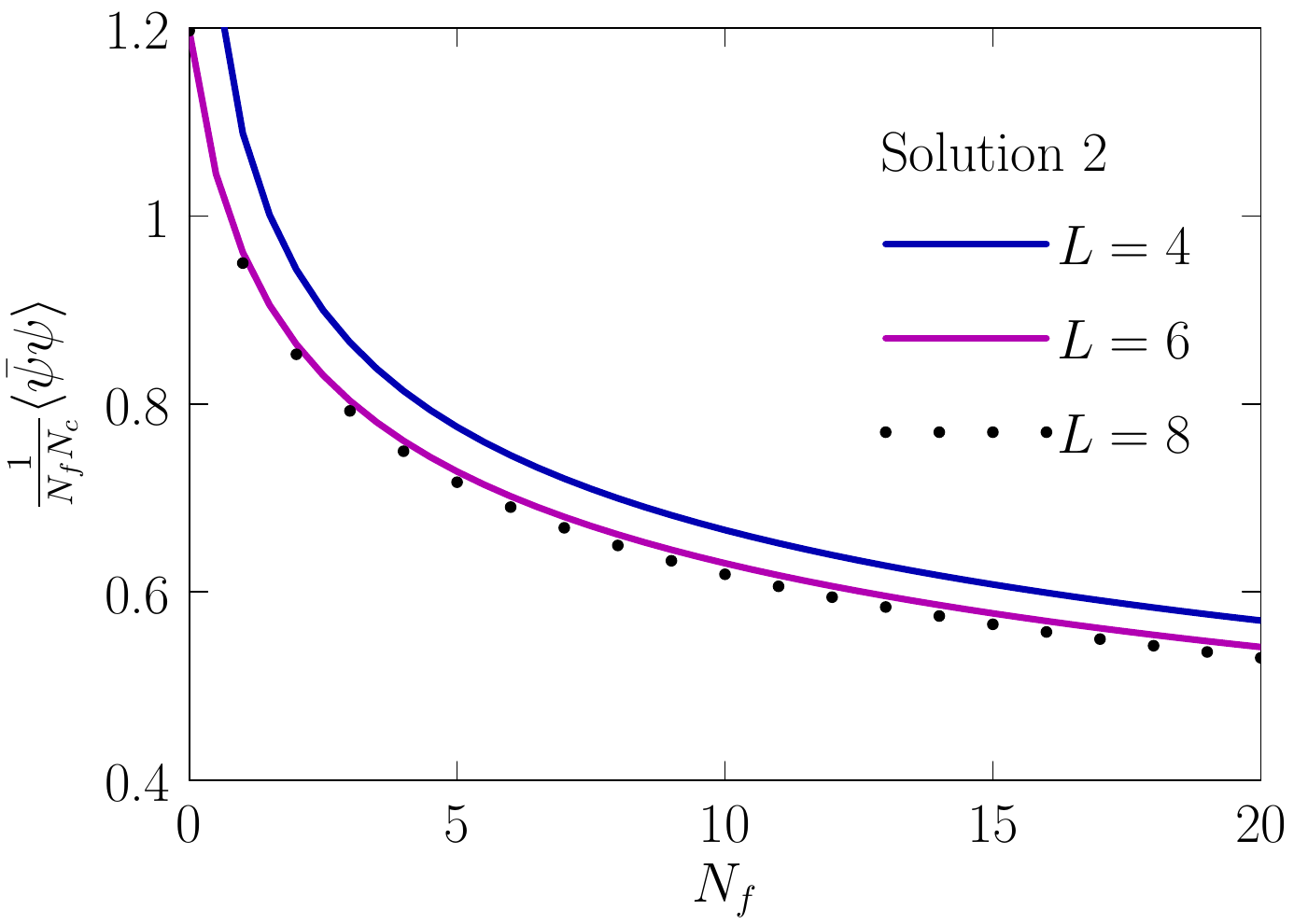}
\end{minipage}
\caption{$\frac{1}{N_f N_c} \langle {\bar \psi} \psi \rangle$ vs. $N_f$ including area $1$ diagrams up to order $L = 4, 6, 8$ for solution $1$ (left), and solution $2$ (right).}
\label{sols}
\end{figure}

The results for each solution of $\frac{1}{N_f N_c}\langle {\bar \psi} \psi \rangle$ as a function of $N_f$  from (\ref{ga-all}) - (\ref{gg-all}) for the $L = 8$ truncation are reproduced in Figure \ref{sols}, along with those from the $L = 6$ truncation in (\ref{ga-2}) - (\ref{gd-2}), and from the $L = 4$ truncation in (\ref{ga-1}) - (\ref{gb-1}), showing how the solutions change as a function of the truncation order $L$. In both cases the solution appears to be converging, at least for the smaller values of $N_f$.

\begin{figure}[h!]
\begin{minipage}{0.5\textwidth}
\includegraphics[scale=0.55]{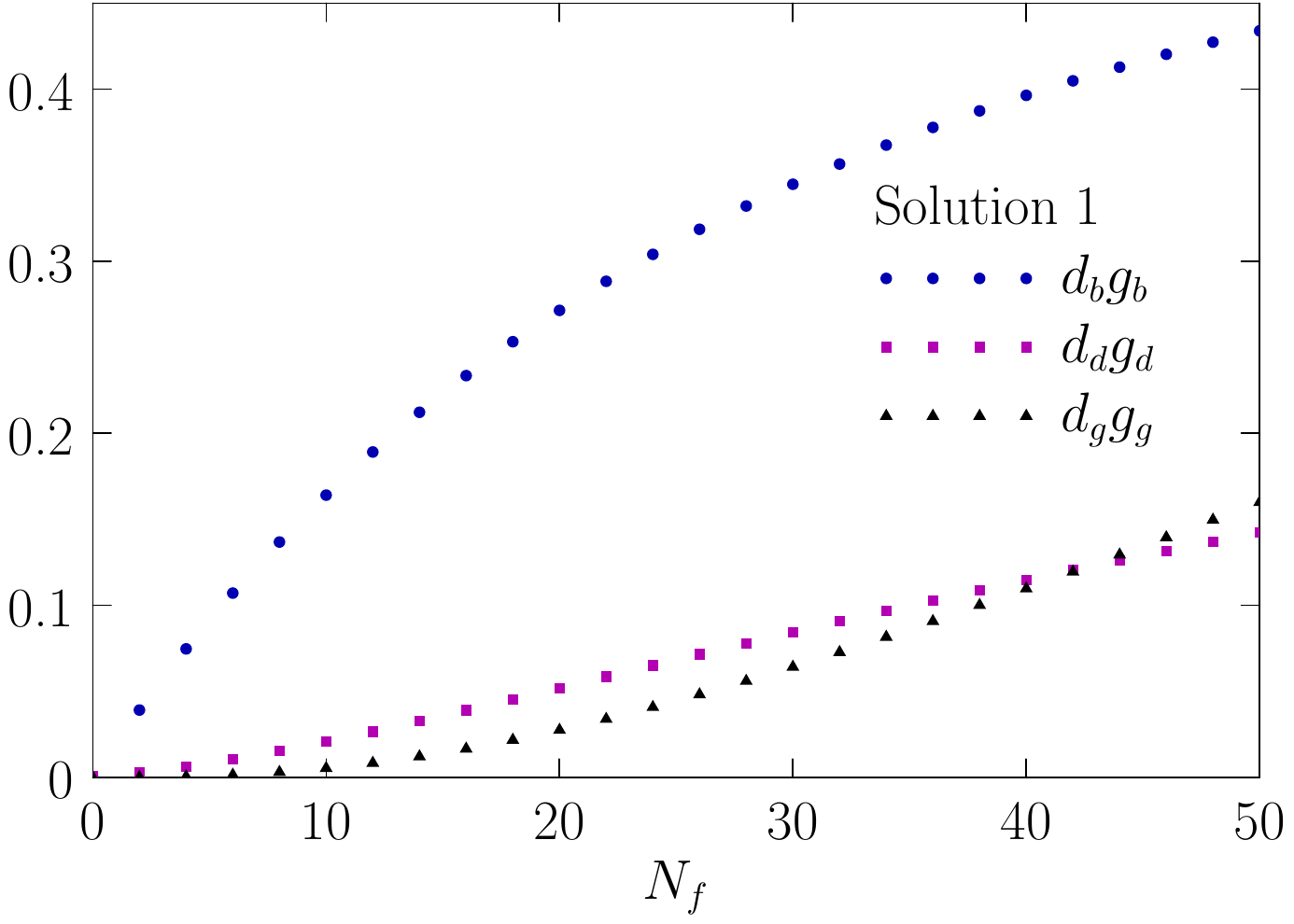}
\end{minipage}
\begin{minipage}{0.5\textwidth}
\includegraphics[scale=0.55]{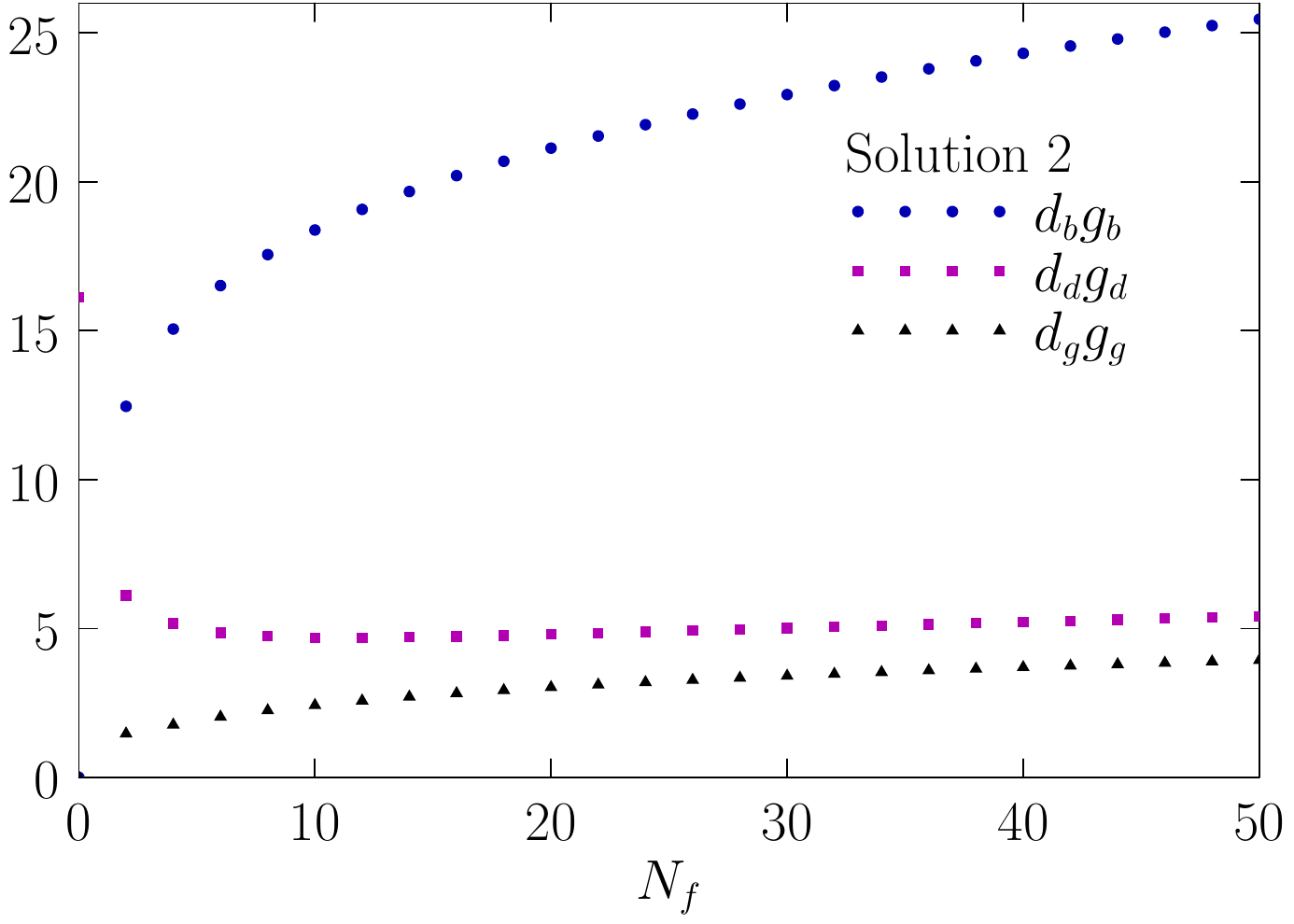}
\end{minipage}
\caption{$g_x$ vs. $N_f$ including area $1$ diagrams up to order $L = 8$ for solution $1$ (left), and solution $2$ (right).}
\label{g-figs}
\end{figure}

Finally, to check convergence, the values of each contribution $d_b g_b$, $d_d g_d$, $d_g g_g$ to (\ref{gser}), that solve the system of equations in (\ref{ga-all}) - (\ref{gg-all}), are plotted in Figure \ref{g-figs}. While in both solutions the higher order contributions from $d_d g_d$ and $d_g g_g$ are smaller in magnitude, the contributions have the potential to become more significant at larger values of $N_f$, since $g_d$ goes like $N_f^2$ (\ref{gd-all}), and $g_g$ goes like $N_f^3$ (\ref{gg-all}).

\subsection{Restricting to reduced graphs}

In order to compare with \cite{Tomboulis:2012nr} we now examine the effects of allowing only reduced graphs, i.e. graphs where each closed loop is separated from all other closed loops as well as the origin by at least one double link. The set of reduced graphs can be obtained by modifying the diagrams used in the construction by inserting extra double links separating the loops. Since reduced graphs are already included in the building of graphs, there is no reason to discard the unreduced graphs in our approach. Furthermore, due to the extra double links the reduced diagrams will have higher powers of $(a_n g_a + b_n g_b + ...)^{-1}$ and could as such be subdominant compared to the corresponding unreduced diagrams, by the arguments at the end of Section \ref{martin-siu}. Regardless of this we will calculate the chiral condensate with a restriction to reduced graphs in order to compare with the results of \cite{Tomboulis:2012nr}. 

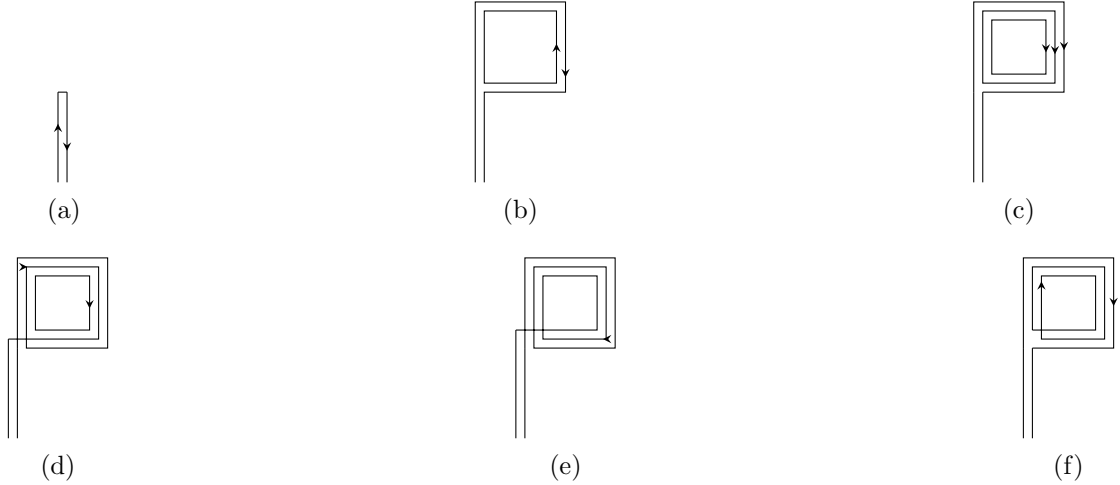
\begin{figure}
\centering
\subfloat[.5\textwidth][]{
\centering
\begin{tikzpicture}[scale=0.6]
\draw [directed] (0.0,0.0) -- (0.0,2.0);
\draw (0.0,2.0) -- (0.2,2.0);
\draw [directed] (0.2,2.0) -- (0.2,0.0);
\end{tikzpicture}
\label{reda}}
\hspace{5cm}
\subfloat[.5\textwidth][]{
\centering
\begin{tikzpicture}[scale=0.6]
\draw [directed] (0.0,-2.0) --(0.0,0.0) -- (0.0,2.0) -- (2.0,2.0) -- (2.0,0.0) -- (0.2,0.0)-- (0.2,-2.0);
\draw [reverse directed] (0.2,0.2) rectangle (1.8,1.8);

\end{tikzpicture}
\label{redb}}
\hspace{5cm}
\subfloat[.5\textwidth][]{
\centering
\begin{tikzpicture}[scale=0.6]
\draw [directed] (0.0,0.0) -- (0.0,2.0) -- (2.0,2.0) -- (2.0,0.0) -- (0.2,0.0);
\draw [directed] (0.2,0.2) rectangle (1.8,1.8);
\draw [directed] (0.4,0.4) rectangle (1.6,1.6);

\draw (0.0,-2.0) --(0.0,0.0);
\draw(0.2,0.0)-- (0.2,-2.0);
\end{tikzpicture}
\label{redc}}
\hspace{5cm}
\subfloat[.5\textwidth][]{
\centering
\begin{tikzpicture}[scale=0.6]
\draw [directed] (0.0,0.0) -- (0.0,2.0) -- (2.0,2.0) -- (2.0,0.0) -- (0.2,0.0) -- (0.2,1.8) -- (1.8,1.8) -- (1.8,0.2) -- (-0.2,0.2);
\draw [directed] (0.4,0.4) rectangle (1.6,1.6);

\draw (0.0,-2.0) --(0.0,0.0);
\draw(-0.2,0.2)-- (-0.2,-2.0);
\end{tikzpicture}
\label{redd}}
\hspace{5cm}
\subfloat[.5\textwidth][]{
\centering
\begin{tikzpicture}[scale=0.6]
\draw [directed] (0.0,0.0) -- (0.0,2.0) -- (2.0,2.0) -- (2.0,0.0) -- (0.2,0.0)
                             -- (0.2,1.8) -- (1.8,1.8) -- (1.8,0.2) -- (0.4,0.2)
                             -- (0.4,1.6) -- (1.6,1.6) -- (1.6,0.4) -- (-0.2,0.4);

\draw (0.0,-2.0) --(0.0,0.0);
\draw(-0.2,0.4)-- (-0.2,-2.0);
\end{tikzpicture}
\label{rede}}
\hspace{5cm}
\subfloat[.5\textwidth][]{
\centering
\begin{tikzpicture}[scale=0.6]
\draw [directed] (0.0,0.0) -- (0.0,2.0) -- (2.0,2.0) -- (2.0,0.0) -- (0.2,0.0);
\draw [directed] (0.2,0.4) -- (0.2,1.8) -- (1.8,1.8) -- (1.8,0.2) -- (0.4,0.2) -- (0.4,1.6) -- (1.6,1.6) -- (1.6,0.4) -- (0.2,0.4);

\draw (0.0,-2.0) --(0.0,0.0);
\draw(0.2,0.0)-- (0.2,-2.0);
\end{tikzpicture}
\label{redf}}
\caption{Lowest order diagrams used for constructing the set of reduced graphs.}
\label{reduced}
\end{figure}

\begin{figure}[h!]
\begin{minipage}{0.5\textwidth}
\includegraphics[scale=0.55]{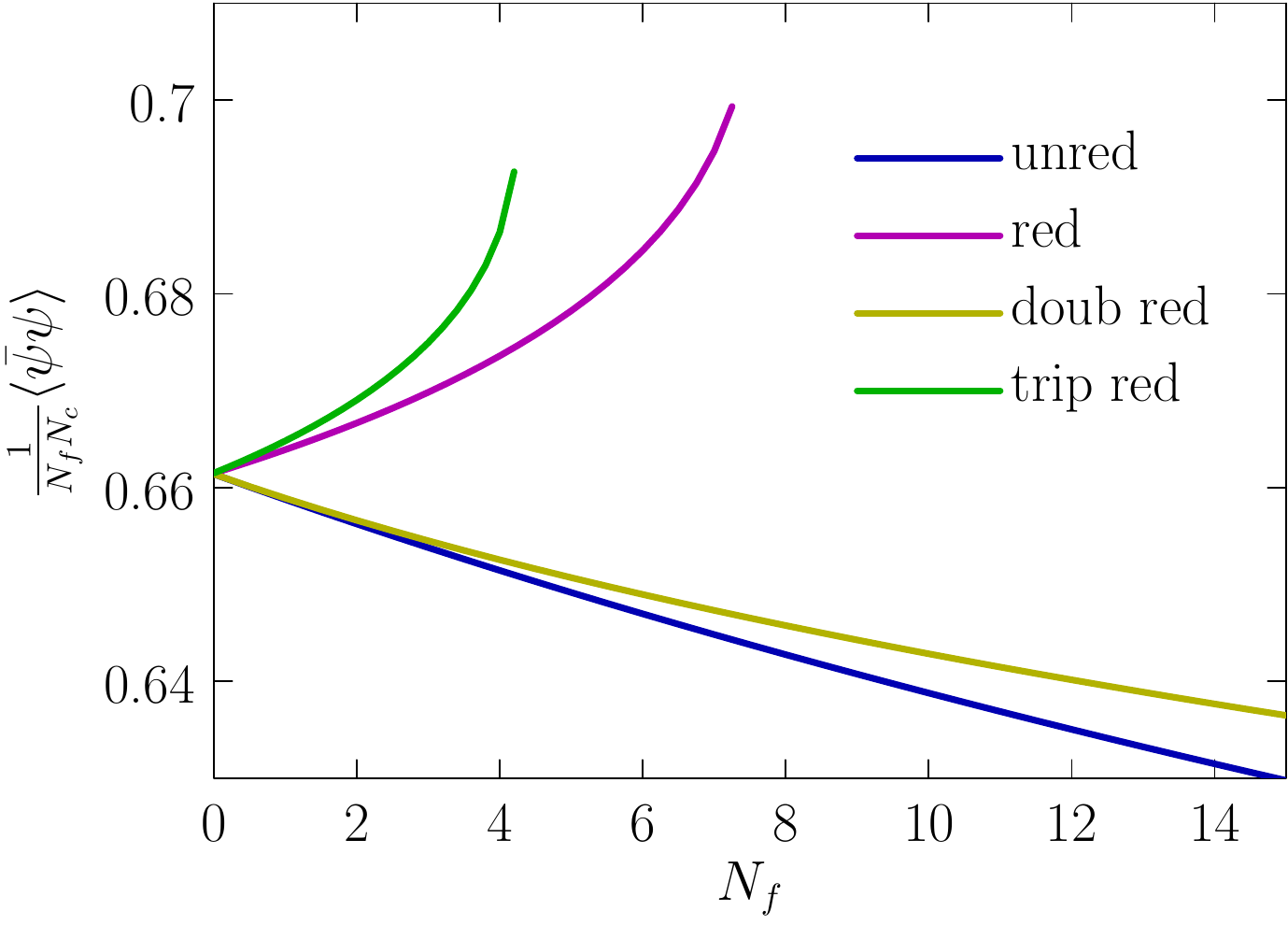}
\end{minipage}
\begin{minipage}{0.5\textwidth}

\end{minipage}
\caption{$\frac{1}{N_f N_c} \langle {\bar \psi} \psi \rangle$ for unreduced, reduced, doubly reduced, and triply reduced graphs.}
\label{reducedplot}
\end{figure}

The lowest order base diagrams in Section \ref{diags} are modified as shown in Figure \ref{reduced}. These imply the following set of equations for the set of reduced graphs (where $g_d=0$ for $N_c\ne3$).
\begin{align}
g_a & = \frac{1}{a_0 g_a + b_0 g_b+b_0g_d} \,,  \label{eq1}\\
g_b & = - \frac{N_f/N_c}{\left(a_1 g_a + d_F g_b+ d_F g_d \right)^8 \left(a_1' g_a + d_F g_b+ d_F g_d \right)} \label{eq2} \,,\\
g_d & = - \frac{\frac{1}{3}(\frac{1}{2}N_f^2+2N_f+1)}{\left(a_1 g_a + d_F g_b+ d_F g_d \right)^{12} \left(a_1' g_a + d_F g_b+ d_F g_d \right)} \label{eq3} \,,\end{align}
and
\SP{\frac{\VEV{\bar{\psi \psi}}}{N_cN_f}=\frac{2}{d_a g_a + d_F g_b + d_F g_d}\, ,}
where $d_F=8d^2(d-1)$ is the dimensionality of a flag type diagram (Figures \ref{redb}-\ref{redf}). Solving these equations for $N_c=3$ and $d=4$ we find the chiral condensate for reduced graphs as shown in Figure \ref{reducedplot} compared to the condensate including the (partially) unreduced graphs built from the base diagrams of Section \ref{diags} \footnote{Note that Figure \ref{reducedplot} only shows one of the two solutions. For the second solution we do not observe a clear trend but the critical $N_f$ are the same in the cases where there is one.}. We see that the $N_f\to0$ limit is unchanged, as expected since the only diagrams that have been changed are those that depend on $N_f$. What is different is that when excluding unreduced graphs, the chiral condensate increases with $N_f$, and at $N_f\gapprox8$ staggered flavours it turns complex. 

In order to examine closer the effects of excluding graphs in the recursive building we define a doubly (triply) reduced graph as a graph where each closed loop is separated from any other closed loop and the origin by at least two (three) tree segments and so on. Now if reduced graphs were in fact dominant, then by the same arguments, doubly reduced graphs (which are clearly also reduced) would be dominant among the reduced graphs. The set of doubly reduced graphs is generated by attaching an extra double link on the diagrams in Figures \ref{redb}-\ref{redf}, which leads to a change of the sign in equations \ref{eq2} and \ref{eq3} as well as an increase in the dimensionality $d_F\to16d^3(d-1)$. Restricting to doubly reduced graphs results in
\begin{align}
g_a & = \frac{1}{a_0 g_a + b_0 g_b+b_0g_d} \,, \\
g_b & = \frac{N_f/N_c}{\left(a_1 g_a + d_F g_b+ d_F g_d \right)^{10} \left(a_1' g_a + d_F g_b+ d_F g_d \right)} \,,\\
g_d & = \frac{\frac{1}{3}(\frac{1}{2}N_f^2+2N_f+1)}{\left(a_1 g_a + d_F g_b+ d_F g_d \right)^{14} \left(a_1' g_a + d_F g_b+ d_F g_d \right)} \,.\end{align}
The chiral condensate including only doubly reduced graphs (see Figure \ref{reducedplot}) is again a decreasing function of $N_f$, which is real for all values of $N_f$, as was the case including unreduced diagrams, built from the base diagrams in Section \ref{diags}. 

Going one step further and restricting to triply reduced graphs, the sign of $g_b$ and $g_d$ changes again,
\begin{align}
g_a & = \frac{1}{a_0 g_a + b_0 g_b+b_0g_d} \,, \\
g_b & = - \frac{N_f/N_c}{\left(a_1 g_a + d_F g_b+ d_F g_d \right)^{12} \left(a_1' g_a + d_F g_b+ d_F g_d \right)} \,,\\
g_d & = - \frac{\frac{1}{3}(\frac{1}{2}N_f^2+2N_f+1)}{\left(a_1 g_a + d_F g_b+ d_F g_d \right)^{16} \left(a_1' g_a + d_F g_b+ d_F g_d \right)} \,,\end{align}
with $d_F=32d^4(d-1)$. As shown in Figure \ref{reducedplot}, the chiral condensate becomes an increasing function of $N_f$ turning complex already for $N_f\ge5$. This trend continues, such that for graphs reduced any even number of times we obtain a real decreasing condensate for all $N_f$, while for any odd number of double links separating the loops, the condensate turns complex at some critical $N_f$. This critical value of $N_f$ decreases with the number of required links, such that for quintuply reduced graphs and beyond the condensate is complex for all $N_f\gapprox1$. 

This suggests that the existence of a critical $N_f$ above which the chiral condensate is complex as found in \cite{Tomboulis:2012nr} is a direct consequence of the reduced approximation due to the change of sign in the equations for the $N_f$-dependent $g_x$ for type $x$-diagrams with an uneven number of double links attached to the loop diagrams. This is further supported by including (partially) unreduced graphs in the recursive building formulated instead as in \cite{Tomboulis:2012nr}. This leads to a normalized chiral condensate that decreases with $N_f$ and remains real for all numbers of flavours.

\section{Discussion and conclusions} \label{concl}

Overall, what we can conclude from our results for $\frac{1}{N_f N_c}\langle {\bar \psi} \psi \rangle$ is that the diagrammatic expansion appears to converge at smaller values of $N_f$, giving two real solutions in which the chiral condensate slowly approaches zero as a function of $N_f$, but does not exhibit discontinuities, other than the non-existence of solutions for non-integer $N_f$ in the $L = 8$ truncation. At each order, the contributions from new sub-diagrams come with the same sign \footnote{In general, we have not been able to find a diagram (which is not a superposition) that has a different sign.}. Our results indicate that the chiral condensate always decreases when more contributions are included, such that the solutions obtained appear to provide an upper bound. 

We have dealt with various sources of error resulting from over-counting, however certain errors remain difficult to avoid. In particular, we note that the contribution of mistakes due to non-factorisation of integrations of overlapping diagrams could be important (see Section \ref{overlap}) and we have not accounted for this effect in these results. In addition the effect of over-counting resulting from symmetries in winding diagrams (see Section \ref{winding}) should be investigated more thoroughly. This effect comes in at $L=6$ for $N_c = 3$ and at $L=8$ for $N_c > 3$. Finally, higher order graphs can become important at larger $N_f$ so there is still room for interesting behaviour in this regime. We leave the precise quantification of these errors for future research.

We believe the differences from \cite{Tomboulis:2012nr} are as follows. The most clear difference is that we have included more contributions. Our calculations include higher order contributions up to $L=8$, \cite{Tomboulis:2012nr} includes contributions up to $L=4$. Another difference is that we allow area $1$ diagrams to attach directly to each other, resulting in ``unreduced" and ``partially reduced" graphs, using the terminology in \cite{Tomboulis:2012nr}. Another notable difference is that the type of attachments we use all come with a negative sign. We note, however, that additional counter-diagrams could be added to correct for mis-counted overlapping diagrams and some of these would come in with a positive sign. 

Higher dimensional representation fermions such as the symmetric, antisymmetric, and adjoint can also be considered however the calculations of diagrams with gauge fields in higher dimensional representations is not simply a replacement of all instances of $N_c$ with $d_R$. This is an interesting topic which we are currently investigating.

\section*{Acknowledgements}

We would like to thank Poul Damgaard, Matti J\"{a}rvinen, Seyong Kim, Kim Splittorff, and Ben Svetitsky for useful discussions. JCM would like to thank the Sapere Aude program of the Danish Council for Independent Research for supporting this work. The work of JR is supported by the START project Y 435-N16 of the Austrian Science Fund (FWF).

\appendix

\section{Dimensionalities}
\label{A}

The dimensionalities $x_n$ are the number of ways to attach a diagram of type $x$ to a graph with area $n$ where the appropriate dimensionality is subtracted off $d_x$ to prevent over-counting, as explained in Section \ref{over-counting}. The dimensionalities in this section are relevant for the diagrams obtained in Section \ref{diags}. 

\begin{minipage}[t]{0.5\textwidth}

\begin{tabular}[h]{|l|l|}
\hline
$a_0$ & $2d-1$ \\
\hline
$b_0$ & $4(d-1)^2$ \\
\hline
$d_0$ & $4(d-1)^2$ \\
\hline
$g_0$ & $4 (d-1)^2$ \\ 
\hline
\end{tabular}

\end{minipage}
\begin{minipage}[t]{0.5\textwidth}

\begin{tabular}[h]{|l|l|}
\hline
$a_1$ & $2d-1$ \\
\hline
$a_1'$ & $2(d-1)$ \\
\hline
$b_1$ & $4(d-1)^2$ \\
\hline
$b_1'$ & $4d^2 - 6d + 1$ \\
\hline
$d_1$ &  $4(d-1)^2$ \\ 
\hline
$d_1'$ &  $4d^2 - 6d + 1$ \\ 
\hline
$g_1$ &  $4(d-1)^2$ \\ 
\hline
$g_1'$ &  $4d^2 - 6d + 1$ \\ 
\hline

\end{tabular}

\end{minipage}\\ \vspace{3mm}






The dimensional prefactors, $d_x$, correspond to the total number of ways an $x$-type diagram can be placed on the lattice. These are listed in the following table:

\noindent
\begin{tabular}[h]{|l|l|}
\hline
$d_a$ & $2d$ \\
\hline
$d_b$ & $4d(d-1)$ \\
\hline
$d_c$ & $12d(d-1)(2d-3)$ \\
\hline
$d_d$ & $4d(d-1)$ \\
\hline
$d_e$ & $12d(d-1)(2d-3)$ \\
\hline
$d_f$ & $48d(d-1)(2d-3)^2$ \\
\hline
$d_g$ & $4d(d-1)$ \\
\hline
\end{tabular}

\section{Calculation of $I_4$}
\label{B}
\allowdisplaybreaks
The integral $I_4$
\begin{equation}
I_4 = \int_{\mathrm{SU}(N_c)} dU \, \Useries{U}{4}{a}{b} \Useries{(U^\dag)}{4}{c}{d} \,,
\end{equation}
can be calculated as explained in section \ref{groupsection}. One makes use of the decomposition 
\begin{align}
\ytableausetup{aligntableaux=center}
&  \ytableaushort[a_]{1} \otimes \ytableaushort[a_]{2} \otimes  \ytableaushort[a_]{3} \otimes \ytableaushort[a_]{4} = \ytableaushort[a_]{1234}\ (S)  \oplus \ytableaushort[a_]{134,2}\ (H_1)\oplus \ytableaushort[a_]{124,3}\ (H_2) \nonumber \\ & \quad \oplus \ytableaushort[a_]{123,4}\ (H_3) \oplus \ytableaushort[a_]{13,24}\ (B_1) \oplus \ytableaushort[a_]{12,34}\ (B_2) \oplus \ytableaushort[a_]{14,2,3}\ (V_1) \oplus \ytableaushort[a_]{13,2,4}\ (V_2) \nonumber \\ & \quad \oplus \ytableaushort[a_]{12,3,4} \ (V_3) \oplus \ytableaushort[a_]{1,2,3,4}\ (A)\,.
\end{align}
In the following, the representations that appear in the right hand side of this equation will be denoted by the symbols in brackets (following their respective Young tableaux).

We can then define the Young projectors, that project onto the standard Young tableaux in the right hand side of the above equation. They are explicitly given by 
\begin{align}
\proj{S}{4}{a}{b}& = \delta_{a_1}^{(b_1} \delta_{a_2}^{b_2} \delta_{a_3}^{b_3} \delta_{a_4}^{b_4)} \,, \nonumber \\
\proj{H_1}{4}{a}{b} & = \frac{1}{8} \Big(\delta_{a _1}^{b _4} \delta_{a _2}^{b _2} \delta_{a _3}^{b _3} \delta_{a _4}^{b _1}-\delta_{a _1}^{b _2} \delta_{a _2}^{b _4} \delta_{a _3}^{b _3} \delta_{a _4}^{b _1}+\delta_{a _1}^{b _3} \delta_{a _2}^{b _2} \delta_{a _3}^{b _4} \delta_{a _4}^{b _1}-\delta_{a _1}^{b _2} \delta_{a _2}^{b _3} \delta_{a _3}^{b _4} \delta_{a _4}^{b _1}\nonumber \\ & \ \ \ +\delta_{a _1}^{b _4} \delta_{a _2}^{b _2} \delta_{a _3}^{b _1} \delta_{a _4}^{b _3}-\delta_{a _1}^{b _2} \delta_{a _2}^{b _4} \delta_{a _3}^{b _1} \delta_{a _4}^{b _3}-\delta_{a _1}^{b _2} \delta_{a _2}^{b _1} \delta_{a _3}^{b _4} \delta_{a _4}^{b _3}  +\delta_{a _1}^{b _1} \delta_{a _2}^{b _2} \delta_{a _3}^{b _4} \delta_{a _4}^{b _3} \nonumber \\ & \ \ \ +\delta_{a _1}^{b _3} \delta_{a _2}^{b _2} \delta_{a _3}^{b _1} \delta_{a _4}^{b _4}-\delta_{a _1}^{b _2} \delta_{a _2}^{b _3} \delta_{a _3}^{b _1} \delta_{a _4}^{b _4}-\delta_{a _1}^{b _2} \delta_{a _2}^{b _1} \delta_{a _3}^{b _3} \delta_{a _4}^{b _4}+\delta_{a _1}^{b _1} \delta_{a _2}^{b _2} \delta_{a _3}^{b _3} \delta_{a _4}^{b _4}\Big) \,, \nonumber \\ 
\proj{H_2}{4}{a}{b}& = \frac{1}{8} \Big(-\delta_{a _1}^{b _3} \delta_{a _2}^{b _4} \delta_{a _3}^{b _2} \delta_{a _4}^{b _1}+\delta_{a _1}^{b _4} \delta_{a _2}^{b _2} \delta_{a _3}^{b _3} \delta_{a _4}^{b _1}+\delta_{a _1}^{b _2} \delta_{a _2}^{b _4} \delta_{a _3}^{b _3} \delta_{a _4}^{b _1}-\delta_{a _1}^{b _3} \delta_{a _2}^{b _2} \delta_{a _3}^{b _4} \delta_{a _4}^{b _1}\nonumber \\ & \ \ \ -\delta_{a _1}^{b _3} \delta_{a _2}^{b _4} \delta_{a _3}^{b _1} \delta_{a _4}^{b _2}+\delta_{a _1}^{b _4} \delta_{a _2}^{b _1} \delta_{a _3}^{b _3} \delta_{a _4}^{b _2}+\delta_{a _1}^{b _1} \delta_{a _2}^{b _4} \delta_{a _3}^{b _3} \delta_{a _4}^{b _2}-\delta_{a _1}^{b _3} \delta_{a _2}^{b _1} \delta_{a _3}^{b _4} \delta_{a _4}^{b _2}\nonumber \\ & \ \ \ -\delta_{a _1}^{b _3} \delta_{a _2}^{b _2} \delta_{a _3}^{b _1} \delta_{a _4}^{b _4}-\delta_{a _1}^{b _3} \delta_{a _2}^{b _1} \delta_{a _3}^{b _2} \delta_{a _4}^{b _4}+\delta_{a _1}^{b _2} \delta_{a _2}^{b _1} \delta_{a _3}^{b _3} \delta_{a _4}^{b _4}+\delta_{a _1}^{b _1} \delta_{a _2}^{b _2} \delta_{a _3}^{b _3} \delta_{a _4}^{b _4}\Big)\,, \nonumber \\
\proj{H_3}{4}{a}{b} & = \frac{1}{8} \Big(-\delta_{a _1}^{b _4} \delta_{a _2}^{b _3} \delta_{a _3}^{b _2} \delta_{a _4}^{b _1}-\delta_{a _1}^{b _4} \delta_{a _2}^{b _2} \delta_{a _3}^{b _3} \delta_{a _4}^{b _1}-\delta_{a _1}^{b _4} \delta_{a _2}^{b _3} \delta_{a _3}^{b _1} \delta_{a _4}^{b _2}-\delta_{a _1}^{b _4} \delta_{a _2}^{b _1} \delta_{a _3}^{b _3} \delta_{a _4}^{b _2}\nonumber \\ & \ \ \ -\delta_{a _1}^{b _4} \delta_{a _2}^{b _2} \delta_{a _3}^{b _1} \delta_{a _4}^{b _3}-\delta_{a _1}^{b _4} \delta_{a _2}^{b _1} \delta_{a _3}^{b _2} \delta_{a _4}^{b _3}+\delta_{a _1}^{b _3} \delta_{a _2}^{b _2} \delta_{a _3}^{b _1} \delta_{a _4}^{b _4}+\delta_{a _1}^{b _2} \delta_{a _2}^{b _3} \delta_{a _3}^{b _1} \delta_{a _4}^{b _4}\nonumber \\ & \ \ \ +\delta_{a _1}^{b _3} \delta_{a _2}^{b _1} \delta_{a _3}^{b _2} \delta_{a _4}^{b _4}+\delta_{a _1}^{b _1} \delta_{a _2}^{b _3} \delta_{a _3}^{b _2} \delta_{a _4}^{b _4}+\delta_{a _1}^{b _2} \delta_{a _2}^{b _1} \delta_{a _3}^{b _3} \delta_{a _4}^{b _4}+\delta_{a _1}^{b _1} \delta_{a _2}^{b _2} \delta_{a _3}^{b _3} \delta_{a _4}^{b _4}\Big) \,, \nonumber \\
\proj{B_1}{4}{a}{b} & = 
\frac{1}{12} \Big(\delta_{a _1}^{b _4} \delta_{a _2}^{b _3} \delta_{a _3}^{b _2} \delta_{a _4}^{b _1}-\delta_{a _1}^{b _3} \delta_{a _2}^{b _4} \delta_{a _3}^{b _2} \delta_{a _4}^{b _1}-\delta_{a _1}^{b _3} \delta_{a _2}^{b _2} \delta_{a _3}^{b _4} \delta_{a _4}^{b _1}+\delta_{a _1}^{b _2} \delta_{a _2}^{b _3} \delta_{a _3}^{b _4} \delta_{a _4}^{b _1}\nonumber \\ & \ \ \ -\delta_{a _1}^{b _4} \delta_{a _2}^{b _3} \delta_{a _3}^{b _1} \delta_{a _4}^{b _2}+\delta_{a _1}^{b _3} \delta_{a _2}^{b _4} \delta_{a _3}^{b _1} \delta_{a _4}^{b _2}-\delta_{a _1}^{b _4} \delta_{a _2}^{b _1} \delta_{a _3}^{b _3} \delta_{a _4}^{b _2}+\delta_{a _1}^{b _1} \delta_{a _2}^{b _4} \delta_{a _3}^{b _3} \delta_{a _4}^{b _2}\nonumber \\ & \ \ \ +\delta_{a _1}^{b _4} \delta_{a _2}^{b _1} \delta_{a _3}^{b _2} \delta_{a _4}^{b _3}-\delta_{a _1}^{b _1} \delta_{a _2}^{b _4} \delta_{a _3}^{b _2} \delta_{a _4}^{b _3}+\delta_{a _1}^{b _2} \delta_{a _2}^{b _1} \delta_{a _3}^{b _4} \delta_{a _4}^{b _3}-\delta_{a _1}^{b _1} \delta_{a _2}^{b _2} \delta_{a _3}^{b _4} \delta_{a _4}^{b _3}\nonumber \\ & \ \ \ +\delta_{a _1}^{b _3} \delta_{a _2}^{b _2} \delta_{a _3}^{b _1} \delta_{a _4}^{b _4}-\delta_{a _1}^{b _2} \delta_{a _2}^{b _3} \delta_{a _3}^{b _1} \delta_{a _4}^{b _4}-\delta_{a _1}^{b _2} \delta_{a _2}^{b _1} \delta_{a _3}^{b _3} \delta_{a _4}^{b _4}+\delta_{a _1}^{b _1} \delta_{a _2}^{b _2} \delta_{a _3}^{b _3} \delta_{a _4}^{b _4}\Big) \,, \nonumber \\
\proj{B_2}{4}{a}{b} & = \frac{1}{12} \Big(\delta_{a _1}^{b _4} \delta_{a _2}^{b _3} \delta_{a _3}^{b _2} \delta_{a _4}^{b _1}+\delta_{a _1}^{b _3} \delta_{a _2}^{b _4} \delta_{a _3}^{b _2} \delta_{a _4}^{b _1}-\delta_{a _1}^{b _2} \delta_{a _2}^{b _4} \delta_{a _3}^{b _3} \delta_{a _4}^{b _1}-\delta_{a _1}^{b _2} \delta_{a _2}^{b _3} \delta_{a _3}^{b _4} \delta_{a _4}^{b _1}\nonumber \\ & \ \ \ +\delta_{a _1}^{b _4} \delta_{a _2}^{b _3} \delta_{a _3}^{b _1} \delta_{a _4}^{b _2}+\delta_{a _1}^{b _3} \delta_{a _2}^{b _4} \delta_{a _3}^{b _1} \delta_{a _4}^{b _2}-\delta_{a _1}^{b _1} \delta_{a _2}^{b _4} \delta_{a _3}^{b _3} \delta_{a _4}^{b _2}-\delta_{a _1}^{b _1} \delta_{a _2}^{b _3} \delta_{a _3}^{b _4} \delta_{a _4}^{b _2}\nonumber \\ & \ \ \ -\delta_{a _1}^{b _4} \delta_{a _2}^{b _2} \delta_{a _3}^{b _1} \delta_{a _4}^{b _3}-\delta_{a _1}^{b _4} \delta_{a _2}^{b _1} \delta_{a _3}^{b _2} \delta_{a _4}^{b _3}+\delta_{a _1}^{b _2} \delta_{a _2}^{b _1} \delta_{a _3}^{b _4} \delta_{a _4}^{b _3}+\delta_{a _1}^{b _1} \delta_{a _2}^{b _2} \delta_{a _3}^{b _4} \delta_{a _4}^{b _3}\nonumber \\ & \ \ \ -\delta_{a _1}^{b _3} \delta_{a _2}^{b _2} \delta_{a _3}^{b _1} \delta_{a _4}^{b _4}-\delta_{a _1}^{b _3} \delta_{a _2}^{b _1} \delta_{a _3}^{b _2} \delta_{a _4}^{b _4}+\delta_{a _1}^{b _2} \delta_{a _2}^{b _1} \delta_{a _3}^{b _3} \delta_{a _4}^{b _4}+\delta_{a _1}^{b _1} \delta_{a _2}^{b _2} \delta_{a _3}^{b _3} \delta_{a _4}^{b _4}\Big)\,, \nonumber \\ 
\proj{V_1}{4}{a}{b} & = \frac{1}{8} \Big(-\delta_{a _1}^{b _4} \delta_{a _2}^{b _3} \delta_{a _3}^{b _2} \delta_{a _4}^{b _1}+\delta_{a _1}^{b _3} \delta_{a _2}^{b _4} \delta_{a _3}^{b _2} \delta_{a _4}^{b _1}+\delta_{a _1}^{b _4} \delta_{a _2}^{b _2} \delta_{a _3}^{b _3} \delta_{a _4}^{b _1}-\delta_{a _1}^{b _2} \delta_{a _2}^{b _4} \delta_{a _3}^{b _3} \delta_{a _4}^{b _1}\nonumber \\ & \ \ \ -\delta_{a _1}^{b _3} \delta_{a _2}^{b _2} \delta_{a _3}^{b _4} \delta_{a _4}^{b _1}+\delta_{a _1}^{b _2} \delta_{a _2}^{b _3} \delta_{a _3}^{b _4} \delta_{a _4}^{b _1}-\delta_{a _1}^{b _3} \delta_{a _2}^{b _2} \delta_{a _3}^{b _1} \delta_{a _4}^{b _4}+\delta_{a _1}^{b _2} \delta_{a _2}^{b _3} \delta_{a _3}^{b _1} \delta_{a _4}^{b _4}\nonumber \\ & \ \ \ +\delta_{a _1}^{b _3} \delta_{a _2}^{b _1} \delta_{a _3}^{b _2} \delta_{a _4}^{b _4}-\delta_{a _1}^{b _1} \delta_{a _2}^{b _3} \delta_{a _3}^{b _2} \delta_{a _4}^{b _4}-\delta_{a _1}^{b _2} \delta_{a _2}^{b _1} \delta_{a _3}^{b _3} \delta_{a _4}^{b _4}+\delta_{a _1}^{b _1} \delta_{a _2}^{b _2} \delta_{a _3}^{b _3} \delta_{a _4}^{b _4}\Big)\,, \nonumber \\
\proj{V_2}{4}{a}{b} & = \frac{1}{8} \Big(-\delta_{a _1}^{b _4} \delta_{a _2}^{b _2} \delta_{a _3}^{b _3} \delta_{a _4}^{b _1}+\delta_{a _1}^{b _2} \delta_{a _2}^{b _4} \delta_{a _3}^{b _3} \delta_{a _4}^{b _1}+\delta_{a _1}^{b _4} \delta_{a _2}^{b _3} \delta_{a _3}^{b _1} \delta_{a _4}^{b _2}-\delta_{a _1}^{b _3} \delta_{a _2}^{b _4} \delta_{a _3}^{b _1} \delta_{a _4}^{b _2}\nonumber \\ & \ \ \ +\delta_{a _1}^{b _4} \delta_{a _2}^{b _1} \delta_{a _3}^{b _3} \delta_{a _4}^{b _2}-\delta_{a _1}^{b _1} \delta_{a _2}^{b _4} \delta_{a _3}^{b _3} \delta_{a _4}^{b _2}-\delta_{a _1}^{b _4} \delta_{a _2}^{b _2} \delta_{a _3}^{b _1} \delta_{a _4}^{b _3}+\delta_{a _1}^{b _2} \delta_{a _2}^{b _4} \delta_{a _3}^{b _1} \delta_{a _4}^{b _3}\nonumber \\ & \ \ \ +\delta_{a _1}^{b _3} \delta_{a _2}^{b _2} \delta_{a _3}^{b _1} \delta_{a _4}^{b _4}-\delta_{a _1}^{b _2} \delta_{a _2}^{b _3} \delta_{a _3}^{b _1} \delta_{a _4}^{b _4}-\delta_{a _1}^{b _2} \delta_{a _2}^{b _1} \delta_{a _3}^{b _3} \delta_{a _4}^{b _4}+\delta_{a _1}^{b _1} \delta_{a _2}^{b _2} \delta_{a _3}^{b _3} \delta_{a _4}^{b _4}\Big) \,, \nonumber \\
\proj{V_3}{4}{a}{b} & =\frac{1}{8} \Big(-\delta _{a _1}^{b _4} \delta _{a _2}^{b _2} \delta _{a _3}^{b _3} \delta _{a _4}^{b _1}+\delta _{a _1}^{b _3} \delta _{a _2}^{b _2} \delta _{a _3}^{b _4} \delta _{a _4}^{b _1}-\delta _{a _1}^{b _4} \delta _{a _2}^{b _1} \delta _{a _3}^{b _3} \delta _{a _4}^{b _2}+\delta _{a _1}^{b _3} \delta _{a _2}^{b _1} \delta _{a _3}^{b _4} \delta _{a _4}^{b _2}\nonumber \\ & \ \ \ +\delta_{a _1}^{b _4} \delta_{a _2}^{b _2} \delta_{a _3}^{b _1} \delta_{a _4}^{b _3}+\delta_{a _1}^{b _4} \delta_{a _2}^{b _1} \delta_{a _3}^{b _2} \delta_{a _4}^{b _3}-\delta_{a _1}^{b _2} \delta_{a _2}^{b _1} \delta_{a _3}^{b _4} \delta_{a _4}^{b _3}-\delta_{a _1}^{b _1} \delta_{a _2}^{b _2} \delta_{a _3}^{b _4} \delta_{a _4}^{b _3}\nonumber \\ & \ \ \ -\delta_{a _1}^{b _3} \delta_{a _2}^{b _2} \delta_{a _3}^{b _1} \delta_{a _4}^{b _4}-\delta_{a _1}^{b _3} \delta_{a _2}^{b _1} \delta_{a _3}^{b _2} \delta_{a _4}^{b _4}+\delta_{a _1}^{b _2} \delta_{a _2}^{b _1} \delta_{a _3}^{b _3} \delta_{a _4}^{b _4}+\delta_{a _1}^{b _1} \delta_{a _2}^{b _2} \delta_{a _3}^{b _3} \delta_{a _4}^{b _4}\Big) \,, \nonumber \\
\proj{A}{4}{a}{b} & = \delta_{a_1}^{[b_1} \delta_{a_2}^{b_2} \delta_{a_3}^{b_3} \delta_{a_4}^{b_4]} \,,
\end{align}
where $(\cdots)$ and $[\cdots]$ in $\mathbb{P}^S$ and $\mathbb{P}^A$ denote complete symmetrization and antisymmetrization (with weight 1, i.e. each term appears with a prefactor $1/24$) respectively.

Using these projectors, the matrix representatives of the irreducible representations $S$, $H_1$, $H_2$, $H_3$, $B_1$, $B_2$, $V_1$, $V_2$, $V_3$, $A$ can be constructed in terms of the matrices $U_a^{\ b}$ in the fundamental representation
\begin{align}
\tens{S}{4}{a}{b} & = \proj{S}{4}{a}{c}\left(\Useries{U}{4}{c}{d}\right) \proj{S}{4}{d}{b}\,, \nonumber \\
\tens{{H_1}}{4}{a}{b} & = \proj{{H_1}}{4}{a}{c}\left(\Useries{U}{4}{c}{d}\right) \proj{{H_1}}{4}{d}{b}\,, \nonumber \\
\tens{{H_2}}{4}{a}{b} & = \proj{{H_2}}{4}{a}{c}\left(\Useries{U}{4}{c}{d}\right) \proj{{H_2}}{4}{d}{b} \,, \nonumber \\
\tens{{H_3}}{4}{a}{b} & = \proj{{H_3}}{4}{a}{c}\left(\Useries{U}{4}{c}{d}\right) \proj{{H_3}}{4}{d}{b} \,, \nonumber \\
\tens{{B_1}}{4}{a}{b} & = \proj{{B_1}}{4}{a}{c}\left(\Useries{U}{4}{c}{d}\right) \proj{{B_1}}{4}{d}{b} \,, \nonumber \\
\tens{{B_2}}{4}{a}{b} & = \proj{{B_2}}{4}{a}{c}\left(\Useries{U}{4}{c}{d}\right) \proj{{B_2}}{4}{d}{b} \,, \nonumber \\
\tens{{V_1}}{4}{a}{b} & = \proj{{V_1}}{4}{a}{c}\left(\Useries{U}{4}{c}{d}\right) \proj{{V_1}}{4}{d}{b}\,, \nonumber \\
\tens{{V_2}}{4}{a}{b} & = \proj{{V_2}}{4}{a}{c}\left(\Useries{U}{4}{c}{d}\right) \proj{{V_2}}{4}{d}{b} \,, \nonumber \\
\tens{{V_3}}{4}{a}{b} & = \proj{{V_3}}{4}{a}{c}\left(\Useries{U}{4}{c}{d}\right) \proj{{V_3}}{4}{d}{b}\,, \nonumber \\
\tens{{A}}{4}{a}{b} & = \proj{{A}}{4}{a}{c}\left(\Useries{U}{4}{c}{d}\right) \proj{{A}}{4}{d}{b} \,.
\end{align}
One can then replace 
\begin{align}
\Useries{U}{4}{a}{b} &  = \tens{S}{4}{a}{b} + \tens{{H_1}}{4}{a}{b} + \tens{{H_2}}{4}{a}{b}+  \tens{{H_3}}{4}{a}{b} \nonumber \\ & \ \ \ + \tens{{B_1}}{4}{a}{b}  + \tens{{B_2}}{4}{a}{b}  +  \tens{{V_1}}{4}{a}{b} + \tens{{V_2}}{4}{a}{b} \nonumber \\ & \ \ \ + \tens{{V_3}}{4}{a}{b}  + \tens{{A}}{4}{a}{b}\,.
\end{align}
Using this in the original integral, one finds that it reduces to a sum of integrals of the form 
\begin{equation}
\int_{\mathrm{SU}(N_c)} dU \,  \tens{{R_1}}{4}{a}{b} \tens{{({R_2})^\dag}}{4}{c}{d} \,.
\end{equation}
This integral is zero when $R_1$ and $R_2$ do not have the same Young tableau structure, so a lot of the terms vanish automatically. In particular, one gets the following non-zero contributions:
\begin{align}
& \int_{\mathrm{SU}(N_c)} dU \, \tens{S}{4}{a}{b} \tens{{(S)^\dag}}{4}{c}{d}  = \frac{24}{N_c(N_c+1)(N_c+2)(N_c+3)} \proj{S}{4}{a}{d}\proj{S}{4}{c}{b} \,,  \nonumber \\ 
& \int_{\mathrm{SU}(N_c)} dU \,\tens{{H_1}}{4}{a}{b}\tens{{(H_1)^\dag}}{4}{c}{d}   = \frac{8}{N_c(N_c^2-1)(N_c+2)} \proj{H_1}{4}{a}{d} \proj{H_1}{4}{c}{b}\,,  \nonumber \\ 
& \int_{\mathrm{SU}(N_c)} dU \, \tens{{H_2}}{4}{a}{b}\tens{{(H_2)^\dag}}{4}{c}{d}  = \frac{8}{N_c(N_c^2-1)(N_c+2)} \proj{H_2}{4}{a}{d} \proj{H_2}{4}{c}{b}\,,  \nonumber \\ 
& \int_{\mathrm{SU}(N_c)}t dU \, \tens{{H_3}}{4}{a}{b}\tens{{(H_3)^\dag}}{4}{c}{d}  = \frac{8}{N_c(N_c^2-1)(N_c+2)} \proj{H_3}{4}{a}{d} \proj{H_3}{4}{c}{b}\,,  \nonumber \\ 
& \int_{\mathrm{SU}(N_c)} dU  \, \tens{{H_1}}{4}{a}{b} \tens{{(H_2)^\dag}}{4}{c}{d} = \frac{8}{N_c(N_c^2-1)(N_c+2)} \projnc{H_1}{4}{a}{d}{{1,3,2,4}} \projnc{H_2}{4}{c}{b}{{1,3,2,4}}\,,  \nonumber \\ 
& \int_{\mathrm{SU}(N_c)} dU \, \tens{{H_1}}{4}{a}{b} \tens{{(H_3)^\dag}}{4}{c}{d}  = \frac{8}{N_c(N_c^2-1)(N_c+2)} \projnc{H_1}{4}{a}{d}{{1,4,2,3}}\projnc{H_3}{4}{c}{b}{{1,3,4,2}}\,,  \nonumber \\ 
& \int_{\mathrm{SU}(N_c)} dU \,  \tens{{H_2}}{4}{a}{b} \tens{{(H_1)^\dag}}{4}{c}{d}  = \frac{8}{N_c(N_c^2-1)(N_c+2)} \projnc{H_2}{4}{a}{d}{{1,3,2,4}} \projnc{H_1}{4}{c}{b}{{1,3,2,4}}\,,  \nonumber \\ 
& \int_{\mathrm{SU}(N_c)} dU \, \tens{{H_2}}{4}{a}{b} \tens{{(H_3)^\dag}}{4}{c}{d}  = \frac{8}{N_c(N_c^2-1)(N_c+2)} \projnc{H_2}{4}{a}{d}{{1,2,4,3}} \projnc{H_3}{4}{c}{b}{{1,2,4,3}} \,,  \nonumber \\ 
& \int_{\mathrm{SU}(N_c)} dU \, \tens{{H_3}}{4}{a}{b} \tens{{(H_1)^\dag}}{4}{c}{d}  = \frac{8}{N_c(N_c^2-1)(N_c+2)} \projnc{H_3}{4}{a}{d}{{1,3,4,2}} \projnc{H_1}{4}{c}{b}{{1,4,2,3}} \,,  \nonumber \\ 
& \int_{\mathrm{SU}(N_c)} dU \, \tens{{H_3}}{4}{a}{b} \tens{{(H_2)^\dag}}{4}{c}{d} = \frac{8}{N_c(N_c^2-1)(N_c+2)} \projnc{H_3}{4}{a}{d}{{1,2,4,3}} \projnc{H_2}{4}{c}{b}{{1,2,4,3}} \,,  \nonumber \\ 
& \int_{\mathrm{SU}(N_c)} dU \, \tens{{B_1}}{4}{a}{b} \tens{{(B_1)^\dag}}{4}{c}{d}  = \frac{12}{N_c^2(N_c^2-1)} \proj{B_1}{4}{a}{d} \proj{B_1}{4}{c}{b}\,,  \nonumber \\ 
& \int_{\mathrm{SU}(N_c)} dU \, \tens{{B_2}}{4}{a}{b} \tens{{(B_2)^\dag}}{4}{c}{d}  = \frac{12}{N_c^2(N_c^2-1)} \proj{B_2}{4}{a}{d} \proj{B_2}{4}{c}{b}\,,  \nonumber \\ 
& \int_{\mathrm{SU}(N_c)} dU \,  \tens{{B_1}}{4}{a}{b} \tens{{(B_2)^\dag}}{4}{c}{d}  = \frac{12}{N_c^2(N_c^2-1)} \projnc{B_1}{4}{a}{d}{{1,3,2,4}} \projnc{B_2}{4}{c}{b}{{1,3,2,4}} \,,  \nonumber \\ 
& \int_{\mathrm{SU}(N_c)} dU \,  \tens{{B_2}}{4}{a}{b} \tens{{(B_1)^\dag}}{4}{c}{d}  = \frac{12}{N_c^2(N_c^2-1)} \projnc{B_2}{4}{a}{d}{{1,3,2,4}} \projnc{B_1}{4}{c}{b}{{1,3,2,4}}\,,  \nonumber \\ 
& \int_{\mathrm{SU}(N_c)} dU \, \tens{{V_1}}{4}{a}{b} \tens{{(V_1)^\dag}}{4}{c}{d} = \frac{8}{N_c(N_c^2-1)(N_c-2)} \proj{V_1}{4}{a}{d} \proj{V_1}{4}{c}{b}\,,  \nonumber \\ 
& \int_{\mathrm{SU}(N_c)} dU \, \tens{{V_2}}{4}{a}{b} \tens{{(V_2)^\dag}}{4}{c}{d}  = \frac{8}{N_c(N_c^2-1)(N_c-2)} \proj{V_2}{4}{a}{d} \proj{V_2}{4}{c}{b}\,,  \nonumber \\ 
& \int_{\mathrm{SU}(N_c)} dU \, \tens{{V_3}}{4}{a}{b} \tens{{(V_3)^\dag}}{4}{c}{d}  = \frac{8}{N_c(N_c^2-1)(N_c-2)} \proj{V_3}{4}{a}{d} \proj{V_3}{4}{c}{b}\,,  \nonumber \\ 
& \int_{\mathrm{SU}(N_c)} dU \, \tens{{V_1}}{4}{a}{b} \tens{{(V_2)^\dag}}{4}{c}{d}  = \frac{8}{N_c(N_c^2-1)(N_c-2)} \projnc{V_1}{4}{a}{d}{{1,2,4,3}}\projnc{V_2}{4}{c}{b}{{1,2,4,3}} \,,  \nonumber \\ 
& \int_{\mathrm{SU}(N_c)} dU \, \tens{{V_1}}{4}{a}{b} \tens{{(V_3)^\dag}}{4}{c}{d}  = \frac{8}{N_c(N_c^2-1)(N_c-2)} \projnc{V_1}{4}{a}{d}{{1,3,4,2}} \projnc{V_3}{4}{c}{b}{{1,4,2,3}} \,,  \nonumber \\ 
& \int_{\mathrm{SU}(N_c)} dU \, \tens{{V_2}}{4}{a}{b} \tens{{(V_1)^\dag}}{4}{c}{d} = \frac{8}{N_c(N_c^2-1)(N_c-2)} \projnc{V_2}{4}{a}{d}{{1,2,4,3}} \projnc{V_1}{4}{c}{b}{{1,2,4,3}} \,,  \nonumber \\ 
& \int_{\mathrm{SU}(N_c)} dU \, \tens{{V_2}}{4}{a}{b} \tens{{(V_3)^\dag}}{4}{c}{d}  = \frac{8}{N_c(N_c^2-1)(N_c-2)} \projnc{V_2}{4}{a}{d}{{1,3,2,4}} \projnc{V_3}{4}{c}{b}{{1,3,2,4}} \,,  \nonumber \\ 
& \int_{\mathrm{SU}(N_c)} dU \, \tens{{V_3}}{4}{a}{b} \tens{{(V_1)^\dag}}{4}{c}{d}  = \frac{8}{N_c(N_c^2-1)(N_c-2)} \projnc{V_3}{4}{a}{d}{{1,4,2,3}} \projnc{V_1}{4}{c}{b}{{1,3,4,2}} \,,  \nonumber \\
& \int_{\mathrm{SU}(N_c)} dU \, \tens{{V_3}}{4}{a}{b} \tens{{(V_2)^\dag}}{4}{c}{d}   = \frac{8}{N_c(N_c^2-1)(N_c-2)} \projnc{V_3}{4}{a}{d}{{1,3,2,4}} \projnc{V_2}{4}{c}{b}{{1,3,2,4}} \,,  \nonumber \\ 
& \int_{\mathrm{SU}(N_c)} dU \, \tens{A}{4}{a}{b} \tens{A}{4}{c}{d}  = \frac{24}{N_c(N_c-1)(N_c-2)(N_c-3)}\proj{A}{4}{a}{d} \proj{A}{4}{c}{b}\,.
\end{align}
The final result for $I_4$ is then given by the sum of all the above terms.





\begin{thebibliography}{10}

\bibitem{Deuzeman:2012ee}
A.~Deuzeman, M.~P. Lombardo, T.~Nunes Da~Silva  and E.~Pallante, \emph{{The
  bulk transition of QCD with twelve flavors and the role of improvement}},
  Phys.Lett. {\bf B720} (2013) 358--365,
\href{http://www.arXiv.org/abs/1209.5720}{{\tt 1209.5720}}

\bibitem{Cheng:2013xha}
A.~Cheng, A.~Hasenfratz, Y.~Liu, G.~Petropoulos  and D.~Schaich, \emph{{Finite
  size scaling of conformal theories in the presence of a near-marginal
  operator}}, Phys.Rev. {\bf D90} (2014) 014509,
\href{http://www.arXiv.org/abs/1401.0195}{{\tt 1401.0195}}

\bibitem{Fodor:2011tu}
Z.~Fodor, K.~Holland, J.~Kuti, D.~Nogradi, C.~Schroeder  {\em et al.},
  \emph{{Twelve massless flavors and three colors below the conformal window}},
  Phys.Lett. {\bf B703} (2011) 348--358,
\href{http://www.arXiv.org/abs/1104.3124}{{\tt 1104.3124}}

\bibitem{Lin:2012iw}
C.-J.~D. Lin, K.~Ogawa, H.~Ohki  and E.~Shintani, \emph{{Lattice study of
  infrared behaviour in SU(3) gauge theory with twelve massless flavours}},
  JHEP {\bf 1208} (2012) 096,
\href{http://www.arXiv.org/abs/1205.6076}{{\tt 1205.6076}}

\bibitem{Itou:2013faa}
E.~Itou, \emph{{The twisted Polyakov loop coupling and the search for an IR
  fixed point}},
\href{http://www.arXiv.org/abs/1311.2676}{{\tt 1311.2676}}

\bibitem{Bursa:2010xn}
F.~Bursa, L.~Del~Debbio, L.~Keegan, C.~Pica  and T.~Pickup, \emph{{Mass
  anomalous dimension in SU(2) with six fundamental fermions}}, Phys.Lett. {\bf
  B696} (2011) 374--379,
\href{http://www.arXiv.org/abs/1007.3067}{{\tt 1007.3067}}

\bibitem{KlubergStern:1982bs}
H.~Kluberg-Stern, A.~Morel  and B.~Petersson, \emph{{Spectrum of Lattice Gauge
  Theories with Fermions from a 1/D Expansion at Strong Coupling}}, Nucl.Phys.
  {\bf B215} (1983)
527

\bibitem{Blairon:1980pk}
J.~Blairon, R.~Brout, F.~Englert  and J.~Greensite, \emph{{Chiral Symmetry
  Breaking in the Action Formulation of Lattice Gauge Theory}}, Nucl.Phys. {\bf
  B180} (1981)
439

\bibitem{Martin:1982tb}
O.~Martin and B.~Siu, \emph{{Chiral Symmetry Breaking in Strongly Coupled
  Lattice Gauge Theory}}, Phys.Lett. {\bf B131} (1983)
419

\bibitem{deForcrand:2012vh}
P.~de~Forcrand, S.~Kim  and W.~Unger, \emph{{Conformality in many-flavour
  lattice QCD at strong coupling}}, JHEP {\bf 1302} (2013) 051,
\href{http://www.arXiv.org/abs/1208.2148}{{\tt 1208.2148}}

\bibitem{Damgaard:1985bn}
P.~Damgaard, D.~Hochberg  and N.~Kawamoto, \emph{{Effective Lagrangian Analysis
  of the Chiral Phase Transition at Finite Density}}, Phys.Lett. {\bf B158}
  (1985)
239

\bibitem{Tomboulis:2012nr}
E.~Tomboulis, \emph{{Absence of chiral symmetry breaking in multi-flavor
  strongly coupled lattice gauge theories}}, Phys.Rev. {\bf D87} (2013) 034513,
\href{http://www.arXiv.org/abs/1211.4842}{{\tt 1211.4842}}

\bibitem{Bars:1979xb}
I.~Bars and F.~Green, \emph{{Complete Integration of U($N$) Lattice Gauge
  Theory in a Large $N$ Limit}}, Phys.Rev. {\bf D20} (1979)
3311

\bibitem{Creutz:1984mg}
M.~Creutz, {\em {Quarks, Gluons and Lattices}}.
\newblock Cambridge Monographs on Mathematical Physics. Cambridge University
  Press,
1985

\bibitem{Cvitanovic:2008zz}
P.~Cvitanovic, {\em {Group theory: Birdtracks, Lie's and exceptional groups}}.
\newblock Princeton University Press,
2008

\bibitem{Wilson:1975id}
K.~G. Wilson,
\emph{{Quarks and Strings on a Lattice}}, CLNS-321, 1975

\bibitem{SimmondsMann}
J.~G. Simmonds and J.~E. Mann, {\em {A First Look at Perturbation Theory}}.
\newblock Dover Publications,
1986

\bibitem{BenderBook}
C.~M. Bender and S.~A. Orszag, {\em {Advanced Mathematical Methods for
  Scientists and Engineers}}.
\newblock McGraw-Hill,
1978

\end{thebibliography}

\providecommand{\href}[2]{#2}\begingroup\raggedright\endgroup

\end{document}